\documentclass[useAMS,usenatbib]{mn2e}
\usepackage{times}
\usepackage{graphicx}
\usepackage{amssymb}
\usepackage{longtable}

\title[What shapes a galaxy?]{What shapes a galaxy? -- Unraveling the role of mass, environment and star formation in forming galactic structure}
\author[Asa F. L. Bluck et al.]{Asa F. L. Bluck$^{1,2,3,*}$, Connor Bottrell$^4$, Hossen Teimoorinia$^5$, Bruno M. B. Henriques$^3$, \newauthor J. Trevor Mendel$^{6,7}$, Sara L. Ellison$^4$, Karun Thanjavur$^4$, Luc Simard$^5$, David R. Patton$^8$, \newauthor Christopher J. Conselice$^9$, Jorge Moreno$^{10}$ \& Joanna Woo$^4$ 
\\\\$^1$ Kavli Institute for Cosmology \& Cavendish Astrophysics, University of Cambridge, Madingley Road, Cambridge, CB3 0HA, UK
\\$^2$ Hughes Hall College, University of Cambridge, Wollaston Road, CB1 2EW, UK
\\$^3$ Swiss Federal Institute of Technology: ETH Zurich, Department of Physics, Wolfgang-Pauli Strasse 27, CH-8097 Zurich, Switzerland
\\$^4$ University of Victoria, Department of Physics \& Astronomy, 3800 Finnerty Road, Victoria, BC V8P 1A1, Canada
\\$^5$ National Research Council of Canada, Herzberg Institute of Astrophysics, 5071 West Saanich Road, Victoria, BC V9E 2E7, Canada
\\$^6$ Australian National University,  Research School of Astronomy \& Astrophysics, Canberra, ACT 2611, Australia
\\$^7$ ARC Centre for Excellence in All-Sky Astrophysics in 3D (ASTRO 3D), Australia
\\$^8$ Trent University, Department of Physics \& Astronomy, 1600 West Bank Drive, Peterborough, ON K9J 7B8, Canada
\\$^9$ University of Nottingham, Centre for Astronomy \& Particle Theory, School of Physics \& Astronomy, Nottingham, NG7 2RD, UK
\\$^{10}$ Pomona College, Department of Physics and Astronomy, Claremont, CA 91711, USA
\\$^*$ Email: asa.bluck@mrao.cam.ac.uk }

\begin{document}

\maketitle

\begin{abstract}

We investigate the dependence of galaxy structure on a variety of galactic and environmental parameters for $\sim$500,000 galaxies at z$<$0.2, taken from the Sloan Digital Sky Survey data release 7 (SDSS-DR7). We utilise bulge-to-total stellar mass ratio, (B/T)$_{*}$, as the primary indicator of galactic structure, which circumvents issues of morphological dependence on waveband. We rank galaxy and environmental parameters in terms of how predictive they are of galaxy structure, using an artificial neural network approach. We find that distance from the star forming main sequence ($\Delta$SFR), followed by stellar mass ($M_{*}$), are the most closely connected parameters to (B/T)$_{*}$, and are significantly more predictive of galaxy structure than global star formation rate (SFR), or any environmental metric considered (for both central and satellite galaxies). Additionally, we make a detailed comparison to the Illustris hydrodynamical simulation and the LGalaxies semi-analytic model. In both simulations, we find a significant lack of bulge-dominated galaxies at a fixed stellar mass, compared to the SDSS. This result highlights a potentially serious problem in contemporary models of galaxy evolution.

\end{abstract}

%
%

\section{Introduction}

Visual morphology is perhaps the earliest measurement of galaxies ever made (e.g., Hubble 1926). From these first observations it became apparent that galaxies come in different morphological types: disk-dominated structures, which frequently possess bright spiral arms; and spheroidal `bulge' structures, which have generally smooth light distributions. Many galaxies contain {\it both} of these components, i.e they are bulge + disk systems. In the decades following this pioneering work, an industry of techniques has emerged to quantify galaxy structure and morphology. Contemporary examples include: by-eye classifications (e.g., Lintott et al. 2008, Nair \& Abraham 2010, Mortlock et al. 2013, Bait et al. 2017); fitting mathematical functions to galaxy light profiles (e.g., S\'{e}rsic 1963, Peng et al. 2002, Simard et al. 2002, Buitrago et al. 2008, 2013, Tarsitano et al. 2018); fitting component bulges and disks in galaxies separately (e.g., Gadotti et al. 2009, Simard et al. 2011, Lang et al. 2014); non-parametric computational morphologies (e.g., Conselice 2003, Conselice et al. 2003, Lotz et al. 2004, 2008, Conselice, Yang \& Bluck 2009, Pawlik et al. 2016); measuring the masses of galactic structural components (e.g., Mendel et al. 2014); machine learning techniques (e.g., Huertas-Company et al. 2008, 2011); and, most recently, structural kinematics (e.g., Forster Schreiber et al. 2009, 2011, Falcon-Barroso et al. 2017, van de Sande 2017).

Ultimately, the reason why galaxy morphology and structure are so frequently studied in modern extragalactic astrophysics is because the distribution of light (and even more so mass) reveals significant information about the kinematics, formation mode, and evolutionary history of galaxies (e.g., Cole et al. 2000, Steinmetz \& Navarro 2002, Conselice 2003, Kormendy \& Kennicutt 2004, Forster Schreiber et al. 2009, Brennan et al. 2017). More specifically, disk-dominated morphologies indicate rotationally supported stellar kinematics and most probably quiescent evolutionary histories with few significant mergers. Alternatively, bulge-dominated morphologies indicate pressure supported stellar kinematics (random orientation of stellar orbits) with more active merger histories and/or violent relaxation episodes (e.g., Toomre \& Toomre 1972, Barnes \& Hernquist 1992, Burkert \& Naab 2005, Stewart et al. 2009, Hopkins et al. 2008, 2010, Law et al. 2012).

In $N$-body simulations, the merger of two disk galaxies has long since been known to lead to the growth of a spheroidal component (e.g. Toomre \& Toomre 1972, Barnes \& Hernquist 1992). However, if the merging galaxies are gas rich, disks can reform during and after the merger event as well (e.g., Burkert \& Naab 2005, Stewart et al. 2009, Hopkins et al. 2010, Mitchell et al. 2014). Thus, the relative importance of (major) merging and secular evolution (e.g. violent disk instabilities) in forming galactic bulges and elliptical galaxies remains hotly debated today.

Observationally, the structural type of galaxies varies as a strong function of stellar mass, environment, star formation/ colour, and epoch. Specifically, higher mass galaxies exhibit preferentially bulge-dominated stellar structures, whereas lower mass galaxies are most frequently disk-dominated (e.g., Bernardi et al. 2010,  Vulcani et al. 2011, Huertas-Company et al. 2013, Shibuya et al. 2015, Thanjavur et al. 2016). Galaxies residing in higher density environments are more frequently bulge-dominated than galaxies in lower density environments (Dressler et al. 1980, Postman et al. 1984, Dressler et al. 1994, Postman et al. 2005, 	Cappellari et al. 2011), even at a fixed stellar mass (Goto et al. 2003, Tasca et al. 2009). 

Quenched (passive / non-star forming) and red galaxies are found to have higher S\'{e}rsic indices, light concentrations, and B/T ratios (by light and mass) than star forming/ blue galaxies (Cameron et al 2009, Bell et al. 2012, Cheung et al. 2012, Fang et al. 2013, Bluck et al. 2014, Omand et al. 2014, Lang et al. 2014, Woo et al. 2015, Bluck et al. 2016, Teimoorinia et al. 2016). Furthermore, within a fixed stellar mass range, the fraction of disk-dominated galaxies increases substantially with redshift (e.g., Scarlata et al. 2007, Bundy et al. 2010, Bluck et al. 2012, Buitrago et al. 2013, Mortlock et al. 2013, Shibuya et al. 2015). Thus, the morphological (or structural) type of galaxies is closely connected with a host of other galactic and environmental parameters, and hence it is very difficult to isolate which correlations are causal in origin and which are merely incidental (i.e. the result of other more fundamental relationships). 

A primary focus of this paper is to rank galaxy and environmental parameters in terms of how predictive they are of galactic structure. Although correlation does not imply causation, it is hard to envisage the existence of causation without ensuing correlation of some sort. Hence, exploring how galactic and environmental parameters are correlated with galaxy structure is a logical first step in searching for the causal origins of the features found within the data. Ultimately, the success or failure of specific galaxy evolution models to account for the correlations observed in the galaxy population will be the most constraining aspect of our investigation with respect to causality.

Theoretically, there are currently two principal approaches used for modelling the formation and evolution of galaxies within a cosmological setting: 1) semi-analytic models (SAMs), which simulate galaxy evolution from a  coupled set of differential equations built upon the framework of a dark matter only $N$-body simulation of structure formation (e.g., Cole et al. 2000, Croton et al. 2006, Bower et al. 2006, 2008, Somerville et al. 2008, Guo et al. 2011, Henriques et al. 2015, 2017, Brennan et al. 2017); and 2) cosmological hydrodynamical simulations, which simulate the gravitational physics (solving the Poisson equation) and the hydrodynamics (solving the Euler equation) simultaneously (e.g., Vogelsberger et al. 2014a,b, Schaye et al. 2015). In addition to cosmological simulations and SAMs, idealised hydrodynamical simulations of galaxies, groups and clusters, as well as high-resolution zoom-in simulations are utilised to test and explore the physics of galaxy evolution (e.g., Moreno et al. 2015, Hopkins 2015, Hopkins et al. 2016).

Recently, both SAMs and cosmological hydrodynamical simulations have achieved generally good agreement with observations for the multi-epoch stellar mass function of galaxies, the star forming main sequence, and the star formation history of the Universe (e.g. Vogelsberger et al. 2014a,b, Schaye et al. 2015, Somerville et al. 2015, Henriques et al. 2015, 2017)\footnote{It is important to note that the good agreement with these diagnostics is achieved partly by construction, since the simulations frequently tune their feedback model to recover these trends.}. However, it remains unclear to what extent the detailed distributions of galaxy structure and morphology are reproduced by contemporary models (although see Wilman et al. 2013, Brennan et al. 2017 and Bottrell et al. 2017a,b for some early results along these lines). As such, one of the primary motivations for this paper is to address whether or not the structures of simulated galaxies resemble observations.

The goal of this paper is to make a detailed statistical analysis of the key dependencies of galaxy structure on other physical observables in galaxies, particularly: stellar mass, dark matter halo mass, central - satellite classification, specific star formation rate, local galaxy over-density, and group centric distance (for satellites). We rank these parameters in order of how important they are for determining galaxy structure, via an artificial neural network machine learning technique. We also make a careful comparison to the LGalaxies model (Henriques et al. 2015) and the Illustris hydrodynamical simulation (Vogelsberger et al. 2014a). 

Throughout the paper, we utilise bulge-to-total stellar mass ratio, (B/T)$_*$, as the primary indicator of galactic structure (derived in Mendel et al. 2014). This parameter represents a significant improvement over many previous studies because it circumvents issues related to the dependence of galaxy morphology on observed wavelength. For some analyses we also consider the B/T ratio by optical light in the $u, g, r, i$ SDSS bands (derived in Simard et al. 2011 and Mendel et al. 2014). The stellar mass, bulge mass and disk mass functions for our sample are presented in Thanjavur et al. (2016), along with the dependence of B/T structure on stellar mass. Here we expand on that prior work by incorporating a much larger variety of galactic and environmental parameters into our structural analysis, ranking these parameters in terms of how connected they are to galaxy structure, and performing a detailed comparison to two simulations.

The paper is organised as follows: Section 2 provides a summary of our data sources and the methods used to derive various galactic and environmental parameters. In Section 3 we present our observational results for the SDSS, including general trends between structure and stellar mass, halo mass, local density and specific star formation rate. In Section 4 we rank the galaxy and environmental parameters considered in this paper in terms of how predictive they are of galaxy structure. In Section 5 we compare our primary observational results to the LGalaxies SAM (Henriques et al. 2015) and the Illustris hydrodynamical simulation (Vogelsberger et al. 2014a). In Section 6 we provide a discussion of our results within the broader scientific context, and in Section 7 we give a brief summary of the paper. In the appendix we provide more details on the reliability of the bulge-disk decompositions; a comparison of structure measured in optical wavebands to structure measured by mass; and show some additional scientific analyses. Throughout the paper we assume a $\Lambda$CDM cosmology with the following parameters: \{$\Omega_{M}$, $\Omega_{\Lambda}$, $H_{0}$\} = \{0.3, 0.7, 70 km s$^{-1}$ / Mpc\}, and adopt AB magnitude units.

%
%

\section{Data Overview \& Parameter Estimation}

\subsection{SDSS -- Structural Parameters, Stellar Masses \& Statistics}

We use as our primary observational sample the SDSS DR7 (Abazajian et al. 2009). We select galaxies at $0.02 \leq z \leq 0.2$ with stellar masses in the range $8 \leq \log(M_{*} / M_{\odot}) \leq 12$. Additionally, we require that all galaxies have successful measurements of stellar mass, B/T structures, star formation rates  and group designation (true for $>$ 90\% of the full dataset). This yields a sample of 538 046 galaxies (423 480 centrals and 114 566 satellites). This is the same sample as analysed in Bluck et al. (2014, 2016) and it is thoroughly discussed in these prior publications, so we give only an overview of the most relevant aspects of the data and parameter estimation here.

We utilise stellar masses for galaxies, bulges and disks from the publicly available Mendel et al. (2014) component mass catalogue\footnote{http://vizier.cfa.harvard.edu/viz-bin/VizieR?-source=J/ApJS/210/3} and utilise B/T morphologies (by light) and other structural parameters from the Simard et al. (2011) {\small GIM2D} photometric catalogue\footnote{http://cdsarc.u-strasbg.fr/viz-bin/Cat?J/ApJS/196/11}. Bulge - disk decompositions are performed using the {\small GIM2D} package (Simard et al. 2002) assuming an exponential ($n$ = 1) disk and a de Vaucouleurs ($n$ = 4) bulge. Attempts to fit the SDSS data with a free S\'{e}rsic bulge component led to a lack of convergence in a significant fraction of cases, hence the choice of fixing the bulge S\'{e}rsic index. Calibrations to deal with potential biases incurred from utilising a fixed bulge component are discussed below (and in Appendix A). The masses of galaxies and their structural sub-components (bulges and disks) are derived through SED fitting to model spectra. Full details on the bulge-disk decompositions in optical SDSS bands are provided in Simard et al. (2011) and full details on the stellar mass derivation of galaxies and structural sub-components are provided in Mendel et al. (2014). 

Considerable effort is expended in the above references, and also in the appendices of Bluck et al. (2014), to test the reliability of the bulge - disk decompositions through the recovery of simulated galaxy light profiles. From this we find a mean error of $\langle$(B/T)$_{\rm err}$$\rangle$ $\sim$ 0.2 per galaxy, which is more than sufficient to separate galaxies into structural types. Furthermore, given that we typically consider the average structures of groupings of thousands of galaxies, this leads to a very accurate determination of the group structure (from the $1/\sqrt{N}$ improvement in accuracy of the error on the mean). A summary of prior tests on the reliability of our bulge-disk decompositions is provided in Appendix A1. Additionally, in Appendix A2 we consider the effect of observational depth on the recovered B/T values (see Bottrell et al. 2018 for the full analysis).

The {\it statistical} error on the stellar masses of bulges and disks are estimated to be $\sim$ 0.1 dex in Mendel et al. (2014). The full stellar mass error is estimated to be $\sim$ 0.2 dex in Mendel et al. (2014) and Thanjavur et al. (2016), taking into account a range of potential systematics including the choice of the initial mass function (IMF), extinction law, and the age of the stellar populations. The fact that we use a fixed ($n$ = 4) bulge model leads to two further issues: 1) galaxies with low S\'{e}rsic index ($n$ $\sim$ 2) pseudo-bulges (e.g., Gadotti 2009, Kormendy et al. 2010) may be fit as a disk + a small classical bulge component erroneously; and 2) elliptical galaxies with very high S\'{e}rsic index ($n >$ 4) may result in a spurious disk component being fit alongside the classical bulge to account for extra light at the centre, and in the wings of the galaxy. 

Although the first issue can be very serious in optical light, pseudo-bulges are typically very blue compared to classical bulges and hence have relatively low mass-to-light ratios (e.g., Mendel et al. 2014, Bluck et al. 2014). Thus, in our stellar mass derived B/T structure, pseudo-bulges lead to very low mass classical bulge estimates, which is essentially correct (see Appendix A, Mendel et al. 2014 and Bluck et al. 2014 for further details). Moreover, in our testing of the sub-sample which is fit successfully with a free S\'{e}rsic bulge model, the presence of pseudo-bulges is exclusively associated with disk-dominated systems, alleviating concerns of catastrophic misclassification here (see also Lackner \& Gunn 2012 for more discussion on pseudo-bulges). In the case where both a pseudo-bulge and a classical bulge is present, our techniques will primarily be sensitive to the classical bulge structure, as desired for the science goals of this paper.

The second issue, however, is potentially a much more serious problem, whereby some elliptical galaxies will be erroneously fit as bulge + disk systems. A method to identify, and correct for, `false disks' in the morphological and structural fits was developed in Bluck et al. (2014) and refined in Thanjavur et al. (2016). Specifically, the probability of a given disk component being spurious ($P_{\rm FD}$) is computed as a function of the axis-ratio of the galaxy ($b/a$), the difference in ($g-r$) colour between the bulge and the disk ($\Delta (g-r)_{b,d}$), and the S\'{e}rsic index from a single S\'{e}rsic fit ($n$). Following the findings of Bluck et al. (2014) and Thanjavur et al. (2016), we set the disk mass to zero in any galaxies with $P_{\rm FD} > 0.2$, provided they are bulge dominated (B/T $>$ 0.5) and have a high probability of being a pure S\'{e}rsic profile in the original photometric fit as well ($P_{pS} > 0.32$). Furthermore, we then set the bulge mass equal to the total stellar mass of the galaxy derived from a single (free) S\'{e}rsic fit. This method effectively removes false disks from the sample.

Throughout the paper, we use as our primary definition of galaxy structure the bulge-to-total stellar mass ratio, defined as:

\begin{equation}
{\rm (B/T)}_* \equiv \frac{M_{\rm bulge}}{M_{*}} = \frac{M_{\rm bulge}}{M_{\rm bulge} + M_{\rm disk}} = 1 - {\rm D/T}
\end{equation}

\noindent where $M_{\rm bulge}$ is the stellar mass of the bulge component, $M_{\rm disk}$ is the stellar mass of the disk component, and $M_{*}$ is the total stellar mass of the galaxy (= $M_{\rm bulge}$ + $M_{\rm disk}$ in our bulge-disk decomposition model). In the mass catalogue of Mendel et al. (2014) the total stellar mass of each galaxy is also derived from a joint bulge + disk SED fit ($M_{bd}$). In the vast majority of cases ($>$ 90\%) these two definitions agree within 1$\sigma$ of the statistical error, and hence the masses are internally consistent. In all of our analyses we remove the few cases where this is not the case, as in Bluck et al. (2014, 2016) but unlike in Thanjavur et al. (2016), where B/T ratios were estimated from $i$-band magnitudes for the components in the discrepant cases, in order to achieve completeness for the mass functions. The galaxies with discrepant masses are not obviously biased to any mass or structure and thus removing them from the sample will not engender any significant systematic effects on the results of this work (which we confirm by checking that our results are insensitive to this minor population).

Our definition of galactic structure has a number of important advantages over similar parameters used in prior studies. Crucially, we use a definition of galaxy structure which is sensitive to stellar mass, not optical light as has most frequently been used in the past (e.g., Driver et al. 2006, Forster-Schreiber et al. 2011, Bell et al. 2012, Buitrago et al. 2013, Lange et al. 2015, Bait et al. 2017). This is critical when comparing to parameters related to star formation and colour, since a small fraction of bright / blue massive stars can dominate the optical morphology of entire galaxies, obscuring the underlying mass distribution. Although clearly a significant improvement over photometric morphologies in a single waveband, our stellar mass B/T ratios may still be affected by errors in the SED fitting of highly star forming systems (see Sorba et al. 2015). However, since the vast majority of SDSS galaxies are forming stars at main sequence levels or (substantially) below, these issues will not dominate any of our results or conclusions.

The majority of galaxies in our sample are found to host substantial disk {\it and} bulge components (each contributing at least 20\% of the stellar mass) and hence a single parameter, e.g. S\'{e}rsic index, would miss much of the complexity contained in the structures and mass-distributions of galaxies. Moreover, in Mendel et al. (2014) and Simard et al. (2011) the $F$-statistic is used to construct the probability of a galaxy being fit better with a two-component model than a single S\'{e}rsic profile, taking into account the statistical expectation to do better with more free parameters. Here again, the majority of our sample is found to host two structural components.    

For a few analyses (particularly in comparing to the Illustris simulation) we also consider the B/T ratios of galaxies in optical wavebands. These are defined as:

\begin{equation}
{\rm (B/T)}_{X} \equiv \frac{L_{{\rm bulge,} X}}{L_{{\rm gal,} X}} = \frac{L_{{\rm bulge,} X}}{L_{{\rm bulge,} X} + L_{{\rm disk,} X}} = 1 - {\rm (D/T)_{X}}
\end{equation}

\noindent where $X$ = \{$u,g,r,i$\} optical SDSS wavebands, and, for example, $L_{{\rm bulge,} X}$ indicates the `$X$'-band luminosity of the bulge component. In general, B/T values measured in optical wavebands are more closely connected with optical morphology (as studied classically by the Hubble sequence, Hubble 1926), whereas B/T structures by mass are more closely connected with the kinematics of galaxies. A comparison of our results for optical wavebands and mass ratio structure is given in Appendix B.

In order to account for the flux limit of the SDSS, we weight all statistics by the inverse of the volume over which any given galaxy would be visible in the survey ($1/V_{\rm max}$). The volume of visibility is a strong function of both stellar mass and colour (see Mendel et al. 2014). This weighting effectively removes the flux limit selection bias of the SDSS, and yields results on statistics as expected for a volume limited sample (see, e.g., Thanjavur et al. 2016 for the stellar mass functions). As an example, one of the main statistics we use to probe the structural dependence on other galaxy properties is the fraction of bulge dominated galaxies, which is defined as:

\begin{equation}
f_{\rm bd, j} = \frac{\sum\limits_i \big( (1/V_{\rm max})_i [{\rm (B/T)_{i}} \geq 0.5 ] \big) }{\sum\limits_i  \big( (1/V_{\rm max})_i [{\rm ALL}] \big)}
\end{equation}

\noindent which represents the ratio of the sum of volume weights for bulge dominated galaxies and the sum of volume weights for all galaxies in each binning ($j$). Similarly, the fraction of disk dominated galaxies is defined as:

\begin{equation}
f_{\rm dd, j} = \frac{\sum\limits_i \big( (1/V_{\rm max})_i [{\rm (B/T)_{i}} < 0.5 ] \big) }{\sum\limits_i  \big( (1/V_{\rm max})_i [{\rm ALL}] \big)}  = 1 - f_{\rm bd, j} 
\end{equation}

\noindent We compute all errors on the structural fractions using a Monte Carlo simulation, where we re-sample the data by taking 100 random draws from a Gaussian distribution centred on each B/T value in the bin, with a dispersion set by the error on each B/T value. We also re-sample the data of whichever galaxy property the structural fraction is plot against (e.g., stellar mass, local density etc.). Thus, we perform an $X$-$Y$ axis random Gaussian re-sampling of all the relevant measurements, within their errors. We then recompute, e.g., the $f_{\rm bd}$ and $f_{\rm dd}$ statistics for each re-sampling of the full data, taking the mean of the set as our final bin value and the 1$\sigma$ error as the standard deviation across the set. 

In this study we benefit from an extremely large sample of over half a million galaxies with reliable bulge-disk decompositions, representing a wide range of galaxy masses, star formation rates, and environments. As such, our current sample is ideal for the purpose of investigating precisely which factors matter for the growth of galactic structure in the local Universe.

\subsection{SDSS -- SFRs, Group Catalogues \& Environment}

Star formation rates (SFRs) are calculated for the SDSS sample in Brinchmann et al. (2004) using two approaches: 1) For emission line galaxies (with $S/N >$ 3), which are not selected as active galactic nuclei (AGN) by the Kauffmann et al. (2003) line ratio cut on the Baldwin, Phillips \& Terlevich (1981) emission line diagnostic diagram, SFRs are measured from extinction corrected emission lines (in $H_{\alpha}$, $H_{\beta}$, $O[III]$, and $N[II]$). These galaxies are defined as 'star forming'. For a review on deriving SFRs from emission lines see Kennicutt (1998). 2) For non-emission line galaxies (and emission line galaxies that are identified as AGN), SFRs are deduced from the empirical relationship between specific star formation rate (sSFR = SFR / $M_{*}$) and the strength of the 4000\AA \hspace{0.05cm} break ($D_{n}$4000). For both methods an aperture correction is applied, based on the colour and magnitude of light in each galaxy, outside of the spectroscopic fibre. See also Bluck et al. (2014, 2016) for more discussion on the SFRs of this sample, where we conclude that the aperture correction does not lead to any significant bias on our results by comparison to photometrically derived SFRs and dust corrected colours. SFRs for star forming galaxies are measured directly, whereas SFRs for passive galaxies are only reliably defined to be low in value. This fact must be taken into account when comparing in detail to simulations and models, which have an arbitrarily high accuracy in SFR, even at low values.

In this paper we use group catalogues from Yang et al. (2007, 2008, 2009), from which we define central galaxies to be the most massive galaxy in the group, and satellite galaxies to be any other group member. Groups are inferred from a friends-of-friends algorithm with a linking length tuned by dark matter halo simulations (Springer et al. 2005). Note that groups may contain only one galaxy, whereupon the sole galaxy in the group is dubbed the central of its group of one. In this sense, the terms central and satellite pertain to location within a dark matter halo, not necessarily a group as one would normally define it. In most cases there is a clear most massive galaxy; however, in some cases (most notably for intermediate mass groups) the separation between centrals and satellites is less certain. Ultimately, the impact of this on our results is to slightly reduce the impact of being a satellite relative to a central in this regime. Extensive testing of the group finding algorithms on simulated data demonstrate that at $M_{\rm halo} > 10^{12} M_{\odot}$, over 90\% of galaxies are correctly assigned to groups (Yang et al. 2007). We utilise halo masses from these catalogues, which are derived from an abundance matching technique applied to the total stellar mass in the group (see Yang et al. 2007, 2009 for full details). More specifically, the group halo mass is defined as the $M_{200}$ virial mass, i.e. the mass contained within the virial radius ($R_{200}$), which is inferred by fitting to a Navarro, Frenk \& White (1997, NFW) profile.

As an additional measurement of environment, we consider the projected distance each satellite resides at from its central galaxy, and express this in units of the virial radius: 

\begin{equation}
D_{cc} = \frac{R_{cc} [{\rm kpc}]}{R_{200} [{\rm kpc}]}
\end{equation}

\noindent where $cc$ indicates `cluster centric', but it is understood that this definition applies equally to groups as well.

We also consider the local density of (other) galaxies around each galaxy in our sample as another useful measurement of environment. Specifically we employ the normalised over-density of galaxies (as in Baldry et al. 2006), which is defined as:

\begin{equation}
\delta_{n} = \frac{\Sigma_{n}}{\langle \Sigma_{n}(z\pm \delta z)\rangle}
\end{equation}

\noindent where,

\begin{equation}
\Sigma_{n} = \frac{n}{\pi r^{2}_{p,n}}
\end{equation}

\noindent $r_{p,n}$ is the projected distance to the $n$th nearest neighbour within a 1000 km/s radial velocity bin. $\langle \Sigma_{n}(z\pm \delta z)\rangle$ indicates the mean density of galaxies at each $\delta z$ slice, where we set $\delta z$ = 0.01. In this study we mostly take $n$ = 5, i.e. we compute the distance to the 5th nearest neighbour. However, we also consider values of $n$ = 3 and 10 in the machine learning analysis (Section 4).

\subsection{Simulations -- LGalaxies \& Illustris} 

We compare our observational results from the SDSS to the latest version of the Munich model of galaxy formation (LGalaxies, Henriques et al. 2015; earlier versions - Kauffmann et al. 2000, Croton et al. 2006, De Lucia et al. 2009, Guo et al. 2011), and to the Illustris cosmological hydrodynamical simulation (Vogelsberger et al. 2014a,b). Full details on these datasets are provided in the above references. 

In order to ensure a fair comparison to these simulated data, we select the appropriate z $\sim$ 0 snapshot for each. Considerable effort is made to compare like qualities between the simulations and the SDSS. For example, halo masses are defined as $M_{200}$ in all; distances of satellites to centrals are given in units of the virial radius ($R_{200}$); local densities are computed as relative over-densities based on each sample; and stellar, bulge and disk masses from the SDSS are compared to the same quantities in LGalaxies. However, for Illustris, the masses of structural components are not available via SED fitting at this time. As such, we compare to photometrically derived B/T values from Bottrell et al. (2017a,b), performed in $r$-band for {\it both} the SDSS and Illustris. Furthermore, we also compare the SDSS mass weighted B/T ratios to kinematically derived B/T ratios in Illustris. Our results for both of these approaches are consistent. Further details on the simulations are provided in Section 5.1.

%
%

\section{Observational Results}

The goal of this paper is to explore how galaxy structure is connected to a wide variety of other galactic and environmental parameters. We then aim to assess how well the observed trends are reproduced in modern simulations (see Section 5).

\subsection{Pedagogical Overview}

\begin{figure*}
\includegraphics[width=0.49\textwidth]{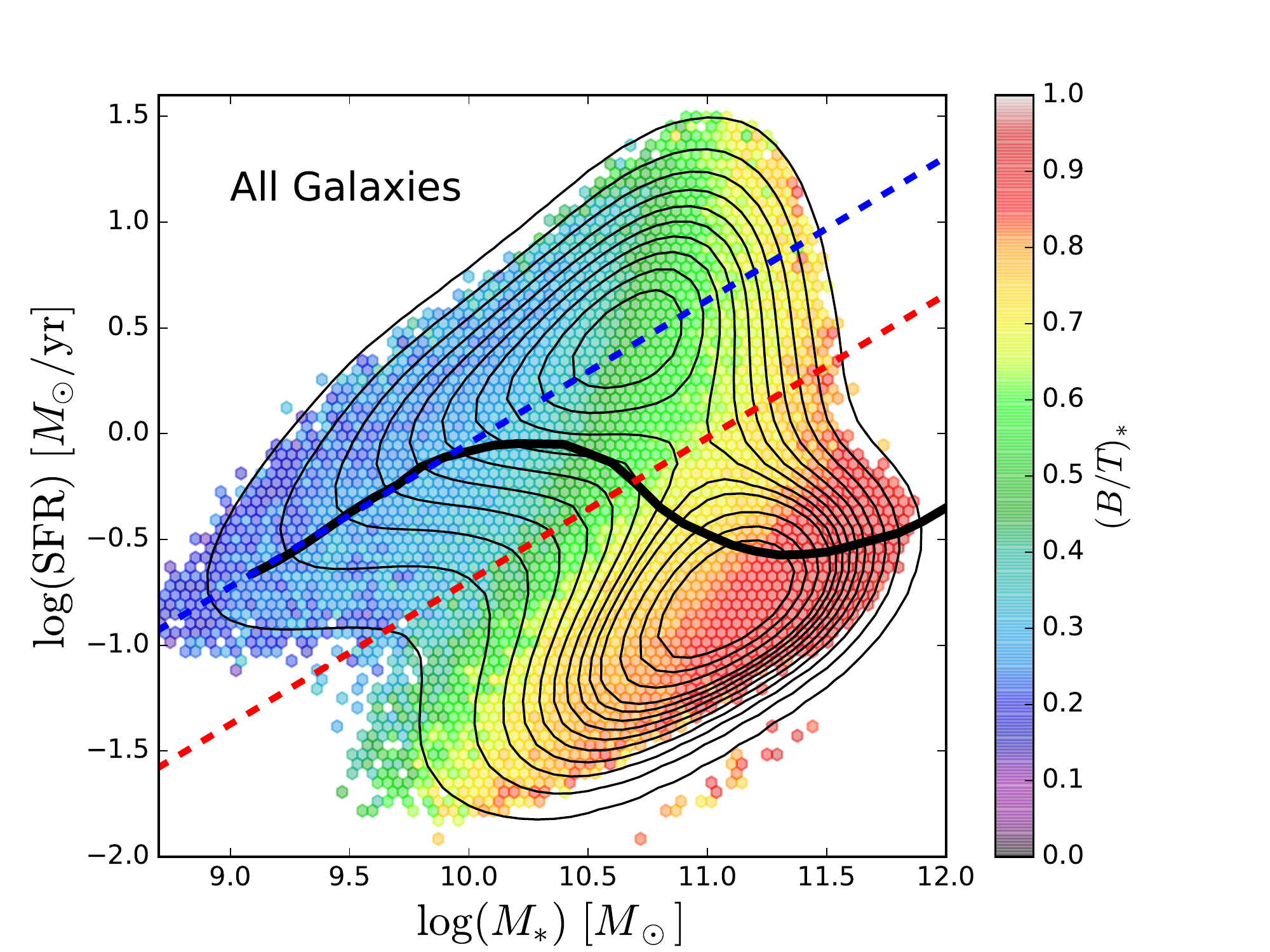}
\includegraphics[width=0.49\textwidth]{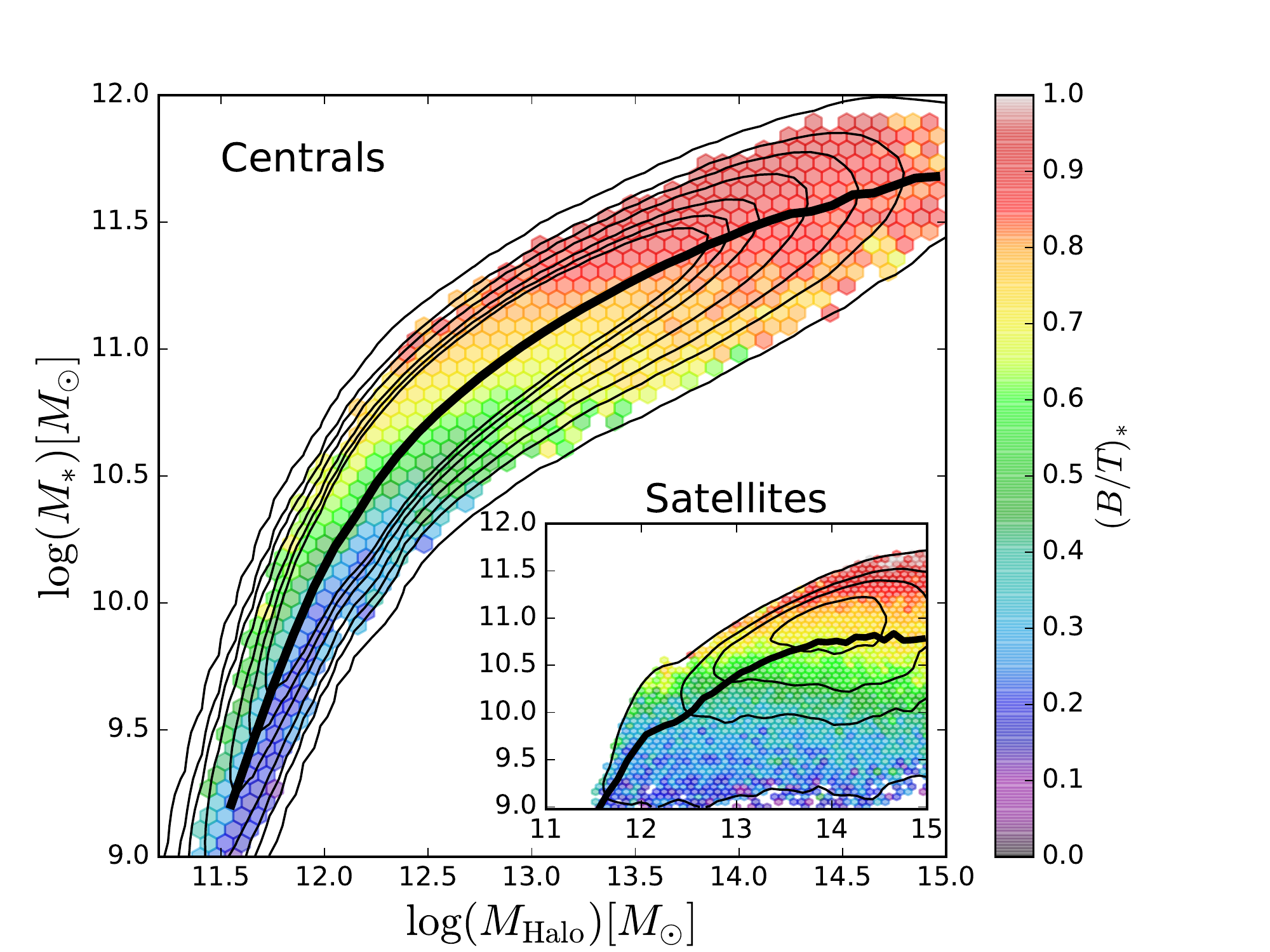}
\caption{{\it Left panel:} The star forming main sequence (${\rm SFR} - M_{*}$ relation) for all local (z $\sim$ 0.1) SDSS galaxies. {\it Right panel:} the stellar mass - group halo mass relationship for central (main plot) and satellite (inset) galaxies. Both panels are colour coded by the mean bulge-to-total stellar mass ratio (B/T) in each hexagonal bin, and we indicate the median relationship with a solid black line. In the left panel, star formation separates out into two distinct regions, with star forming systems having disk-dominated structures (low B/T, bluer colours) and quenched systems having bulge-dominated structures (high B/T, redder colours). We also note a more subtle increase in B/T for each population with increasing stellar mass. In the right panel, increasing stellar and halo mass leads to higher B/T ratios in centrals, whereas in satellites B/T varies predominantly as a function of stellar mass.}
\end{figure*}

\begin{figure}
\includegraphics[width=0.49\textwidth]{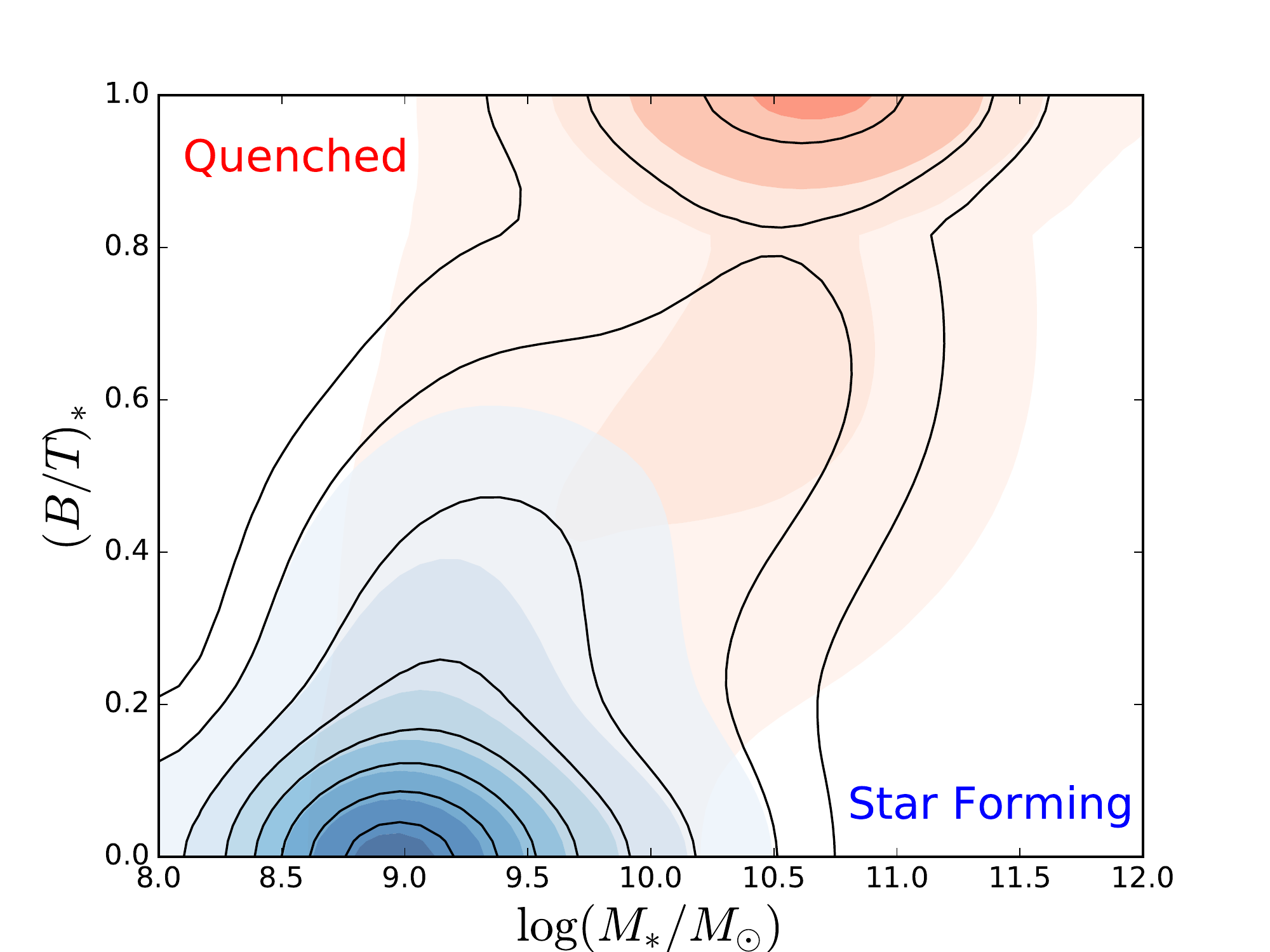}
\caption{The B/T - $M_*$ relationship, shown as density contours for the full population (black line contours), star forming systems (blue shaded contours) and quenched systems (red shaded contours). The full distribution of SDSS galaxies is clearly highly bimodal, whereas the star forming and quenched systems are each uni-model, although for quenched systems there is a pronounced power law tail to lower B/T and $M_{*}$ values.}
\end{figure}

In this sub-section, we begin with a general presentation of how our structural parameter (B/T)\footnote{Throughout the results section we adopt the following convention: ${\rm B/T} \equiv ({\rm B/T})_*$. However, in other sections we occasionally consider alternative photometric definitions of B/T.} varies throughout two key diagnostic plots. This has been investigated before (e.g., Moster et al. 2010, Wuyts et al. 2011). However, our motivation to include it here is twofold: 1) Understanding how B/T structure varies as a function of other well-known relationships is important to interpreting our novel results later in this section and in Sections 4 - 5. As such, it provides a useful pedagogical introduction to readers less familiar with the literature on galaxy structures. And, 2) Although the variation in galaxy structure has been studied across the star forming main sequence and halo mass - stellar mass relationship before, it has never been performed with a stellar mass definition of structure, or for a sample as large and complete as the SDSS DR7.

In Fig. 1 (left panel) we show the star forming main sequence (SFR - $M_{*}$ relation, e.g. Brinchmann et al. 2004) for all galaxies in our sample, colour coded by mean B/T in each hexagonal bin. There is clear bimodality in this plot, as evidenced by the density contours, with an upper star forming `main sequence', and a lower `quenched' population of passive galaxies (as seen in, e.g., Strateva et al. 2001, Baldry et al. 2006, Peng et al. 2010, 2012) . The solid black line indicates the median SFR - $M_{*}$ relation, which transitions from the star forming main sequence to the passive/ quenched population at $M_{*} \sim 10^{10.5} M_{\odot}$. The blue dashed line indicates a least squares fit to the star forming main sequence, and the red dashed line indicates the threshold of quenched systems, located at the minimum of the density contours. 

Star forming galaxies in Fig. 1 have noticeably lower B/T ratios than quenched galaxies at the same stellar mass, with a mean difference of $\Delta(B/T)_{SF | Q} \sim 0.4$ . In fact, the majority of star forming galaxies are disk dominated (B/T $<$ 0.5) and the majority of quenched systems are bulge dominated (B/T $>$ 0.5). Therefore, it is clear that structural transformation of galaxies from disk-dominated to bulge-dominated systems, and quenching from star forming to passive systems are closely related galactic processes (as previously argued for in Cameron et al. 2009, Bell et al. 2012, Bluck et al. 2014, Lang et al. 2014, Omand et al. 2014, Bluck et al. 2016). Crucially, it should be stressed here that this close link between structure and star formation is not a result of differential disk fading due to quenching, because our structural metric is sensitive to stellar mass, not optical light. Hence, quenched galaxies are {\it structurally} different to star forming galaxies, not just in visual morphology, which implies that a different set of physical processes must be responsible for each population.

In Fig. 1 (right panel) we show the stellar mass - halo mass relation (e.g. Moster et al. 2010), for central galaxies (main plot) and satellite galaxies (inset). Central galaxies are defined as the most massive galaxies in each group, with satellites being any other group members. In both plots the solid black line indicates the median relation, and the faint black lines are density contours. As with the star forming main sequence, we colour code each hexagonal bin in the $M_{*} - M_{\rm halo}$ relation by mean B/T. 

For central galaxies, there is a very strong dependence of stellar mass on halo mass up to $M_{\rm halo} \sim 10^{12.5} M_{\odot}$, where the relationship becomes noticeably less steep. The `knee' in this relation is a result of two physical effects - 1) quenching of high mass galaxies and 2) the formation of clusters, where individual merging timescales for low mass satellite galaxies with their high mass centrals reach beyond the Hubble time. Low mass centrals (in stellar and halo mass) are predominantly disk dominated (low B/T), whereas high mass centrals are predominantly bulge dominated (high B/T).  For satellite galaxies (inset), there is a much weaker dependence of stellar mass on group halo mass, which is expected since, in general, satellites form in their own sub-haloes. For satellites, B/T varies most strongly with stellar mass, not halo mass, as can be seen by colours shifting from blue to red vertically and not horizontally on the right panel inset of Fig. 1. This implies that stellar mass is more important than halo mass for establishing bulge structures in satellites (we quantify this in Section 3.2).

In Fig. 2, we show the B/T - $M_{*}$ relation for all galaxies as a density contour map. This 2d-distribution is evidently highly bimodal - with a density peak at low stellar masses and low B/T ratios and another density peak at high stellar masses and high B/T ratios. There is clearly a correlation between stellar mass and B/T but it is very broad, i.e. there is a large range in galaxy structures at any given mass. We separate the sample into actively star forming and quenched systems via a cut in sSFR at $10^{-2}$ Gyr$^{-1}$. Unlike the full galaxy population, star forming and quenched galaxies, separately, are uni-model in their distribution in the  B/T - $M_{*}$ plane. However, the quenched systems do exhibit a power-law like tail to lower stellar masses and B/T ratios. Given that the quenched contours are essentially co-located with the high B/T mode, and that the star forming contours are essentially co-located with the low B/T mode, it is reasonable to identify (to first order at least) the two modes in the B/T - $M_{*}$ plane with star forming and quenched systems (see also Cameron et al. 2009, Bell et al. 2011 \& Bluck et al. 2014).

\subsection{Main Observational Trends}

\begin{figure*}
\includegraphics[width=0.49\textwidth]{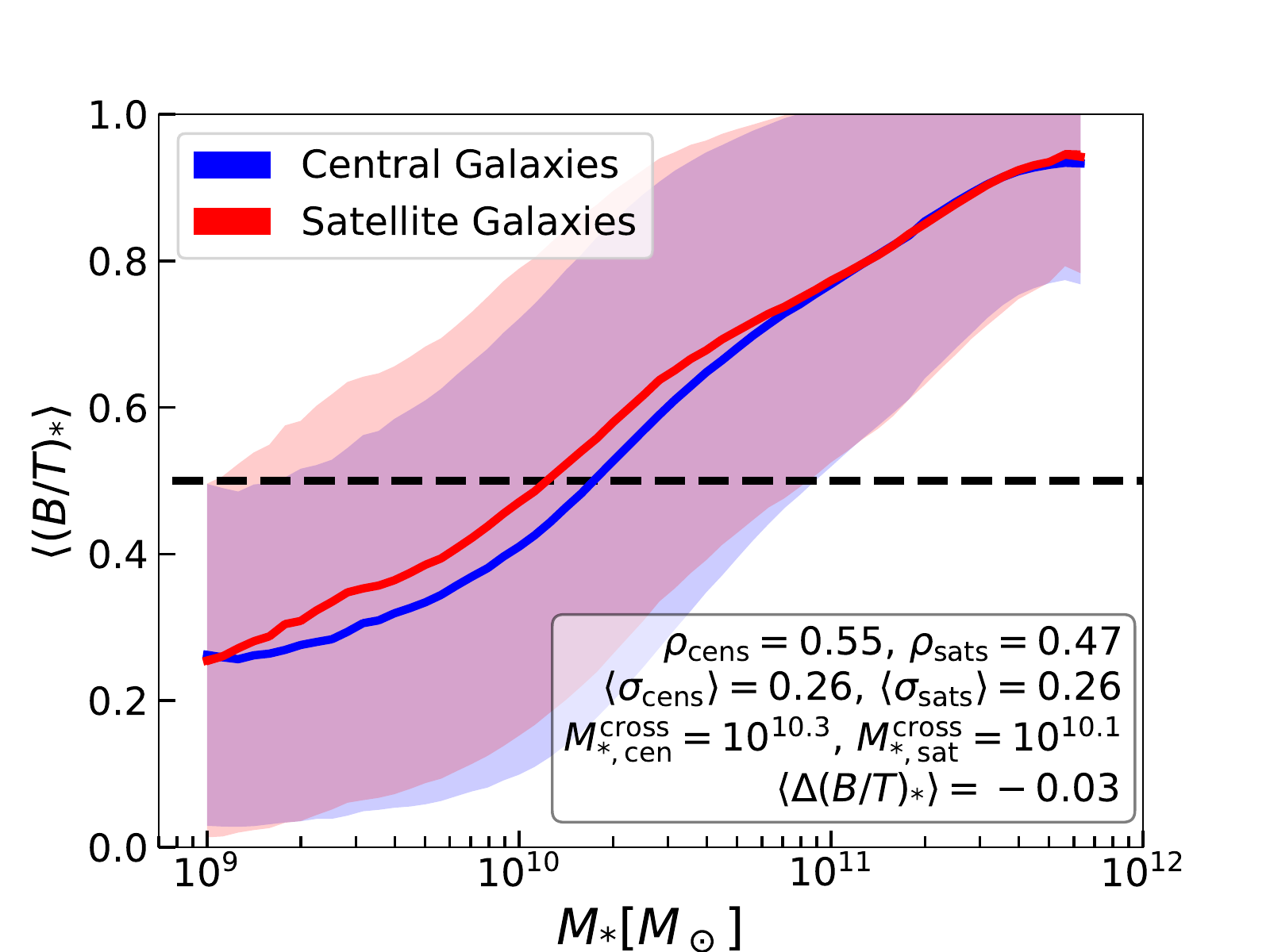}
\includegraphics[width=0.49\textwidth]{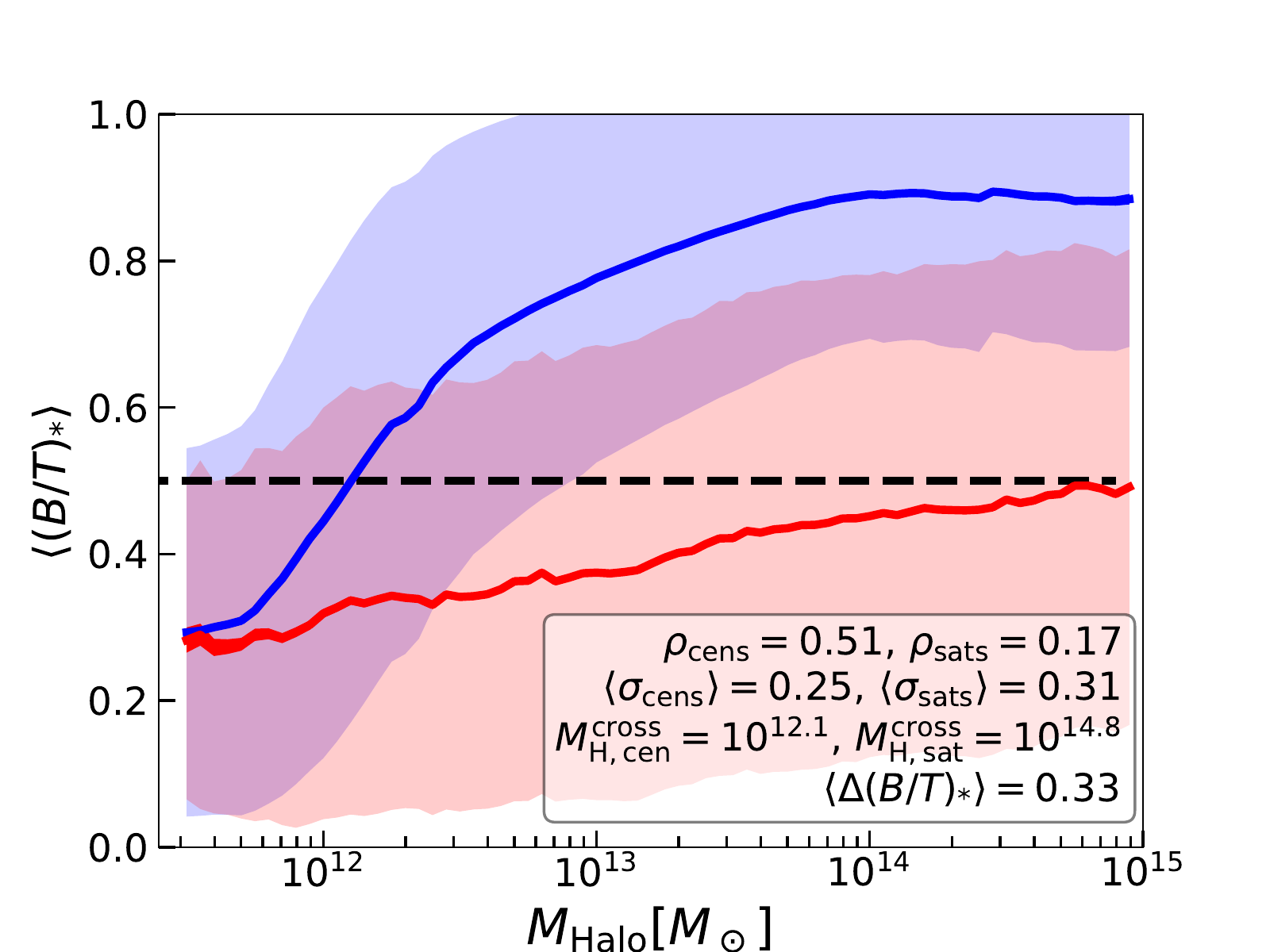}
\includegraphics[width=0.49\textwidth]{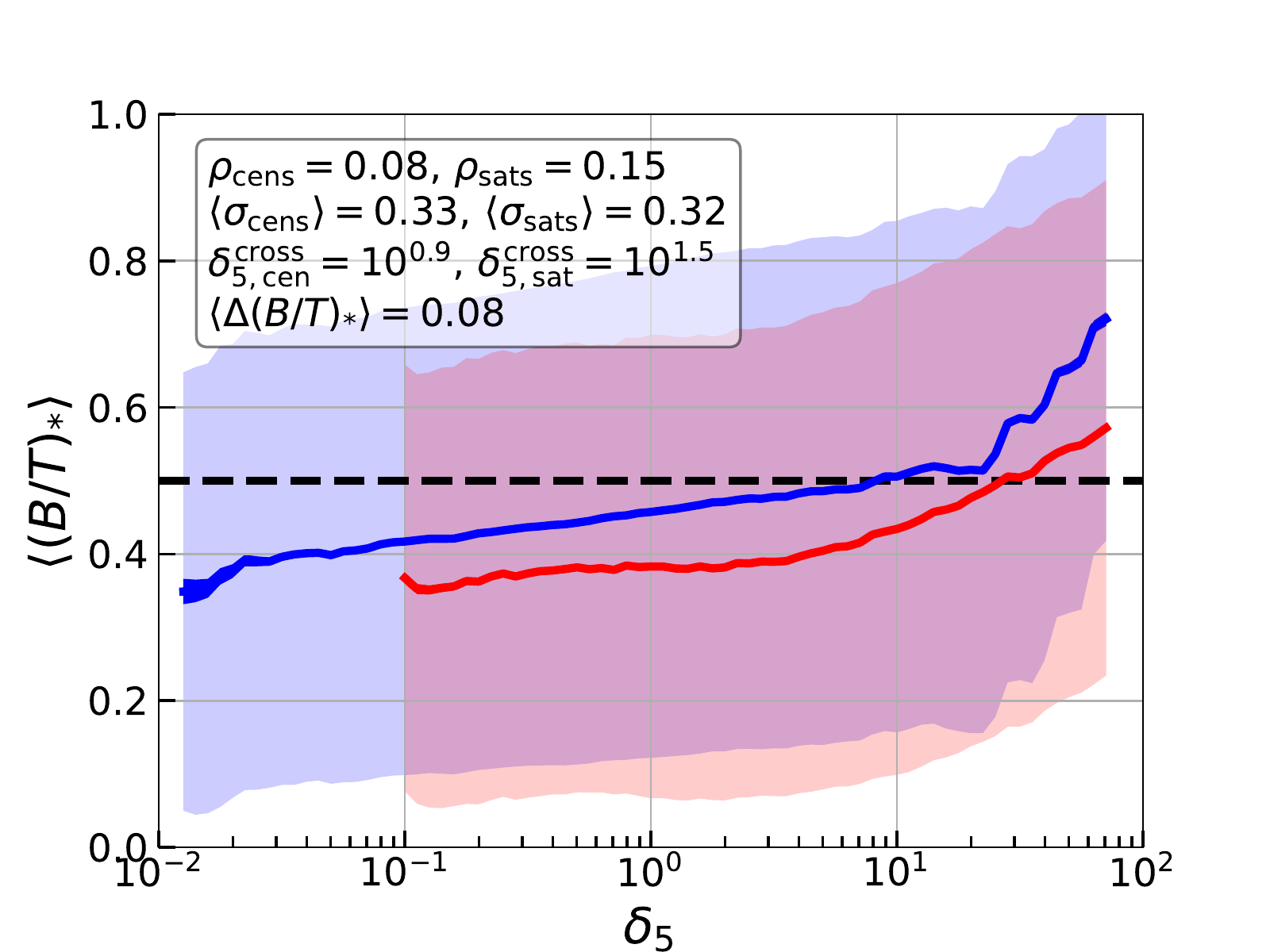}
\includegraphics[width=0.49\textwidth]{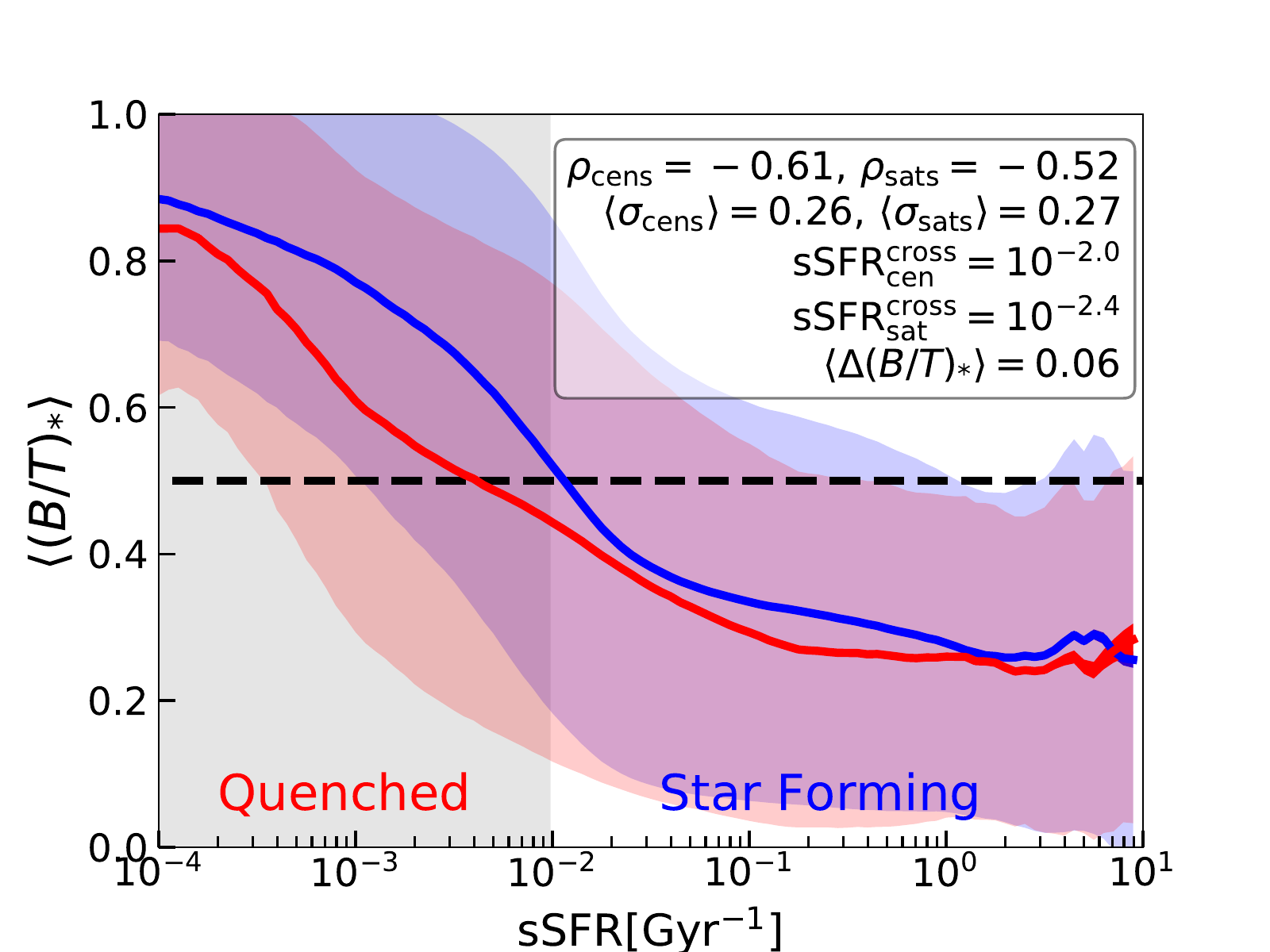}
\caption{The relationship between stellar mass derived B/T ratio and, from top-left to bottom-right: stellar mass, group halo mass, local density evaluated at the 5th nearest neighbour, and sSFR. Each panel shows the volume corrected mean B/T ratio in each bin (centering of the lines), the standard error on the mean (solid shading) and standard deviation (light shading), for central galaxies (shown in blue) and satellite galaxies (shown in red). The strength of correlation, average dispersion, crossover value at B/T = 0.5, and mean difference in B/T between centrals and satellites is shown on each panel for each relationship. In all panels galaxies are selected consistently at $M_{*} > 10^{9} M_{\odot}$.}
\end{figure*}

In this sub-section we present the main observational trends of B/T with various galaxy properties. We sub-divide our analysis into central and satellite galaxies throughout, which exposes the similarities and differences in these populations' structural dependence on other galactic and environmental parameters.

In Fig. 3 we show the mean B/T ratio as a function of, from top left to bottom right: stellar mass, halo mass, local density and sSFR. The width of each line represents the 1$\sigma$ error from the standard error on the mean, and the light shaded region indicates the standard deviation at each binning of the X-axis variable. We additionally present in Fig. 3 a set of statistics computed separately for centrals and satellites. Specifically, we compute the strength of correlation ($\rho$) from the Spearman rank correlation coefficient, the average dispersion ($\langle \sigma \rangle$), the value of the X-coordinate parameter at which B/T = 0.5 (indicating values above or below which galaxies are typically bulge or disk dominated), and finally, the average difference in B/T between centrals and satellites at fixed X-coordinate variables.

In the top left panel of Fig. 3, we find a positive dependence of B/T on stellar mass for both centrals and satellites (with a moderately strong correlation of $\rho = 0.55$ for centrals and slightly weaker correlation of $\rho = 0.47$ for satellites). The crossover mass (i.e. the stellar mass at which there is an average B/T ratio of 0.5, and hence higher masses lead to typically bulge-dominated systems and lower masses lead to typically disk-dominated systems) occurs for centrals at $M^{\rm cens.}_{*\rm{, cross}} \sim 2 \times 10^{10} M_{\odot}$ and for satellites at a slightly lower mass of $M^{\rm sats.}_{*\rm{, cross}} \sim 1.3 \times 10^{10} M_{\odot}$. Another way of describing this subtle difference between centrals and satellites, is that the fraction of bulge dominated galaxies is slightly higher for satellites than centrals, and on average is $\langle \Delta (B/T)_* \rangle \sim -0.03$.

In the top right panel of Fig. 3, we find a positive dependence of B/T on halo mass for both centrals and satellites. However, the correlation is much stronger for centrals ($\rho$ = 0.51) than for satellites ($\rho$ = 0.17). For centrals, the correlation between halo mass and B/T is comparably strong to that of stellar mass and B/T; whereas, for satellites, the correlation between B/T and halo mass is much weaker than between B/T and stellar mass. This result is interesting because it highlights that the two populations have structures which scale very differently with the mass of the dark matter halo they are in. Furthermore, the mean B/T values of satellites are significantly lower than that of centrals at a fixed halo mass (on average by $\sim$ 0.3). 

An explanation for these differences can be found by noting the strong relationship between stellar and halo mass for centrals, which is absent for satellites (see Fig. 1). If it is ultimately stellar, rather than halo, mass which drives the growth of bulge structure, a strong correlation between halo mass and B/T for centrals and a weak correlation for satellites is expected. Furthermore, given that centrals are by definition the most massive galaxies in a given halo mass, it is natural to expect they will have higher B/T values than satellites in the same dark matter halo, if stellar mass is the primary delineator of galaxy structure. The crossover halo mass for centrals occurs at $\sim 10^{12} M_{\odot}$, below which centrals are predominantly disk-dominated and above which centrals are predominantly bulge-dominated. Satellites, however, never fully cross the B/T = 0.5 line, except possibly at the highest halo masses: $> 10^{14.5} M_{\odot}$, and hence are predominantly disk-dominated in all groups and clusters.

In the bottom left panel of Fig. 3, we show the dependence of B/T on local over-density evaluated at the 5th nearest neighbour for centrals and satellites. Central and satellite galaxies look qualitatively similar on this plot, with both experiencing a gradual increase in B/T at low-to-intermediate over-densities, which steepens markedly in gradient at $\delta_{5} >$ 10. We find that local density is only very weakly correlated with galaxy structure at low to intermediate densities, but becomes significantly more correlated with structure in very high over-density environments. Interestingly, in this plot we notice the same effect as in the B/T - halo mass relation, whereby central galaxies are preferentially more bulge-dominated (by on average $\langle (B/T)_* \rangle \sim$ 0.08) throughout the large dynamic range in over-densities explored here. 

Finally, in the bottom right panel of Fig. 3, we show the dependence of B/T structure on sSFR for central and satellite galaxies. Both populations of galaxies exhibit the strongest correlations witnessed in Fig. 3 with sSFR ($\rho$ = -0.61, -0.52 for centrals and satellites, respectively). In this case the correlations are negative, such that higher sSFR values lead to lower B/T values. As with stellar mass, the correlation between sSFR and B/T is stronger for centrals than satellites. Moreover, central galaxies are offset to higher bulge fractions than satellite galaxies, by on average $\sim$ 0.06. The region of this plot where quenched galaxies reside is shown as shaded because the SFRs for this sub-population are inferred indirectly (from the strength of the 4000 $\AA$ break in Brinchmann et al. 2004\footnote{Note that although the actual values of sSFR for quenched systems are not known accurately, the fact that they are low (relative to star forming systems) is highly robust (see, e.g., Brinchmann et al. 2004).}).

In summary of Fig. 3, we find strong correlations between stellar mass determined B/T ratio and stellar mass, sSFR and halo mass (for centrals), with significantly weaker correlations with halo mass for satellites and local density for all galaxy types. These results suggest that stellar mass and sSFR are closely connected to the growth of bulge structures in all galaxy types. Furthermore, central and satellite galaxies vary their structures in a similar manner with stellar mass, but exhibit significant differences as a function of halo mass. These differences may be explained if galaxy structure is better constrained by stellar mass than environment.

\subsection{Structural Dependence at a Fixed Stellar Mass -- Environment \& Star Formation}

\begin{figure*}
\includegraphics[width=0.49\textwidth]{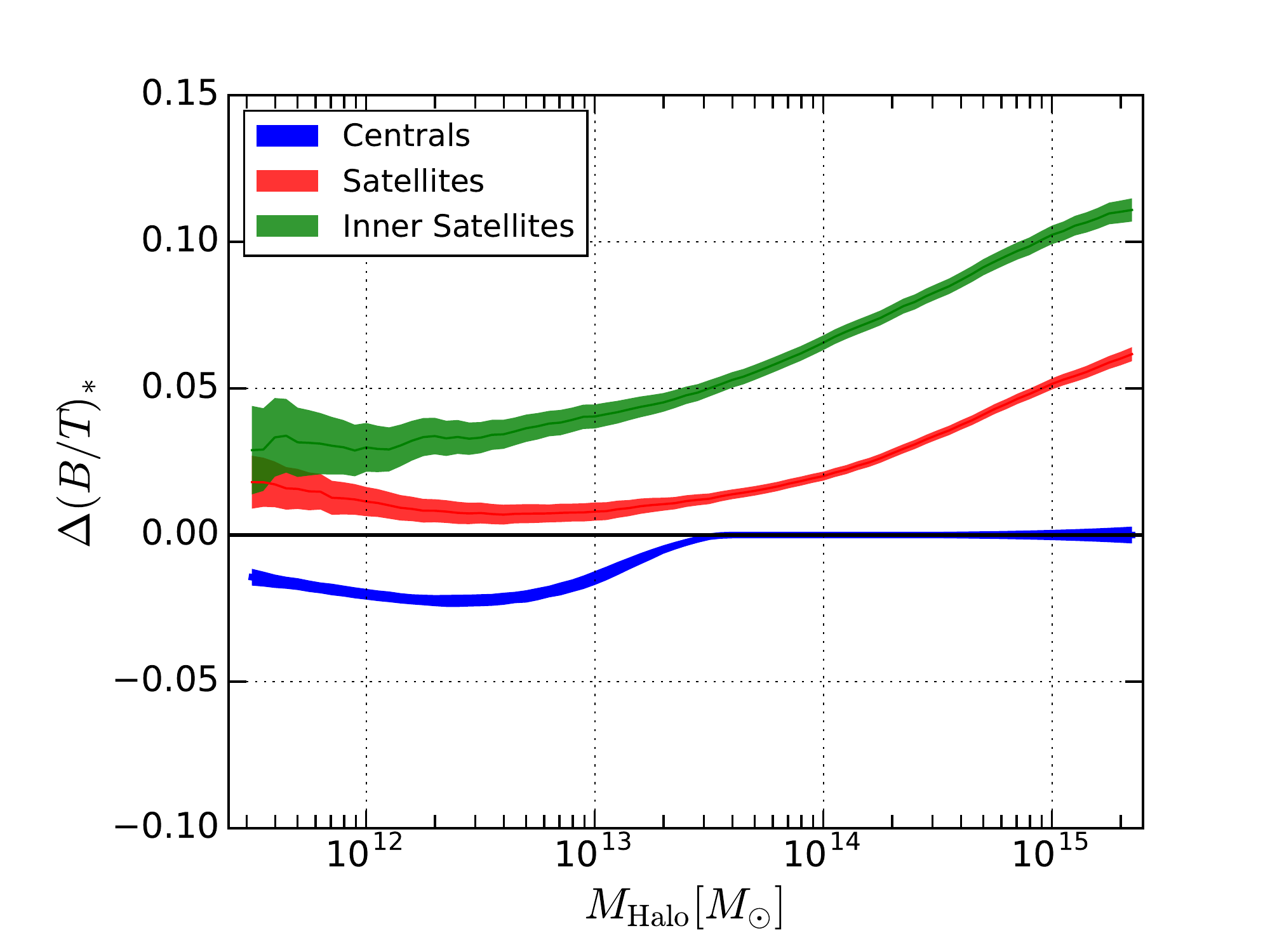}
\includegraphics[width=0.49\textwidth]{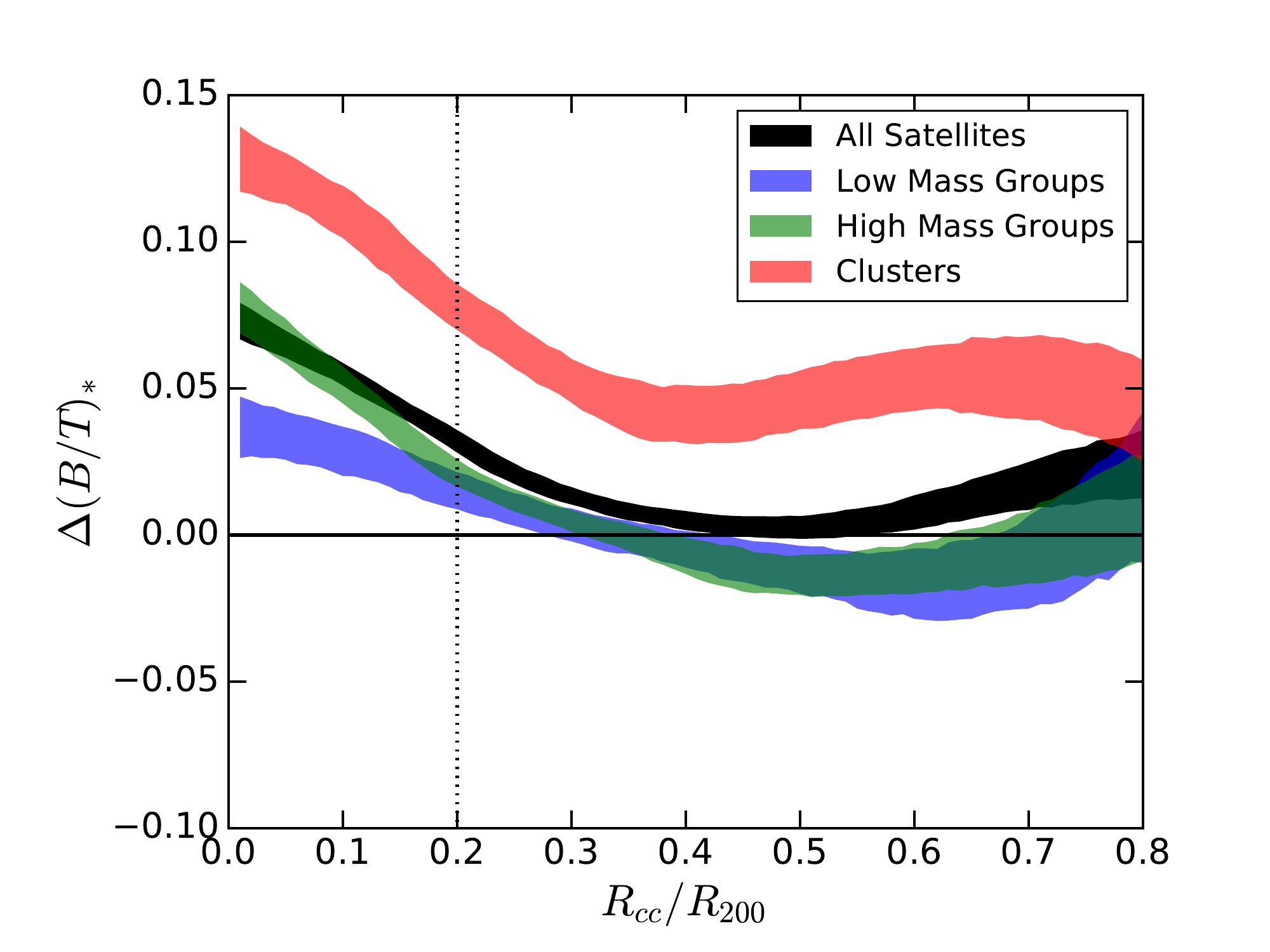}
\includegraphics[width=0.49\textwidth]{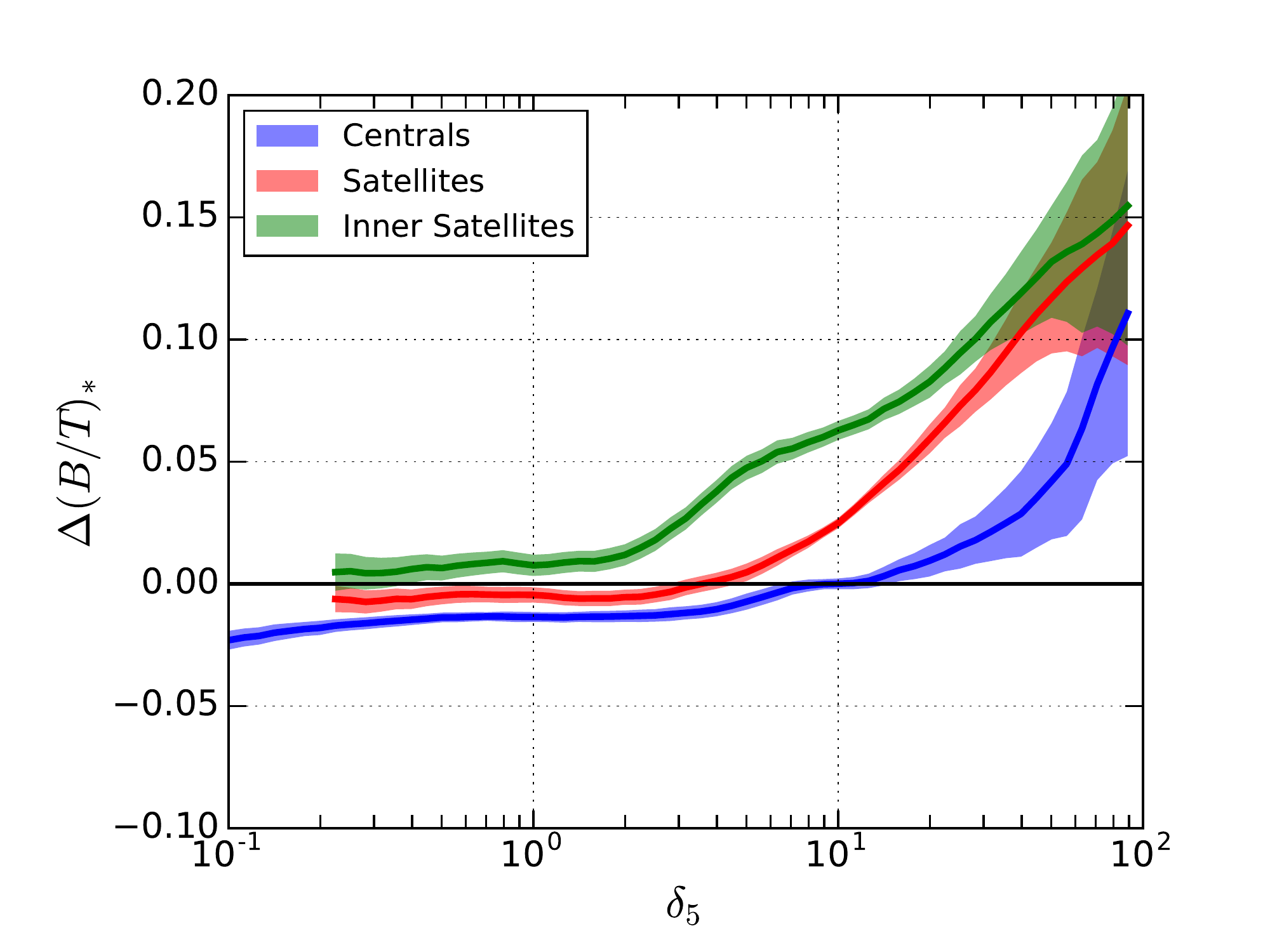}
\includegraphics[width=0.49\textwidth]{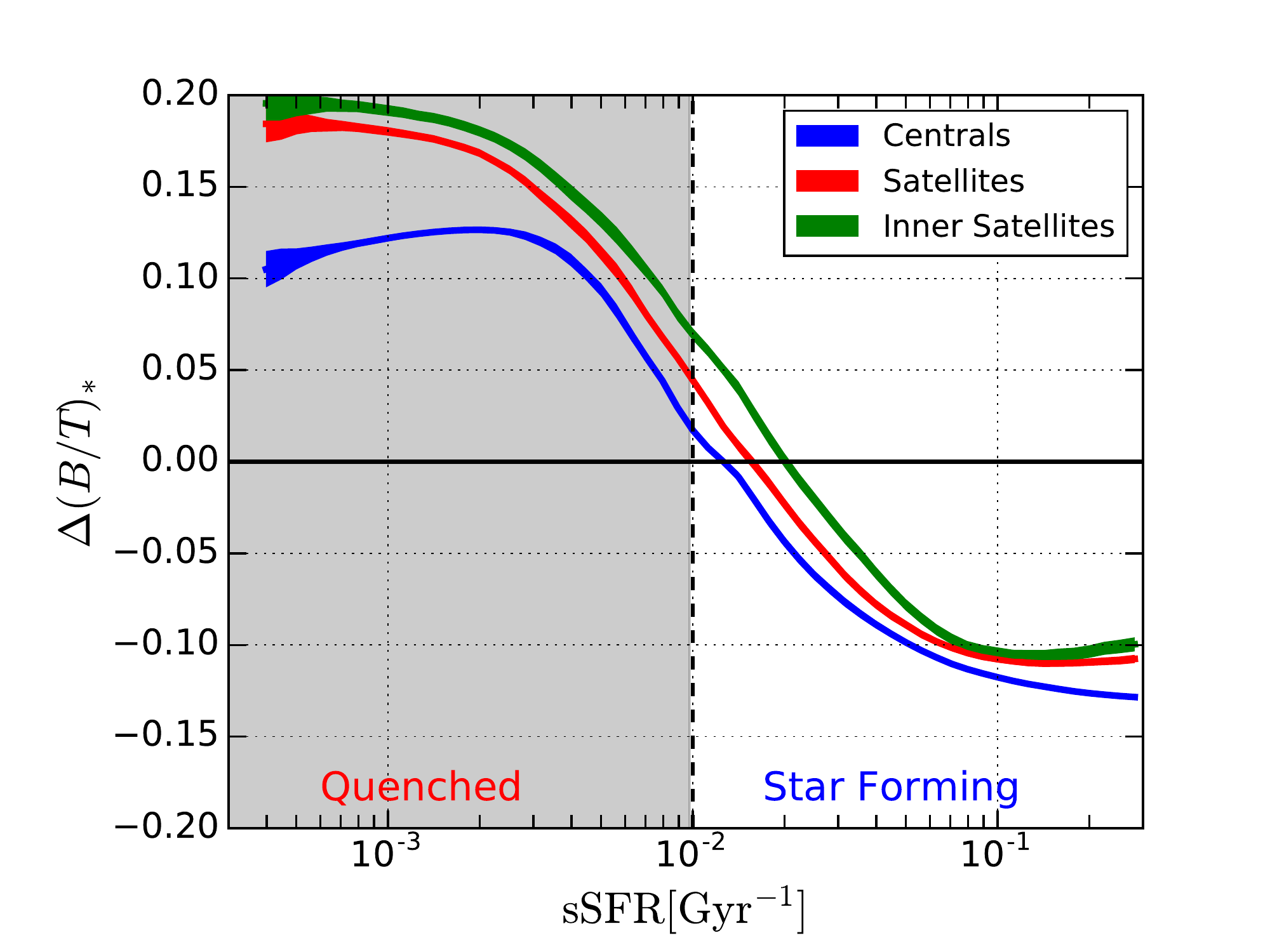}
\caption{The effect of environment and star formation on galaxy structure at a fixed stellar mass. Each panel shows the median enhancement in bulge dominance ($\Delta$(B/T)$_*$) relative to a control sample of all galaxies matched in stellar mass, plot for various galaxy populations (as indicated in the captions) against (from top left to bottom right): halo mass ($M_{\rm halo}$), projected distance from central galaxy in units of the virial radius ($D_{cc} = R_{cc} / R_{200}$), local galaxy over-density ($\delta_{5}$), and sSFR. The width of each coloured line indicates the 1$\sigma$ error estimated from a random Gaussian Monte Carlo re-sampling of the data. Note the different scale in the $Y-$axis between the top and bottom row. The shaded region on the bottom right panel highlights the high uncertainty of low sSFR values, which are measured indirectly.}
\end{figure*}

Given the strong dependence of B/T structure on stellar mass for both central and satellite galaxies (Fig. 3, top-left panel), in this sub-section we look for statistical deviations from the mean B/T - $M_{*}$ relation for different galaxy populations as a function of other parameters. This approach will expose any additional, more subtle, dependencies on galaxy structure, after first carefully removing the strong dependence on stellar mass. 

To investigate the impact of parameters on galactic structure formation at a fixed stellar mass we define a new statistic for each galaxy in our sample as follows:

\begin{equation}
\Delta {\rm (B/T)}_{*,j}  \equiv  {\rm (B/T)}_{*,j} -{\rm med}\big( {\rm (B/T)}_{*, {\rm controls}} \big)_{1/V_{\rm max}}
\end{equation}

\noindent where $({\rm B/T})_{*, j}$ is the bulge-to-total stellar mass ratio for each galaxy $j$. The second term on the right hand side of eq. 8 represents the $1/V_{\rm max}$ volume corrected median\footnote{To compute the volume corrected median we repeat entries of galaxies in our sample $N_{j}$ times (where $N_{j}$ = $V_{\rm max, survey} / V_{{\rm max}, j}$). We then sort the expanded sample and select the median and inter-quartile range as normal.} of a large sample of control galaxies, selected to be at the same stellar mass ($\delta M_{*}$ $\pm$ 0.05 dex) as each galaxy in the full sample in turn. The median number of unique controls per galaxy is $\sim$ 4000, and the $<$ 1\% of galaxies with fewer than 10 controls are excluded from the following analysis. The purpose of this statistic is to ascertain how parameters other than stellar mass impact galactic structure, removing the significant impact of stellar mass on the correlations of Fig. 3. On a galaxy-by-galaxy basis, $\Delta {\rm (B/T)}_{*,j}$ quantifies how offset B/T is relative to the median of all galaxies at the same mass. Additionally, we sub-divide the sample into centrals, satellites and inner satellites to ascertain whether or not these sub-populations behave differently in terms of their structural dependence on parameters other than stellar mass.

In Fig. 4 we plot the volume corrected median $\Delta {\rm (B/T)}_{*}$ for a variety of galaxy populations against the following parameters: halo mass ($M_{\rm halo}$, top left panel), projected distance to central galaxy ($D_{cc} \equiv R_{cc} / R_{200}$, top right panel), local over-density evaluated at the 5th nearest neighbour ($\delta_{5}$, bottom left panel), and sSFR (bottom right panel). The width of each coloured line indicates the 1$\sigma$ error on the  $\Delta {\rm (B/T)}_{*}$ statistic, computed from a random Gaussian re-sampling of the $X$ \& $Y$ axis input data within their intrinsic error values. Each coloured line is centred on the mean value of 100 realisations of the sample space.

In the top left panel of Fig. 4, we subdivide the analysis into centrals (blue), satellites (red) and inner satellites (green). Inner satellites are defined here to be satellites located within 0.2 virial radii of their host central galaxies, projected. A value of  $\Delta {\rm (B/T)}_{*}$ = 0 indicates no change relative to the controls, positive values indicate increased bulge structure, and negative values indicate decreased bulge structure (or equivalently increased disk structure) relative to the controls. Central galaxies have only a very weak dependence of  $\Delta {\rm (B/T)}_{*}$ on halo mass, whereas satellite galaxies have a stronger dependence (particularly at high halo masses), and inner satellites exhibit the strongest dependence (from intermediate to high halo masses). In high mass haloes, satellites have a median increase in bulge dominance up to 0.06, and inner satellites experience increases of up to 0.11. It is instructive to compare this panel with the top right panel of Fig. 3: despite the stronger dependence of galactic structure on halo mass for centrals than satellites globally, at a fixed stellar mass halo mass impacts galactic structure for satellites more than centrals, and moreover impacts inner satellites significantly more than the general satellite population.

In the top right panel of Fig. 4, we show the median $\Delta {\rm (B/T)}_{*}$ - $D_{cc}$ relation for all satellite galaxies (in black), satellites in clusters (defined as $M_{\rm halo} > 10^{14.5} M_{\odot}$, shown in red), satellites in high mass groups (defined as $10^{13.5} M_{\odot} < M_{\rm halo} < 10^{14.5} M_{\odot}$, shown in green), and satellites in low mass groups (defined as $M_{\rm halo} < 10^{13.5} M_{\odot}$, shown in blue). Note that all centrals would have a value of $D_{cc}$ = 0 by definition, and a mean $\Delta {\rm (B/T)}_{*}$ = 0, shown as the solid black line on all panels in this figure. Satellites in clusters have the largest bulge enhancements, relative to their stellar mass-matched controls, of any galaxy sub-population. They also exhibit the most clear trend with distance to the central galaxy. Interestingly, cluster satellites at all distances from their centrals (up to the virial radius) are bulge enhanced, unlike satellites in groups. Satellites in high mass groups have a significant, though weaker, enhancement at close proximities to their centrals. Satellites in low mass groups have the weakest dependence on distance from their centrals, and more or less trace the $\Delta {\rm (B/T)}_{*}$ = 0 line, with some scatter, at $D_{cc} >$ 0.2. However, in the central regions of low mass groups there is still a small enhancement in bulge dominance. These results indicate that it is a combination of halo mass and location within the halo which impacts galactic structure at a fixed stellar mass.

In the bottom left panel of Fig. 4, we show the median $\Delta {\rm (B/T)}_{*}$ - $\delta_{5}$ relation for central galaxies (blue), satellite galaxies (red) and inner satellite galaxies (green). At low to moderate over-densities, there is no dependence on density for any of the galaxy populations. At high over-densities, all galaxy populations exhibit a similar strong positive dependence of $\Delta {\rm (B/T)}_{*}$ on $\delta_{5}$. However, departure from the null dependence occurs at lower $\delta_{5}$ values for inner satellites than satellites, and for satellites than centrals; which suggests that the total integrated impact on B/T structure from local over-density at a fixed stellar mass is greater for inner satellites than any other galaxy population. We note that the highest increases in B/T, relative to the controls, reaches $\sim$ 0.15, indicating on average a 30\% increase in bulge-dominance for galaxies in highly over-dense regions.

Taken together, the first three plots of Fig. 4 indicate that environment plays a subtle, yet significant, role in shaping galactic structure in the local Universe, at a fixed stellar mass. These results are in essence a modification to the classic morphology - density relation (e.g., Dressler 1980, Dressler et al. 1994), updating previous approaches by computing offsets from a carefully matched control sample in stellar mass, and utilising a mass weighted definition of structure (as opposed to a wavelength-dependent structural measurement or morphological classification).

Finally, we look at the effect of varying star formation on B/T structure at a fixed stellar mass. In Fig. 4 bottom right panel, we show the median $\Delta {\rm (B/T)}_{*}$ - sSFR relation for centrals (shown in blue), satellites (shown in red) and inner satellites (shown in green). There is a strong dependence of B/T on sSFR, at a fixed stellar mass, particularly around the threshold of quenching (sSFR $\sim 10^{-2}$ Gyr$^{-1}$, shown as a dashed vertical line on the plot). Once again we note that the exact sSFR values for quenched systems are highly uncertain (and likely over-estimated). As a reminder of this fact, the quenched region is shaded in this panel. Quenched galaxies are preferentially more bulge-dominated, and star forming galaxies are preferentially more disk-dominated, than expected from a control sample of galaxies matched at the same stellar mass. Thus, the star forming state of galaxies significantly affects their structures at a fixed stellar mass, indicating that there are important {\it structural} differences between the population of star forming and quenched systems. This most probably points to different evolutionary tracks for quenched and star forming objects (as discussed in Section 6). For inner satellites and satellites, at all values of sSFR, B/T values are slightly enhanced relative to centrals. This implies that environmental effects impact the structures of galaxies at fixed stellar mass and sSFR, albeit in a relatively subtle manner.

%
%

\section{Ranking of Parameters -- Machine Learning Analysis}

In this section we present a ranking of how closely connected galaxy structure is to a variety of other galactic and environmental properties. To achieve this we follow an approach qualitatively similar to Teimoorinia, Bluck \& Ellison (2016; hereafter TBE), where we utilise an artificial neural network (ANN) to classify star forming and passive galaxies from given galaxy and environmental properties. The key result of TBE is that central velocity dispersion is more predictive of the star forming state of central galaxies than any other property considered, including structure / morphology, stellar mass, halo mass and several environmental parameters. Here we employ the same general methodology as TBE, but apply it to the question of what properties are most closely connected to galactic structure, quantified throughout in this paper with B/T by stellar mass. The general structure of the ANN approach and the key mathematical relations are presented in TBE, particularly in their Section 3. We only depart from this prior work in the technical detail of the implementation (explained below).

\subsection{ANN Method}

We use a set of publicly available and free machine learning tools from the powerful {\small SciKit-Learn} Python package. Specifically, we use {\small MLPRegressor} for a regression analysis (predicting actual B/T values per galaxy) and {\small MLPClassifier} for a classification analysis (predicting whether galaxies are early- or late-types). For both analyses we use a three-layered network, structured with neurons arranged in the sequence 20 : 10 : 5, with a $relu$-function activation, which closely approximates a universal function generator (e.g. Wichchukit \& Mahony 2011).  We use the {\small ADAM} minimisation solver (which is recommended in the {\small SciKit-Learn} documentation). 

For each individual ANN run, the sample is divided into a training set (30\% of the data) and a validation set (70\% of the data). The training set is provided with a truth value for each galaxy (B/T for regression; galaxy type for classification) and then the network constructs the optimal mapping (set of weights for each neuron) to move from the data to the prediction in the training sample. The trained network is then applied (in regression or classification mode) to the validation sample. Finally, the measured B/T values (or galaxy types) can be compared to those predicted from the network. Ultimately, the success or failure of parameters to accurately predict B/T values (or galaxy types) reveals how closely connected those parameters are to galactic structure.

More specifically, for the regression analysis, we use our measured B/T values as a target, (B/T)$_{t}$, in conjunction with a set of other parameters (e.g., $M_{*}$ and SFR) to train the network to construct a mapping from those parameters to a predicted B/T value, (B/T)$_{p}$. For the classification analysis (shown in the appendix as a test), we use our measured B/T values to construct a label (1, 0) to identify, respectively, early-type galaxies (with B/T $\geq$ 0.5) and late-type galaxies (with B/T $<$ 0.5). We then train the network to construct a mapping from the parameters provided to a probability for each classification type (i.e. probability of each galaxy being early- or late-type). 

In both cases (regression and classification), we perform 25 independent training runs (for 30\% of the data, randomly selected) followed by 25 independent regression or classification runs (for the remaining 70\% of the data), using the network to predict B/T values or galaxy morphological types, respectively. For regression, we use a volume weighted sample of galaxies, to best approximate the true distribution of galaxies in the local Universe. For classification, we utilise a `balanced' sample where we train on 50\% early-type galaxies and 50\% late-type galaxies, selected randomly from the volume weighted parent sample. The distinction is made because classification analyses tend to work better, and are more intuitive to interpret, in the case where classes are evenly sampled by the network. 

To quantify the performance, for regression, we compare the predicted B/T values to the measured B/T values and compute the mean square error on the data. For classification, we take the fraction of successfully classified objects as our performance statistic. The final performance of each variable (or set of variables) is given by the mean of the performance statistic from the 25 independent runs (excluding any which do not reach convergence), and the error is taken as the standard deviation across the converged runs\footnote{Typically, $>$20/25 runs converge, and those that do not are easily identified as outliers in the performance statistic. Lack of convergence is a common issue with machine learning approaches but we mitigate this in most cases by employing an early stopping routine to pre-process the network (see TBE for further discussion on this point.)}. A high success rate at predicting galaxy types, or a tight reproduction of B/T values, indicates that the parameter (or set of parameters) used in the training run are closely connected with galaxy structure.

\subsection{ANN Regression Analysis -- Single Variables}

In Fig. 5 we show the results of an ANN regression analysis to predict galaxy B/T values from a large set of other galaxy parameters, taken in isolation (as indicated by the $X$-axis labels).  Results are shown separately for centrals (in blue), satellites (in red) and inner satellites (in green). In the top panel, the $Y$-axis shows the root of the mean squared error, defined as:

\begin{equation}
\sqrt{\rm MSE} \equiv \sqrt{ \big\langle \big({\rm (B/T)}_{p} - {\rm (B/T)}_{t}\big)^{2} \big\rangle}
\end{equation} 

\noindent where ${\rm (B/T)}_{p}$ is the network predicted B/T value, given each individual parameter made available in turn, and ${\rm (B/T)}_{t}$ is the measured B/T truth value. As such, this analysis reveals the maximum potential of each parameter in turn for predicting B/T values for galaxies in the local Universe. Lower values of this statistic correspond to a higher connection between the property under consideration and galactic structure. 

In both panels of Fig. 5 the $X$-axes show each singular parameter under consideration (labelled in the legend). They are ordered from most to least predictive for central galaxies (shown in blue). The network's performance in B/T prediction as a function of galaxy properties is also shown for satellites and inner satellites, as red and green lines (respectively), which can be compared to the central galaxy case. Additionally, we present the performance of the network using simultaneously all eight galaxy and environmental parameters (labelled as `ALL'). This yields the maximum performance possible from the network plus the available input data. The `ALL' run gives a root mean squared error of $\sim$ 0.23, which is very close to the average error on the B/T values used for training and validation ($\langle \sigma_{B/T} \rangle \sim$ 0.2). Thus, the combination of all eight variables leads to a highly accurate prediction of galaxy structure, within the error limitation. 

We also show the results for a random variable (labelled `Rand.'). The performance of the network with a random parameter encodes information about how the distribution of B/T is structured for each galaxy sub-population.  As a very simple example, if there were only galaxies with B/T = 1 in the training set then the network would likely predict B/T values of 1 regardless of any additional parameters made available to it.  Hence, it is the improvement over random which is really interesting for each parameter. This is shown in the bottom panel of Fig. 5, and is defined as:

\begin{equation}
{\rm Improvement (\%)} = \frac{ \sqrt{\rm MSE}_{i} - \sqrt{\rm MSE}_{\rm rand} }  {{0 - \sqrt{\rm MSE}_{\rm rand}}} \times 100\%
\end{equation}

\noindent where $\sqrt{\rm MSE}_{i}$ gives the performance indicator for each variable in turn, $\sqrt{\rm MSE}_{\rm rand}$ gives the performance of the network utilising a random variable, and 0 is the maximum possible performance for regression (i.e. identical predicted and measured values).

From Fig. 5, we find the best single variable for predicting B/T values for centrals is $\Delta$SFR\footnote{ Note that we use $\Delta$SFR as opposed to sSFR in this section to avoid explicitly combining SFR and $M_{*}$ in the ANN analysis. $\Delta$SFR is defined as the logarithmic distance from the star forming main sequence each galaxy resides at (e.g., Bluck et al. 2014, 2016). This encodes {\it different} information about galaxies to SFR, and indeed we find here that it is significantly superior to SFR for predicting B/T values.}, followed by $M_{*}$ and then $M_{\rm Halo}$. $\Delta$SFR and $M_{*}$ are also the most predictive for satellite and inner satellite galaxies as well. However, $M_{\rm halo}$ performs much worse for satellites and inner satellites than for centrals. This is most likely due to the tight correlation between halo mass and stellar mass present for centrals but absent for satellites (which is discussed above in Section 3.1). It is interesting that $\Delta$SFR performs significantly better than SFR for predicting galaxy structure. This fact highlights that it is not SFR per se which is closely connected to B/T (and vice versa) but rather whether or not a galaxy is forming stars on the main sequence. Thus, whether a galaxy is star forming or quenched is a key indicator of galaxy structure, but precisely where on the main sequence a galaxy resides is not strongly connected to galaxy structure.

We emphasize that in this section (as in Section 3.3) we are using simultaneously two qualitatively different approaches for assessing environment - 1) splitting the population into centrals, satellites and inner satellites; and 2) investigating the predictive power of quantitative metrics (i.e. local density, distance from central, halo mass). Our motivation for this dual approach is to expose whether or not the explicit quantitative environmental dependence on B/T is dependent on the type of galaxies under investigation. Perhaps surprisingly, for all galaxy types, explicit environmental parameters perform much worse than intrinsic metrics as predictors of galaxy structure.

\begin{figure*}
\includegraphics[width=0.8\textwidth]{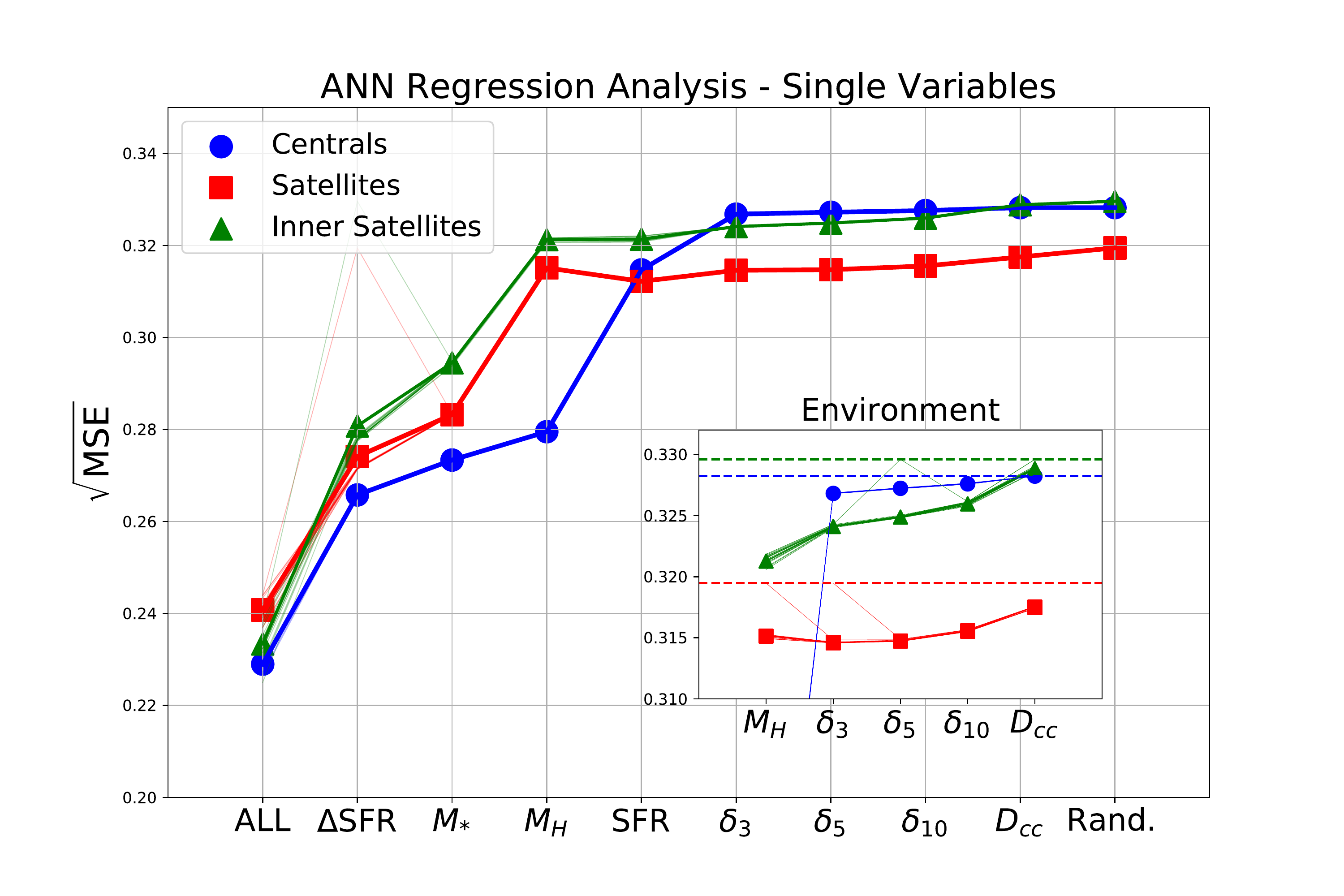}
\includegraphics[width=0.8\textwidth]{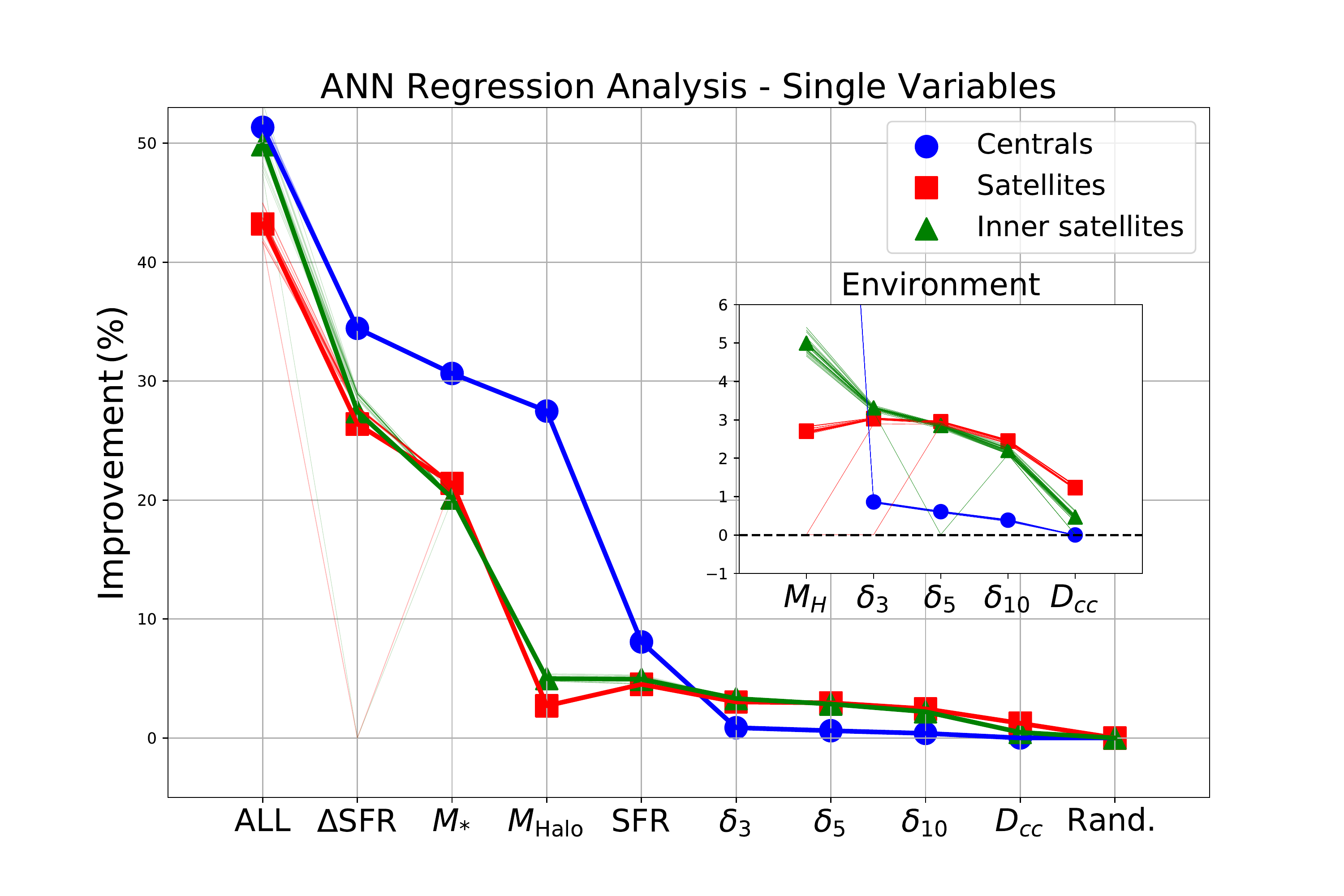}
\caption{Artificial neural network (ANN) regression analysis to determine which parameters are most useful for predicting galaxy structure (B/T). Results are shown separately for centrals (blue), satellites (red) and inner satellites (green). On the top panel, the $Y$-axis shows the root mean square error (defined in eq. 9), which quantifies the performance of each variable in predicting B/T. The $X-$axis labels each of the single variables considered, as well as a run utilising all inputs simultaneously (labelled `All'), and a run with a random variable (labelled `Rand.'). The $X-$axis is ordered for both panels from most to least predictive of galaxy structure for central galaxies. The bottom panel reproduces the same results as the top panel, but shown as a percentage improvement over random (defined in eq. 10). This statistic effectively takes into account differences in the distributions of B/T for the different sub-populations of galaxies considered. On each panel, the results from 25 independent ANN runs are shown as light coloured lines, with their average value shown as a large shape-marker, as indicated on the legend. Typical errors for individual parameters are $\sim 1\%$, as estimated from the variance of the 25 ANN runs.}
\end{figure*}

\begin{figure*}
\includegraphics[width=0.8\textwidth]{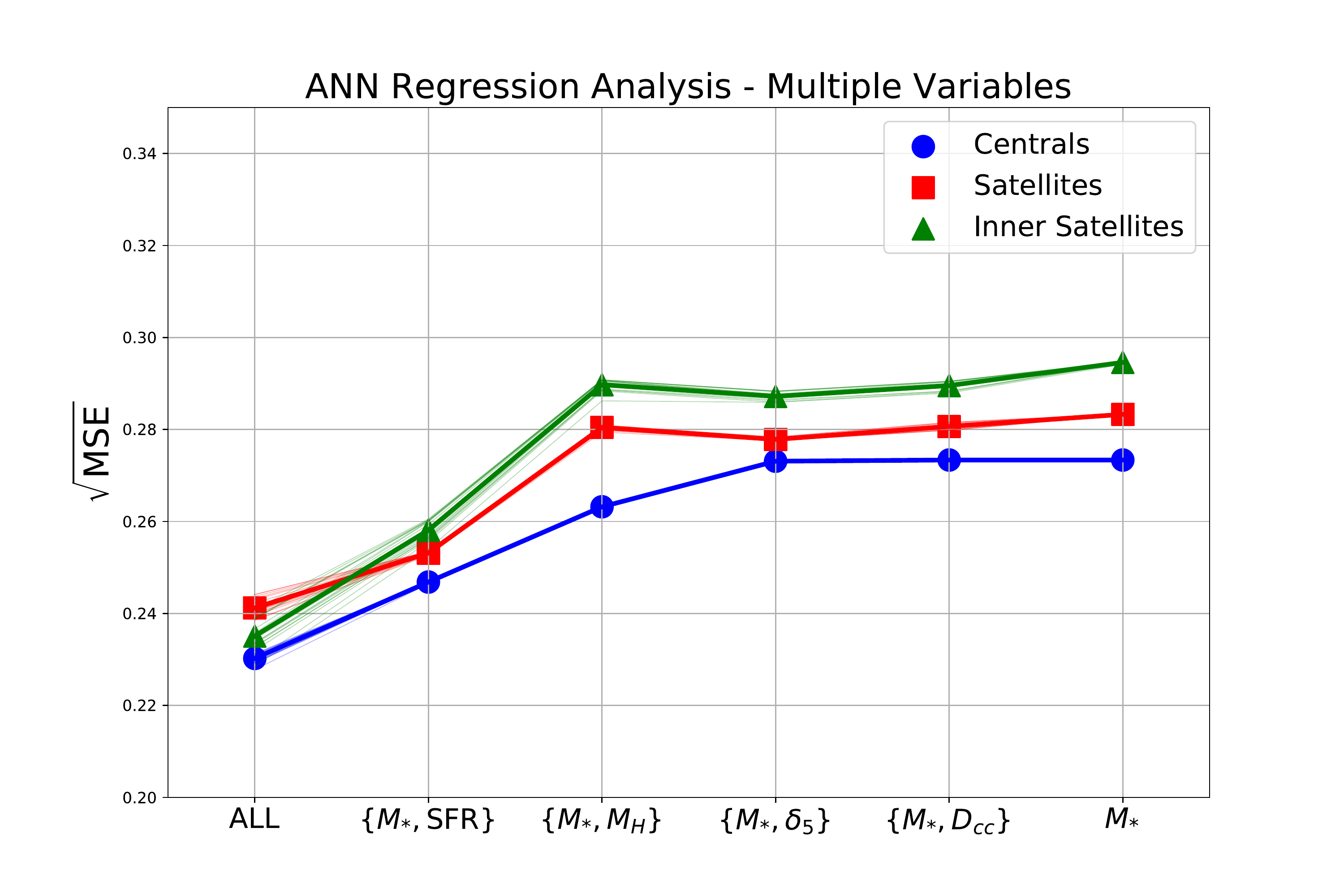}
\includegraphics[width=0.8\textwidth]{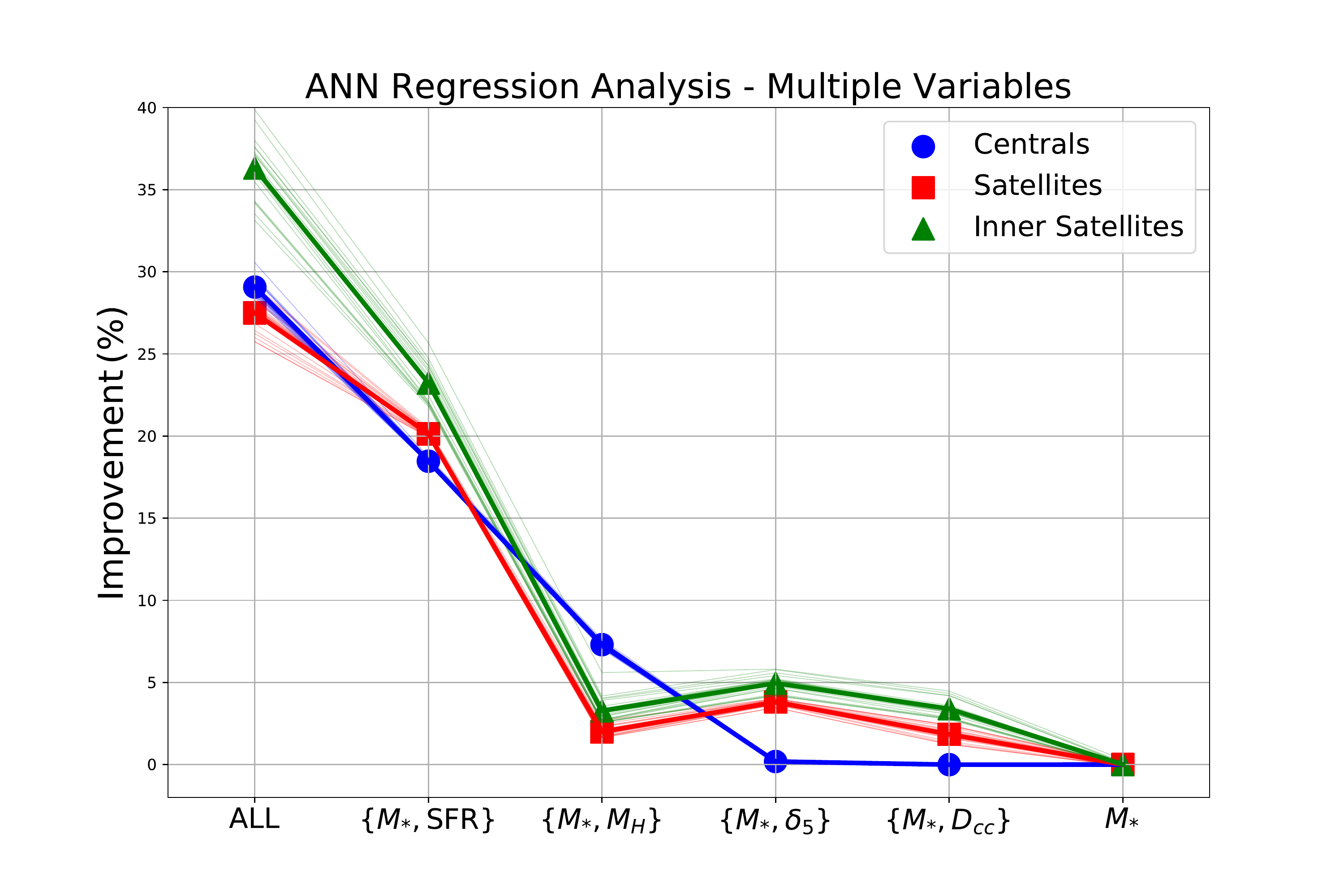}
\caption{Artificial neural network regression analysis to predict B/T from multi-parameter sets of variables (indicated on the $X$-axes). The results for central galaxies are shown in blue, satellite galaxies in red, and inner satellite galaxies in green. The $Y$-axis on the top panel shows the root mean square error of predicted to measured B/T values, and the $Y$-axis on the bottom panel shows the percentage improvement of the performance of the network for the multi-parameter set relative to a regression analysis with stellar mass alone. Hence, this plot shows how useful other parameters are at enhancing B/T determination, after fully accounting for the mass dependence. 25 independent ANN runs are shown as light coloured lines, with the mean of the set shown as a large marker-shape. The error inferred from the variance of these runs is typically $\sim$ 2\% for the 2-variable sets and $\sim$ 4\% for the `ALL' run.}
\end{figure*}

To look at the subtle dependence of environment on galaxy structure, we show the results for quantitative environmental parameters in an enlarged inset on each panel in Fig. 5.  In the top panel, the random parameter result is shown for each galaxy sub-population as an appropriately coloured dashed line. The improvement over random, bottom panel, is relative to the random result for each sub-population. Looking at the inset on the bottom panel of Fig. 5, we see that environmental parameters are significantly more predictive of B/T for satellites and inner satellites than for centrals, with the sole exception of halo mass (which is closely correlated with stellar mass for centrals). For satellites, density parameters are slightly superior to halo mass, which in turn is superior to distance from central as a predictor of B/T. Whereas, for inner satellites, halo mass becomes the most useful quantitative environmental parameter for predicting B/T structure. However, it must be stressed that {\it no} environmental metric is highly predictive of galaxy structure for any galaxy type, with the sole exception of halo mass for centrals.

In general, intrinsic galaxy properties are more predictive of galaxy structure for centrals than for satellites or inner satellites (red and green lines lying below the blue lines in the improvement plot for the first four single variables). Conversely, environmental parameters are slightly more predictive of galaxy structure for satellites and inner satellites than for centrals (red and green lines lying above the blue lines in the improvement plot for the last four single variables). This suggests that, although intrinsic properties (particularly $\Delta$SFR \& $M_{*}$) are superior to environmental properties for establishing galaxy structure in all types of galaxies, the impact of varying environment on B/T ratio is greater for satellites and inner satellites than for centrals. 

From the environmental parameters inset, we do not find any significant difference in predictive power of local density evaluated at 3rd, 5th or 10th nearest neighbour (hence, retrospectively, our choice to focus just on the 5th nearest neighbour in Section 3). However, we note that halo mass is significantly more predictive of galaxy structure for inner satellites than the full satellite population. Taken in isolation, distance from the central galaxy is the least predictive measure of galaxy structure available from our list for satellites and inner satellites, and (as expected) it performs identically to a random variable in the central galaxy case (where $D_{CC}$ = 0 by definition). However, when used in the classification of a sub-population (i.e. inner satellites) it does have the power to affect how B/T correlates with other parameters (as demonstrated by the green lines not being fully synchronous to the red or blue lines in Fig. 5).

In addition to the ANN regression analysis, we also consider the strength of the correlations between each single variable and B/T structure using the Spearman rank correlation coefficient (shown in Appendix C1). The correlation approach yields the exact same ordering of parameters from most to least predictive of galaxy structure, which adds further confidence to our machine learning results. Furthermore, we also consider an ANN classification analysis, which determines the probability of each galaxy being bulge- or disk-dominated from the single variables in turn (shown in Appendix C2). This also yields an almost identical result to the ANN regression ranking shown here. The only significant difference is that the ordering of $\Delta$SFR and $M_*$ is switched. All of the other rankings are essentially the same within their respective errors. We comment on possible reasons for this small differences in the appendix, but note here that the effectiveness of a parameter for estimating the exact B/T value and for predicting which side of the B/T = 0.5 line a galaxy will reside on, though closely connected, are ultimately different questions which may logically lead to different answers. Overall, there is a very high level of consistency between rankings from ANN regression, ANN classification, and correlation coefficient analysis, suggesting the stability of our primary results to statistical method (see Table C1 for a summary).

\subsection{ANN Regression Analysis -- Multiple Variables}

In this sub-section we consider the performance of the network for predicting galaxy structures for a few selected multi-parameter combinations. Since the ANN regression, ANN classification, and correlation analyses all yield such similar results for single variables, for the sake of brevity, we restrict our multi-parameter runs to the regression case. In Fig. 6 we show the root mean square error and percentage improvement over an ANN run with stellar mass alone, for a set of multi-parameter input variables (as indicated on the $X$-axis). Central galaxies are shown in blue, satellites galaxies are shown in red, and inner satellite galaxies are shown in green. The approach of looking for improvement in B/T determination for variables {\it in addition to} $M_{*}$ is qualitatively similar to the $\Delta (B/T)_*$ analysis of Section 3.3. But here we concentrate on ranking the multi-parameter sets, whereas in Section 3.3 we focused on exploring the general trends. For all types of galaxies, the run utilising stellar mass alone yields an improvement of 0\% by definition. The error on the root mean square error, and the percentage improvement, are given by the standard deviation across the 25 independent network runs.

In this sub-section we concentrate on SFR as the sole star forming metric, because a sophisticated multi-layered network, like the one we are employing here, can easily disentangle both sSFR and $\Delta$SFR from an input of $M_{*}$ and SFR, should it prove statistically advantageous to do so. Moreover, given that our implementation of ANN is a universal function generator, any combination of parameters is effectively included with the original parameters in each run. We have checked that the performance of the network with SFR and $M_{*}$ is identical to the performance with sSFR or $\Delta$SFR and $M_{*}$ (within the errors generated by the network variance), and indeed it is. Were this not the case, this would provide evidence that the network was either not fully converged, or that there were serious defects in the data. Happily, this is not the case.

For all types of galaxies (centrals, satellites and inner satellites) the most useful parameter to add to stellar mass for predicting B/T values is SFR, which performs significantly better than any other parameter considered here. This result indicates that the star forming state of galaxies is an important indicator of galactic structure, even at a fixed stellar mass. Thus, the connection between structure and star formation is not reducible to an underlying dependence on mass or environment. The latter point is established by the fact that adding SFR to $M_{*}$ is more effective than adding any quantitative environmental parameter to $M_{*}$. We also checked the converse, adding $M_{*}$ to SFR (as opposed to the other way around) and find a similar result. That is, in addition to SFR, $M_{*}$ adds significant improvement to the accuracy of predicting B/T values. Hence, both SFR and $M_{*}$ are closely connected to B/T structure in such a manner that neither connection is reducible to a correlation with the other variable.

For central galaxies, we find adding each environmental parameter in turn leads to no significant improvement over stellar mass alone, with the sole exception of halo mass, which leads to a significant improvement of B/T determination. Satellites have in general a greater improvement in B/T determination from addition of quantitative environmental parameters to stellar mass than centrals, with inner satellites having the most significant improvement over both centrals and the general satellite population. The most useful environmental parameter to add to stellar mass for determining B/T in satellites is local density (here evaluated at the 5th nearest neighbour only for brevity), followed by a similarly weak performance of halo mass and distance to central galaxy. We find similar results to these multi-parameter runs utilising a partial correlation analysis technique (shown in Appendix C as an additional test on our results).

A discussion on possible interpretations of the rankings presented here is provided in Section 6.1.

%
%

\section{Comparison to Simulations}

In this section we turn our attention to the question of what causes bulge growth and galactic scale structural transformations. Given the immense complexity involved in galaxy evolution it is extremely difficult to deduce the nature of physical processes directly from observations. As such, we employ as an interpretative tool a semi-analytic model of galaxy formation and evolution (LGalaxies: the Munich Model, Henriques et al. 2015), and also make a comparison to a cosmological hydrodynamical simulation (Illustris: Vogelsberger et al. 2014a,b). We begin this section with a summary of the simulations and then progress to a careful comparison with the SDSS observations.

\subsection{Summary of the Models}

\subsubsection{LGalaxies}

The full details for the prescription of galaxy formation and evolution in the Munich Model (LGalaxies) is provided in Henriques et al. (2015), with earlier versions of the model explained in Croton et al. (2006), De Lucia et al. (2009), and Guo et al. (2011). In this subsection we give a summary of the most relevant details of the model, but refer readers to the above publications for a more thorough description.

LGalaxies, like all SAMs, is built upon halo merger trees, in this case taken from the N-body dark-matter-only Millennium Simulation (Springel et al. 2005). The physics of baryons, and ultimately galaxies, is built upon the established dark matter halo tree via a simple set of coupled differential equations, which aim to parameterise the key physics of galaxy formation and evolution. Free parameters in the model are assigned to values based upon comparison to certain key observations (most commonly the multi-epoch stellar mass function and star forming main sequence, and the z=0 fraction of red/ quenched galaxies as a function of stellar mass). In the modern rendering of the LGalaxies SAM, the free parameters are fit via a Markov Chain Monte Carlo (MCMC) simulation, using the above observational diagnostics as a constraint (see Henriques et al. 2013, 2015). 

The resulting predictions for galaxy properties from the LGalaxies model can then be compared to observations, especially those which were not used in the training of the model parameters. If there is consistency between the model and the new observations, the model may be used as an explanatory tool to aid interpretation of the observations. If there is disagreement between the model and the observations, the observations can be used to re-tune the free parameters in the model, and hence improve the model's fidelity. Importantly, in the case where re-sampling the parameter space of free variables in the model cannot improve agreement with observations, new physics must be added to the model. A notable example of this is the inclusion of radio-mode AGN feedback into SAMs to account for the high mass end of the galaxy stellar mass function (e.g., Croton et al. 2006, Bower et al. 2006). Similar physical prescriptions are now utilsed as sub-grid physics in modern cosmological hydrodynamical simulations as well (e.g. Vogelsberger et al. 2014a,b, Schaye et al. 2015).

The two most important parameters for our comparison between the SDSS and LGalaxies are stellar mass and bulge mass. We explain how each are determined in turn. In LGalaxies, the formation of stars in galaxies is split into two channels: 1) a `normal' star formation mode caused by gravitational collapse of gas clouds accreted into a disk component (above a critical density threshold set by the empirical Kennicutt-Schmidt relation, Kennicutt 1983, 1989, 1998); and 2) a `star burst' mode, which originates in gas rich galaxy mergers (of all mass ratios) and deposits new stars into a bulge component. Over 90\% of stars formed in the model at all redshifts are formed through the normal star forming mode, leaving $<$ 10 \% of stars left to form in violent interactions of gas rich systems. Thus, the formation of stellar mass in the model is predominantly undertaken in situ and without violent interactions.

There are also two principal routes available for growing a bulge in LGalaxies: 1) the formation of new stars directly into a bulge component (from the star burst mode of star formation, explained above), and 2) the redistribution of extant disk stars into a bulge component during a galaxy merger. Note that both of these channels are only available during galaxy - galaxy interactions and mergers. The latter channel (redistribution of stars from a merging disk) is by far the most important route for bulge growth in the model, and accounts for the vast majority of bulge stars. Major and minor mergers are treated separately in the model; where major mergers are defined to have a mass ratio $\mu > 1/10$, and minor mergers are defined to have a mass ratio $\mu < 1/10$. In a major merger the stars and gas of both systems are placed into a bulge component. Thereafter, disks may reform if conditions are appropriate. In a minor merger only the gas and stars of the minor companion are placed into a bulge, i.e. the host galaxy disk is left intact. In both cases, gas from the minor companion may be tidally stripped in the early stages of interaction (the satellite phase) leading to the growth of the major companion's hot gas halo. Additionally, in all mergers, a few percent of stars are placed in the stellar halo.

In addition to bulge formation and growth via mergers, disk stars can also be deposited in a bulge component directly in a `violent disk instability' (VDI) mode. If the disk mass becomes large relative to the global dark matter halo potential, fragmentation can occur triggering this mechanism. In the current implementation, when a disk becomes kinematically unstable, the minimum amount of mass needed to achieve stability is transferred from the disk to the bulge. This is unlike some other SAMs which will move up to the entirety of the disk to the bulge in a disk instability (e.g., Somerville et al. 2015). As such, the VDI mode in LGalaxies does not significantly affect galaxies, with $<$ 2\% of bulge stars originating from the parent disk via violent disk instabilities, in the current implementation.

In summary, stars in LGalaxies are formed primarily in disks via gas accretion during secular evolution. Bulges, on the other hand, are formed primarily from extant disk stars, which have their aligned angular momentum vectors randomly disrupted during violent merger events. Major mergers are the most effective process for growing bulges and spheroids, but minor mergers are more frequent. Consequently, major and minor mergers are roughly equally important to bulge growth in the model. Star bursts and VDIs are also present in the model, but are together responsible for only $\sim$10\% of the mass of galaxy bulges.

\subsubsection{Illustris}

The Illustris project simulates the gravitational and hydrodynamical evolution of a $\sim 100 {\rm Mpc}^{3}$ co-moving volume of the Universe, assuming a $\Lambda$CDM cosmology with initial conditions given by the WMAP-9 results (Hinshaw et al. 2013). Full details on the simulation are given in Vogelsberger et al. (2014a,b). 

In Illustris, as with all cosmological hydrodynamical simulations, physical processes on scales below the resolution limit (e.g., star formation, feedback from supernovae and AGN, metal production and the formation and evolution of supermassive black holes) are governed by sub-grid physical prescriptions. The sub-grid physics is handled by simple coupled sets of differential equations exactly as in a SAM; the principal difference is that the large scale hydrodynamical evolution of galaxies is simulated directly (see, e.g., Sijacki et al. 2007,  Vogelsberger et al. 2013, Torrey et al. 2014 for further details).

To compare our observational results on the structures of galaxies with the Illustris simulation, we utilise a bulge-disk decomposition of processed galaxy images from the z $\sim$ 0 Illustris snapshot performed in Bottrell et al. (2017a). Galaxies in Illustris are pre-processed through the {\small SUNRISE} radiative transfer code, dimmed to the appropriate apparent luminosity given the redshift, and inserted into realistic SDSS backgrounds (including appropriate sky noise, neighbours, and multiple orientations). There is considerable effort in the Bottrell et al. (2017a,b) papers to approximate as closely as possible the realistic view of Illustris galaxies, as if observed in the SDSS.

In Bottrell et al. (2017a) photometric bulge-disk decompositions are performed in the $g$- and $r$-bands only. Hence, component stellar masses derived via multi-wavelength SED fitting (i.e. in the same vein as those used in the bulk of this paper) are therefore unavailable at present. As a result of this, in this section we compare B/T fractions in $r$-band between Illustris and the SDSS. It is important to note that these will be different to those computed by stellar mass (see Appendix B for a detailed exploration of structural dependence on waveband). However, the most important thing here is that we make a fair comparison of like with like.

In addition to the photometric B/T ratios from Bottrell et al. (2017a), stellar masses of structural components of Illustris galaxies may be estimated by allocating star particles to a bulge or disk component using their specific angular momenta\footnote{The essential idea is to construct the distribution of star particle angular momenta. For a disk structure, this should be approximately a Gaussian distribution centred around some positive characteristic angular momentum of the galaxy. However, for a bulge component, the random orientation of angular momenta of star particles leads to a Gaussian distribution centred around {\it zero}. The bulge mass is then taken as simply twice the mass of stars with negatively aligned angular momentum relative to the global angular momentum of the galaxy (assuming a symmetric distribution). This works well at high masses, where the global galaxy angular momentum is high relative to zero, but becomes highly biased with low S/N for lower mass galaxies. See Genel et al. (2015) for a more detailed account.}. 
However, accurate kinematic separation of the structural components of Illustris galaxies becomes highly challenging at lower masses ($M_{*} < 10^{11} M_{\odot}$). Thus, restricting our analysis to kinematic measurements alone would significantly reduce the range over which the structures of galaxies can be compared in Illustris to the SDSS. Moreover, the $g-$ and $r-$band decompositions of Illustris galaxies in Bottrell et al. (2017a) are more consistent in methodology to our measurements for SDSS galaxies in any case. Hence, we utilise these as our primary means of comparison to Illustris. Nonetheless, we will also show the kinematic relationships as well as the photometric measurements in some figures for comparison, in the mass range where they are reliably measured.

\subsection{Simulations Results}

\begin{figure}
\includegraphics[width=0.49\textwidth]{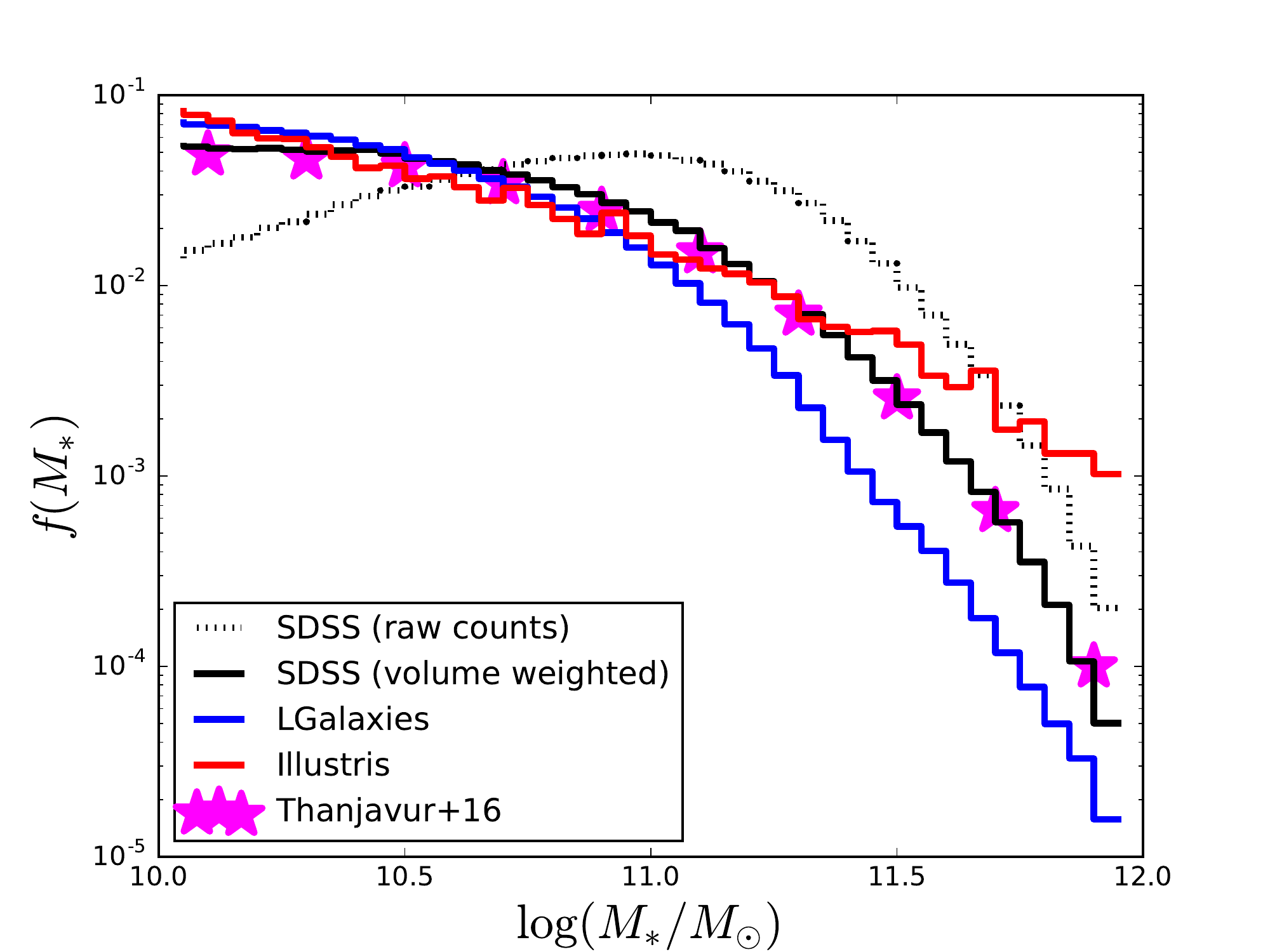}
\includegraphics[width=0.49\textwidth]{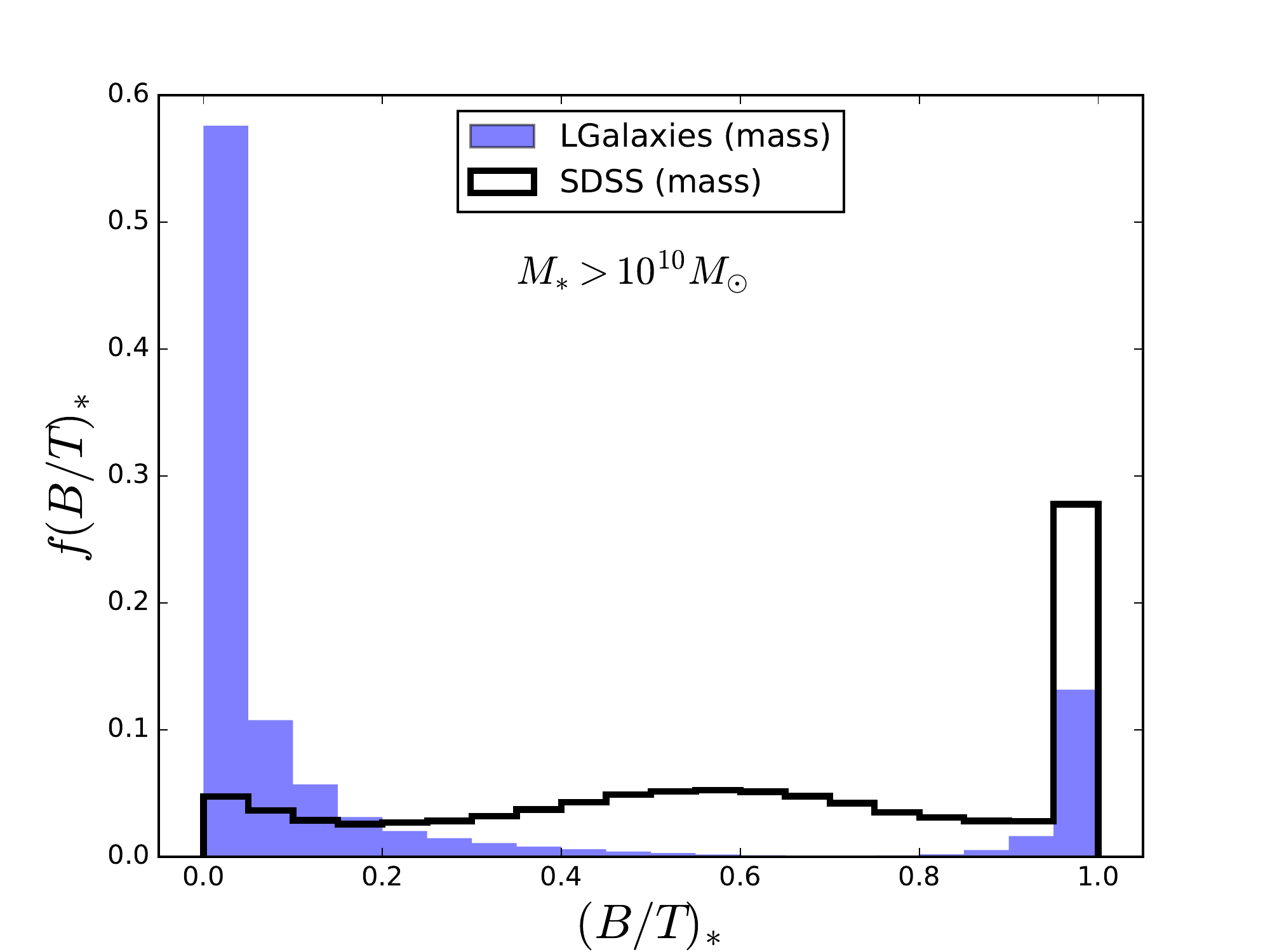}
\includegraphics[width=0.49\textwidth]{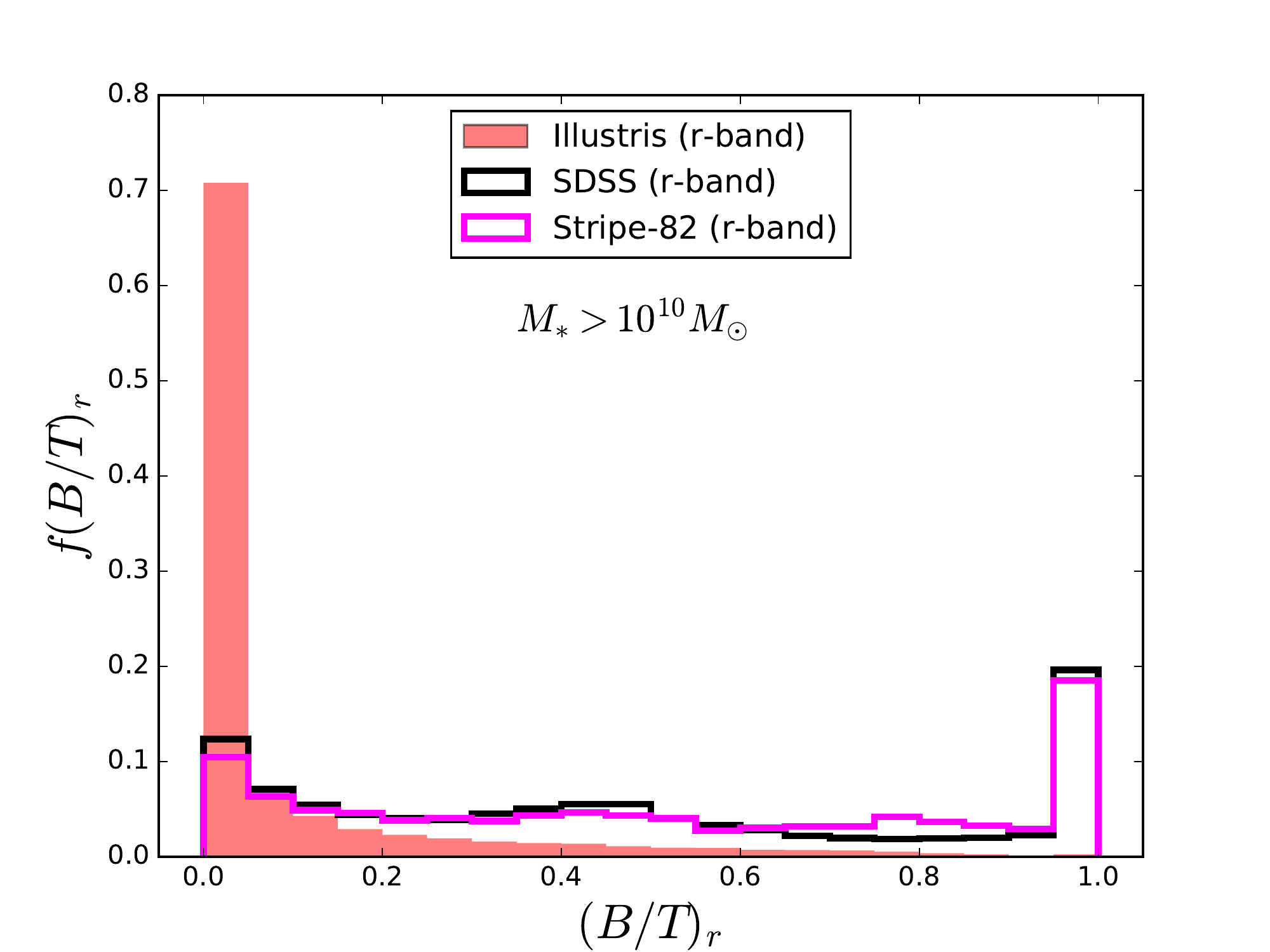}
\caption{{\it Top panel:} Stellar mass distributions for the SDSS (raw counts and volume weighted), Illustris and LGalaxies. Note that this figure indicates a mass distribution rather than a mass function, with the area under each line set equal to unity. The corresponding stellar mass distribution derived from the stellar mass function of Thanjavur et al. (2016) is overlaid as a check for sample biases. {\it Middle panel:} The distribution of B/T ratio (evaluated in stellar mass) for LGalaxies and the SDSS. {\it Bottom panel:} The distribution of B/T ratio (evaluated in $r$-band) for Illustris and the SDSS. Additionally, we show the distribution of B/T ratios in $r$-band with deeper imaging from Stripe-82.}
\end{figure}

\subsubsection{Stellar Mass Distributions}

In the top panel of Fig. 7 we show the stellar mass distribution of our sample of SDSS galaxies (as a light grey dotted line). We also show the volume weighted stellar mass distribution of our SDSS sample (as a solid black histogram). This is essentially the stellar mass function for the SDSS, but is displayed here as a distribution in the sense that the integral over the full mass range is defined to be equal to unity, not the number of galaxies in a given volume. The reason for this distinction is that we have reduced our sample slightly by requiring that there are successful bulge-disk decompositions in all four primary wave-bands and by stellar mass, as well as requiring a variety of other successful measurements (e.g. SFR, local density, and group identification). As such, our sample excludes $\sim10 \%$ of SDSS galaxies, which is necessary in order to achieve the scientific goals of this paper.

To check if our reduced SDSS sample is biased in any way compared to the total sample, we overlay on Fig. 7 (top panel) the volume weighted mass distribution of the full SDSS sample taken from the stellar mass function analysis of Thanjavur et al. (2016, shown on the plot as magenta stars). Since the distributions of stellar masses are almost identical, we conclude that our sample is not biased with respect to stellar mass in any significant manner.

Our primary purpose in this comparison section is to compare the structures of galaxies in Illustris and LGalaxies to that of the SDSS. However, before addressing this question we first check to see how well the distribution of stellar mass is reproduced in each simulation. In Fig. 7 (top panel) the distribution of stellar masses is shown for the Illustris simulation (red histogram) and the LGalaxies model (blue histogram). The general shape of the stellar mass distributions are similar between the simulations and the SDSS, but there is not perfect agreement. In particular, Illustris over-predicts the abundance of massive galaxies relative to the SDSS, and LGalaxies under-predicts the abundance of massive galaxies relative to the SDSS. 

Given that the mass distributions of the simulations are similar to that of the SDSS, it is reasonable to compare directly the structural and morphological B/T distributions of the samples (shown in the next sub-section). However, since the mass distribution agreement is far from perfect, it is also sensible to restrict our final comparison to fixed stellar mass binnings (shown in Sections 5.2.3 \& 5.2.4).

\subsubsection{B/T Distributions}

In the middle panel of Fig. 7 we show the mass weighted distribution of B/T structure for the SDSS (shown as a solid black line) and for LGalaxies (shown as a shaded blue region). Both samples are selected at $M_{*} > 10^{10} M_{\odot}$. It is clear that, relative to the SDSS, LGalaxies under-predicts the abundance of spheroidal and composite (bulge + disk) systems, and consequently over-predicts the abundance of disk-dominated systems. Qualitatively, the agreement between the two distributions is quite poor, which furthermore suggests that a reasonable approximation of the stellar mass distribution does not guarantee a reasonable approximation of the structural distribution of galaxies. Interestingly, the discrepancy between LGalaxies and the SDSS is in the opposite direction as that found by Wilman et al. (2013) using an earlier version of the Munich model, whereby they find the model to over-predict the abundance of spheroidal galaxies with respect to observations. However, it should be noted that the earlier version of the Munich model had a significantly poorer agreement to observations with the stellar mass function, and that the observational sample utilised optical morphology as opposed to mass structure, and analysed a much smaller sample of galaxies than considered in this paper.

In the bottom panel of Fig. 7 we show the $r$-band B/T morphological distribution of Illustris galaxies (shown as a red shaded region), compared to the $r$-band B/T morphological distribution of observed SDSS galaxies (shown as a solid black line). Both samples are selected at $M_{*} > 10^{10} M_{\odot}$. As with LGalaxies, Illustris significantly under-predicts the fraction of bulge-dominated spheroidal galaxies and composite systems, consequently significantly over-predicting the relative abundance of disk-dominated systems (as seen in Bottrell et al. 2017b). Furthermore, in Illustris there are essentially no pure spheroidal galaxies found in $r$-band morphology at $M_{*} > 10^{10} M_{\odot}$. This dearth of spheroids is surprising, and clearly disagrees with observations, both in this study and in the literature (e.g. Bamford et al. 2009, Wilman et al. 2013, Thanjavur et al. 2016).

To investigate whether the shape of the SDSS morphological distribution is affected by depth limitations of the survey, in the bottom panel of Fig. 7 we show the distribution of $r$-band B/T ratio for the SDSS Stripe-82 data, which is deeper by $\sim$ two magnitudes than the full SDSS survey. Although the B/T distribution for Stripe-82 is slightly different (as expected given that it is a different sample), it is qualitatively very similar to the SDSS distribution, and hence very different to the Illustris distribution. Consequently, we find that the depth of SDSS observations does not significantly affect our measured B/T ratios, or their distributions (see Appendix A2 for further tests on depth; and Bottrell et al. 2018 for the full analysis of the impact of depth on B/T measurements in the SDSS).

\subsubsection{Stellar Mass Dependence}

\begin{figure}
\includegraphics[width=0.49\textwidth]{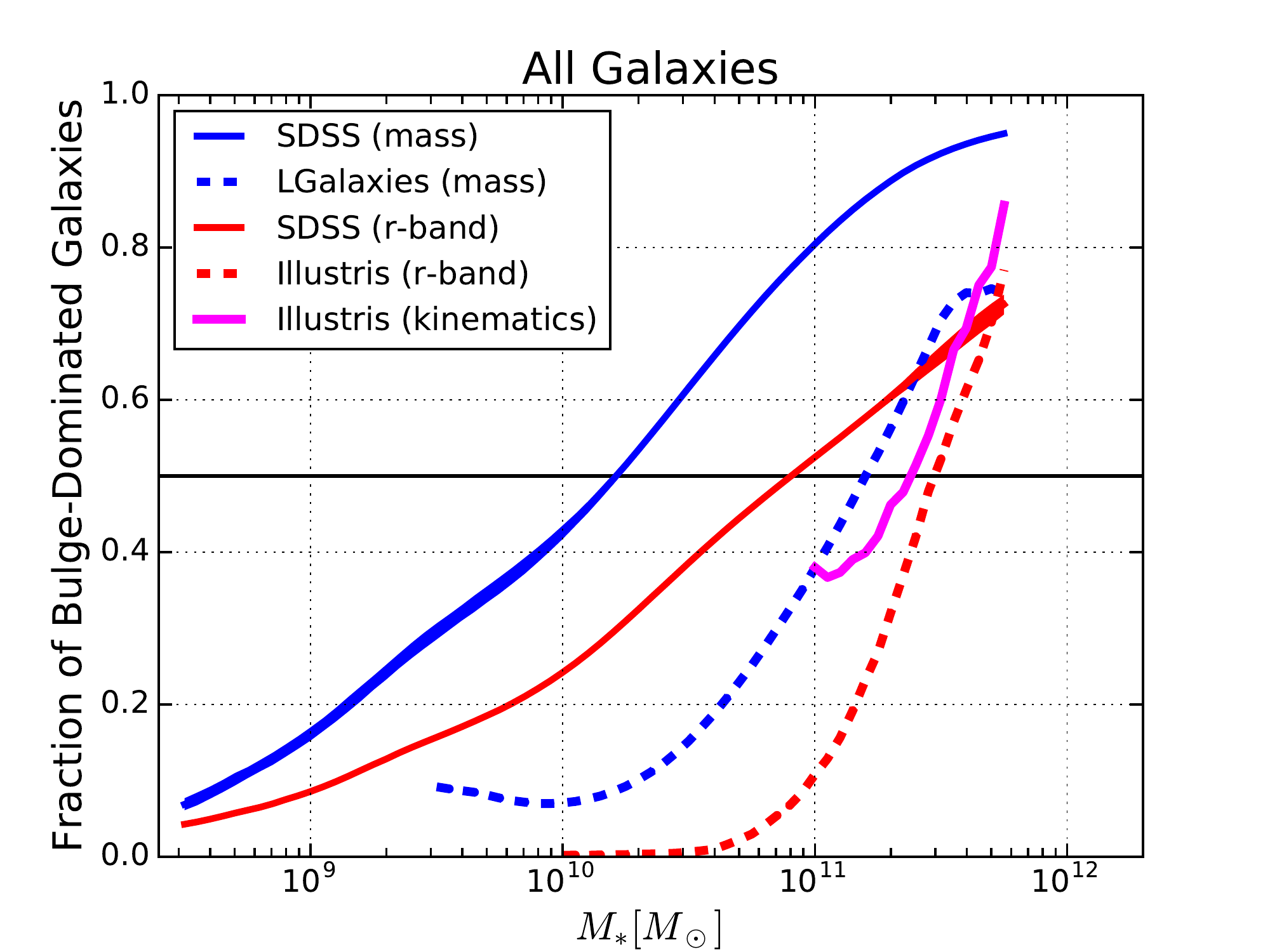}
\caption{The fraction of bulge-dominated galaxies as a function of stellar mass, shown for the SDSS, LGalaxies, and Illustris. Bulge dominated galaxies are defined as having $(B/T)_X \geq 0.5$, where $X$ is defined variously to indicate the stellar mass, $r$-band, or kinematic ratio (as indicated on the legend). Like coloured lines may be compared freely, whereas differing coloured lines should be compared only with caution. For the observational data, the width of each line corresponds to the 1$\sigma$ error on the bulge-dominated fraction, evaluated from an MC re-sampling of the $X$- and $Y$-axis variables. }
\end{figure}

Since the stellar mass functions of the SDSS, Illustris and LGalaxies are not perfectly overlapping, it is more robust to explore the fraction of bulge and disk dominated galaxies as a function of stellar mass (i.e. in small ranges of stellar mass). Moreover, this approach also allows us to look at how well the simulations reproduce some of the key relationships found in this study. 

In Fig. 8 we show the fraction of bulge-dominated galaxies as a function of stellar mass, computed for the SDSS, Illustris and LGalaxies via a number of complementary methods. We compute the bulge-dominated fraction for the SDSS using stellar mass structural component ratios (as in the rest of the paper, shown in blue) and via $r$-band photometry (shown in red). The width of the SDSS lines indicate the 1$\sigma$ error, computed via a Monte Carlo simulation as in Fig. 4. Overlaid on Fig. 8 are the relationships between the fraction of bulge-dominated galaxies and stellar mass for LGalaxies (computed via stellar mass, shown as a blue dashed line) and Illustris (computed via $r$-band photometry, shown as a red dashed line). As such, the red lines should be compared with each other, and the blue lines should be compared with each other.

A positive relationship between fraction of bulge-dominated galaxies and stellar mass is found in the observations and both of the simulations. Hence, qualitatively at least, there is general agreement between the simulations and observations as to the trend between mass growth and bulge dominance in galaxy evolution. However, LGalaxies significantly under-predicts the fraction of bulge-dominated galaxies as a function of stellar mass relative to the SDSS. The most significant discrepancy is found at intermediate stellar masses ($M_{*} \sim 10^{10} - 10^{11} M_{\odot}$). Similarly, Illustris significantly under-predicts the fraction of bulge dominated galaxies as a function of stellar mass relative to the SDSS, as measured in $r$-band. The largest discrepancies between Illustris and the SDSS also occur at $M_{*} \sim 10^{10} - 10^{11} M_{\odot}$.

We additionally show on Fig. 8 the fraction of bulge dominated galaxies as defined through a kinematic analysis of angular momentum in Illustris (as a solid magenta line). Strictly speaking there is no exact corollary in the SDSS measurements to compare this to. However, in the literature (e.g. Genel et al. 2015) these types of measurements are often taken as proxies for by-mass measurements of structure. This is logical given that the kinematics are driven by an underlying mass distribution. Compared to the SDSS mass-weighted fraction of bulge dominated galaxies, the Illustris kinematic relationship significantly under-predicts spheroids. Thus, the photometric $r$-band and kinematic measures of galaxy structure are self-consistent in Illustris, but both disagree with the most relevant observations.

\subsubsection{Comparison of Different Galaxy Populations}

\begin{figure*}
\includegraphics[width=0.49\textwidth]{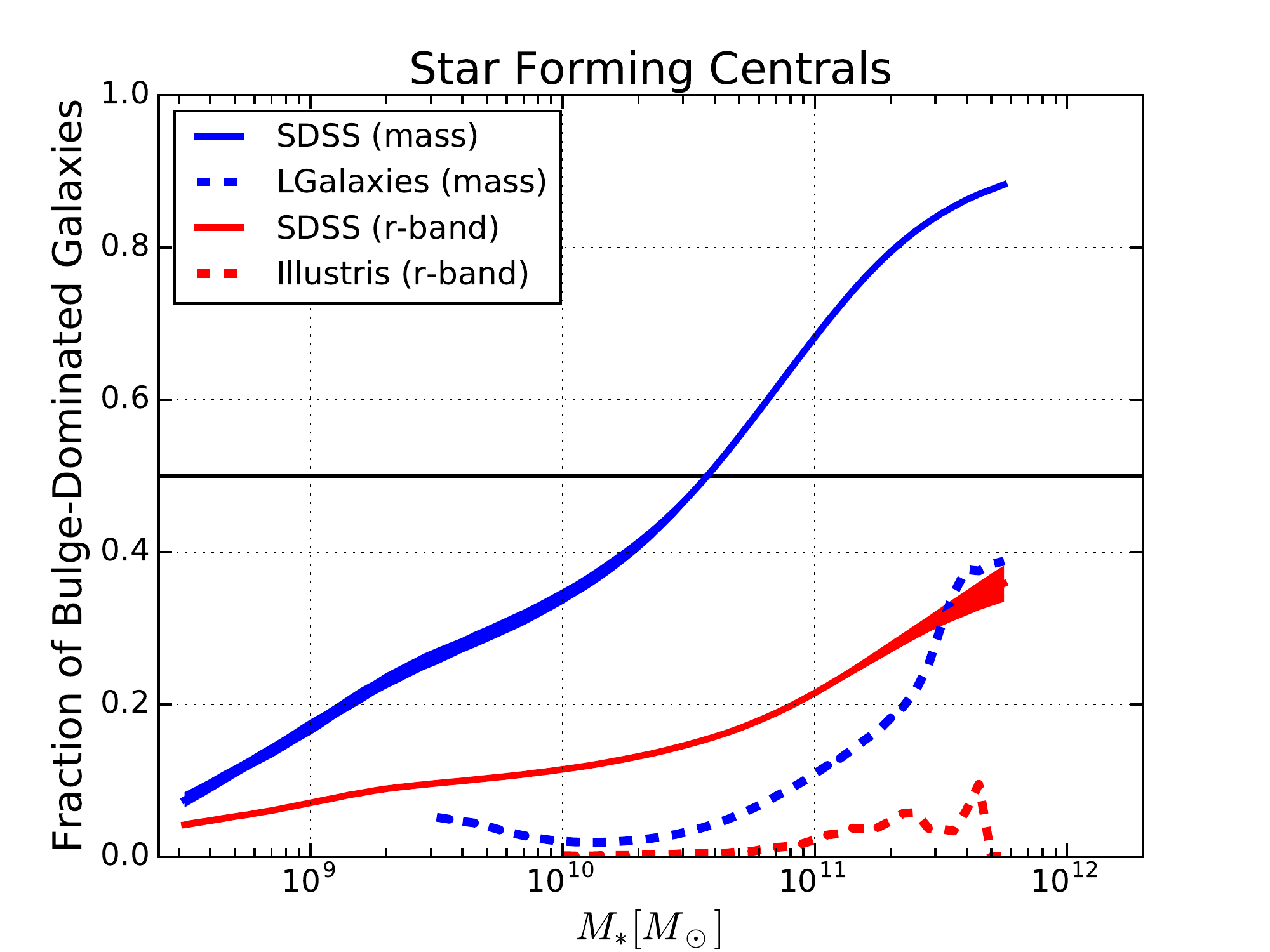}
\includegraphics[width=0.49\textwidth]{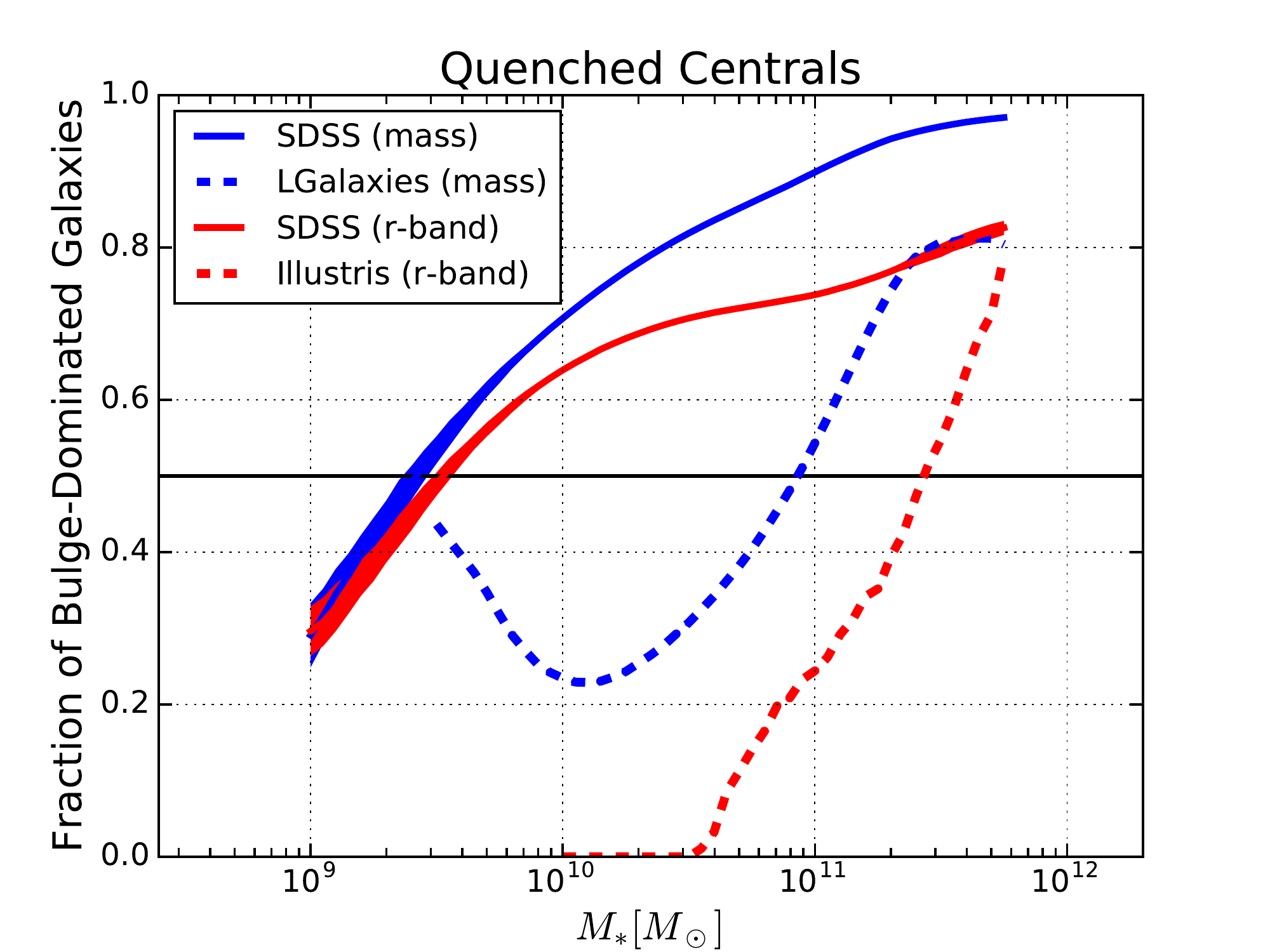}
\includegraphics[width=0.49\textwidth]{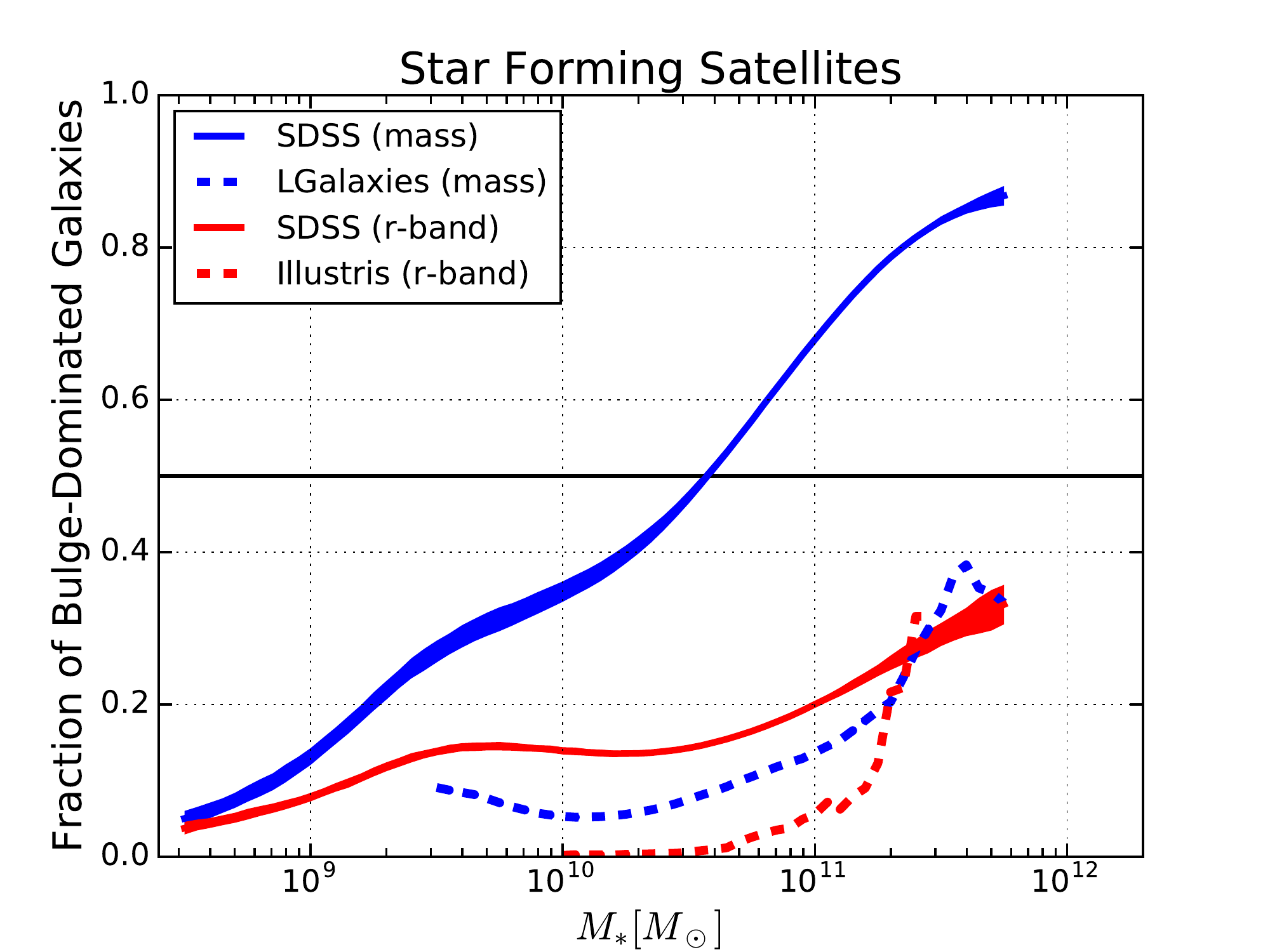}
\includegraphics[width=0.49\textwidth]{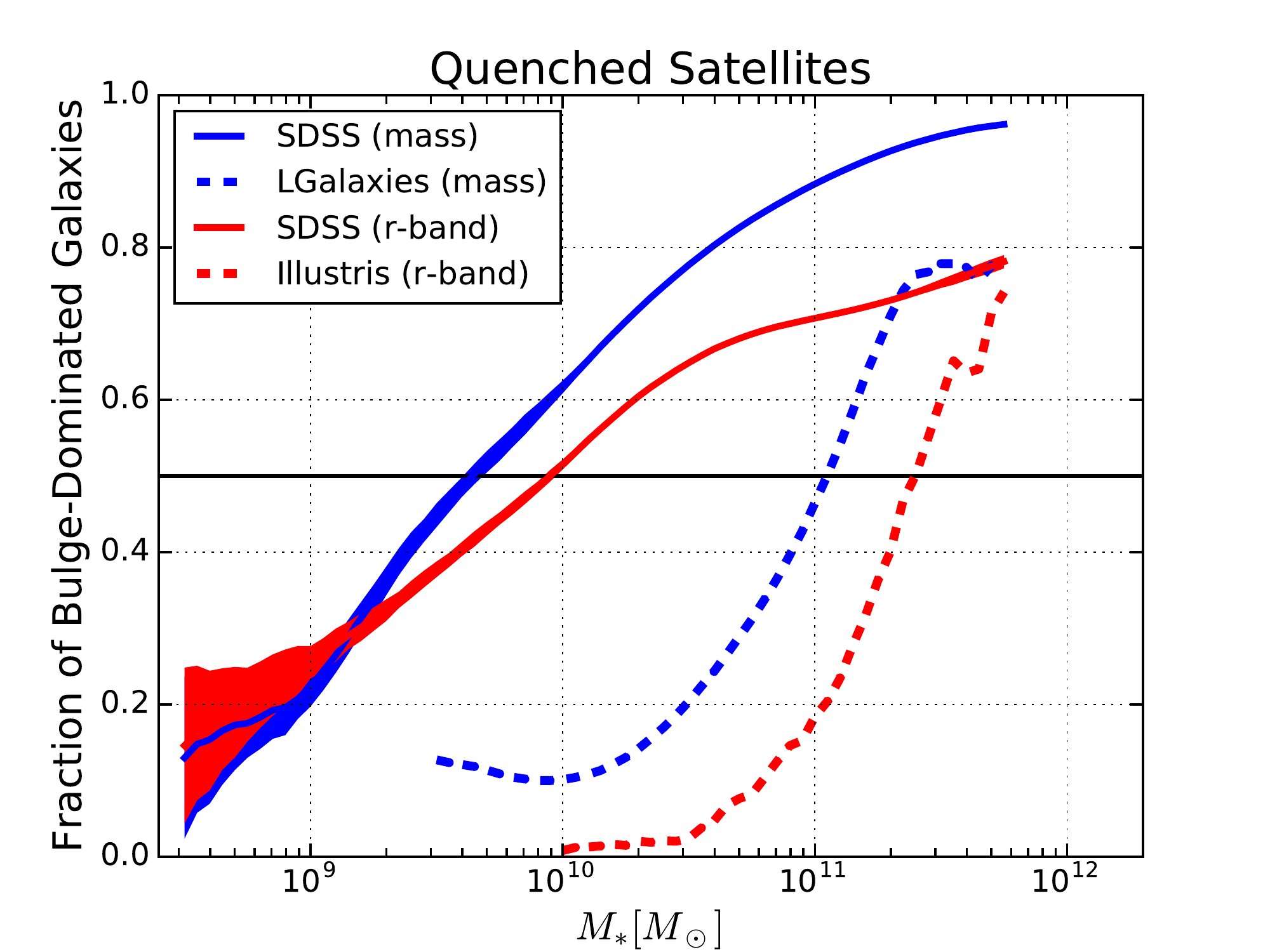}
\caption{The fraction of bulge-dominated galaxies as a function of stellar mass, separated into four galaxy populations (from top left to bottom right): star forming centrals, quenched centrals, star forming satellites, and quenched satellites. Results are shown separately for the SDSS (in mass and $r$-band light), LGalaxies (in mass), and Illustris (in $r$-band light). In each panel like colours should be compared with like colours. Between panels the same line type should be compared with the same line type. For the observational data, the width of each line corresponds to the 1$\sigma$ error on the bulge-dominated fraction, evaluated from an MC re-sampling of the $X$- and $Y$-axis variables. }
\end{figure*}

\begin{table*}
\begin{center}
\caption{Mean Bulge-to-Total Ratio $\langle ({\rm B/T})_{*} \rangle$ comparison between the SDSS, LGalaxies and Illustris.}
\begin{tabular}{c c c c c c}
\hline\hline
 & {\bf All Galaxies} & {\bf Passive Centrals} & {\bf Passive Satellites} & {\bf Star Forming Centrals} & {\bf Star Forming Satellites}\\
\hline \\
{\bf SDSS} ($M_{*}$) & 0.524 $\pm$ 0.001  & 0.787 $\pm$ 0.001 & 0.676  $\pm$ 0.001  & 0.382  $\pm$ 0.001  & 0.347  $\pm$ 0.001 \\\\
{\bf LGalaxies} ($M_{*}$) & 0.159 $\pm$ 0.001  & 0.389 $\pm$ 0.001  & 0.206 $\pm$ 0.001  & 0.073 $\pm$ 0.001  & 0.104 $\pm$ 0.001 \\\\
\hline \\
{\bf SDSS} ($r$-band) & 0.476 $\pm$ 0.001 & 0.699 $\pm$ 0.001  & 0.654 $\pm$ 0.001  & 0.251 $\pm$ 0.001  & 0.250 $\pm$ 0.001  \\\\
{\bf Illustris} ($r$-band) & 0.088 $\pm$ 0.001 & 0.414 $\pm$ 0.006  & 0.239 $\pm$ 0.005  & 0.030 $\pm$ 0.001  & 0.050 $\pm$ 0.001  \\\\
\hline
\end{tabular}
\label{tab-data}
\end{center}
Notes: Passive galaxies are defined to have sSFR$ < 10^{-2} {\rm Gyr}^{-1}$, and star forming galaxies are defined to have sSFR$ > 10^{-2} {\rm Gyr}^{-1}$. Mass weighted samples (LGalaxies \& SDSS) are selected at $M_{*} > 10^{9.5} M_{\odot}$; $r$-band weighted samples (Illustris \& SDSS) are selected at $M_{*} > 10^{10} M_{\odot}$. The observational data has a $1/V_{\rm max}$ weighting applied to all statistics to best approximate the volume complete simulated data. All errors on the mean are computed as $\sigma / \sqrt{N}$ and set to a minimum value of 0.001 in the case where they are computed to be lower.
\end{table*}

In this sub-section we compare the structures of galaxies in Illustris and LGalaxies to the SDSS for four distinct populations of galaxies: 1) star forming centrals, 2) quenched centrals, 3) star forming satellites, and 4) quenched satellites. As before, central galaxies are defined as the most massive galaxy in their group dark matter halo, and satellites are defined as any other group member. Star forming galaxies are defined to have sSFR $> 10^{-2} {\rm Gyr}^{-1}$, and quenched/ passive galaxies are defined to have sSFR $<10^{-2} {\rm Gyr}^{-1}$. In Table 1 we present the mean B/T ratio for each galaxy population, shown for the SDSS in stellar mass and $r$-band light, for LGalaxies in stellar mass, and for Illustris in $r$-band light. Galaxies are selected above the same stellar mass limit for each comparison.

For all galaxies, the difference in B/T, on average, between Illustris and the SDSS (in $r$-band) is $\Delta$B/T $\sim$ 0.4, with a similar offset for LGalaxies and the SDSS (by stellar mass). Even if this shift to higher B/T values were applied equally to all populations of simulated galaxies, this would still not solve all of the problems with morphology and structure. The {\it relative} increase in bulge fraction from star forming to passive centrals is essentially the same in Illustris and LGalaxies to the SDSS, it is just that low mass galaxies start with lower B/T ratios than observed. However, the relative increase in bulge fraction from star forming to passive satellites is $\sim$ a factor of two lower than that observed in both LGalaxies and Illustris. This suggests two, possibly separate, problems: 1) low mass galaxies do not form enough material in a bulge (as seen already for Illustris in Bottrell et al. 2017b), and 2) the processes associated with quenching satellites in both Illustris and LGalaxies do not engender the same magnitude of structural transformation as in real galaxies.

In Fig. 9 we show the fraction of bulge-dominated galaxies as a function of stellar mass for each population of galaxies. This is displayed for the SDSS (in mass and $r$-band), Illustris (in $r$-band) and LGalaxies (in mass). Essentially Fig. 9 is a reproduction of Fig. 8, separating the sample into the four classes under consideration in this sub-section. For the SDSS observations, the fraction of bulge-dominated galaxies is substantially greater for quenched systems than for star forming ones. The crossover mass at which the fraction of bulge- and disk-dominated galaxies are equal occurs at an order of magnitude lower stellar mass for quenched galaxies ($M^{\rm Q}_{*, {\rm cross}} \sim 3 \times 10^{9} M_{\odot}$) than for star forming galaxies ($M^{\rm SF}_{*, {\rm cross}} \sim 4 \times 10^{10} M_{\odot}$). This is approximately the same for both the central and satellite populations. We also find that the bulge-dominated fraction - stellar mass relation is only slightly different between centrals and satellites in general (as in Fig. 3). 

Both Illustris and LGalaxies predict positive correlations between the fraction of bulge-dominated galaxies and stellar mass in all populations of galaxies. Additionally, both simulations predict that quenched galaxies will be preferentially bulge-dominated at a fixed stellar mass, compared with star forming systems. Hence, in these respects, the simulations qualitatively agree with the SDSS observations. Crucially however, Illustris and LGalaxies both seriously underestimate the fraction of bulge-dominated galaxies, relative to the SDSS, for all populations of galaxies. This is most significant for LGalaxies with star forming systems and for Illustris with quenched systems.

One other feature to note from Fig. 9 is the difference between $r$-band morphology and stellar mass determined B/T structure. For quenched systems, the fraction of bulge-dominated galaxies - stellar mass dependence is similar for morphology and structure, whereas for star forming systems these two methods for measuring bulge-dominance are wildly discrepant. The reason for this is that in quenched systems $r$-band light approximately traces the mass distribution of galaxies, yet for star forming systems $r$-band light traces more the bright star forming regions in galaxies than the older and redder stellar populations, which dominate the mass distribution. Further details on the differences between by-mass structures and by-light morphologies are presented in Appendix B.

In summary, although the general directionality of the trends witnessed in observations are reproduced by LGalaxies and Illustris, the details are not. Most importantly, the fraction of bulge-dominated galaxies is significantly underestimated in all populations. In Section 6.2 we speculate on a number of fruitful areas to explore in order to improve the models' reproduction of galaxy structure.

\subsection{Further Simulations Comparisons}

\begin{figure*}
\includegraphics[width=0.49\textwidth]{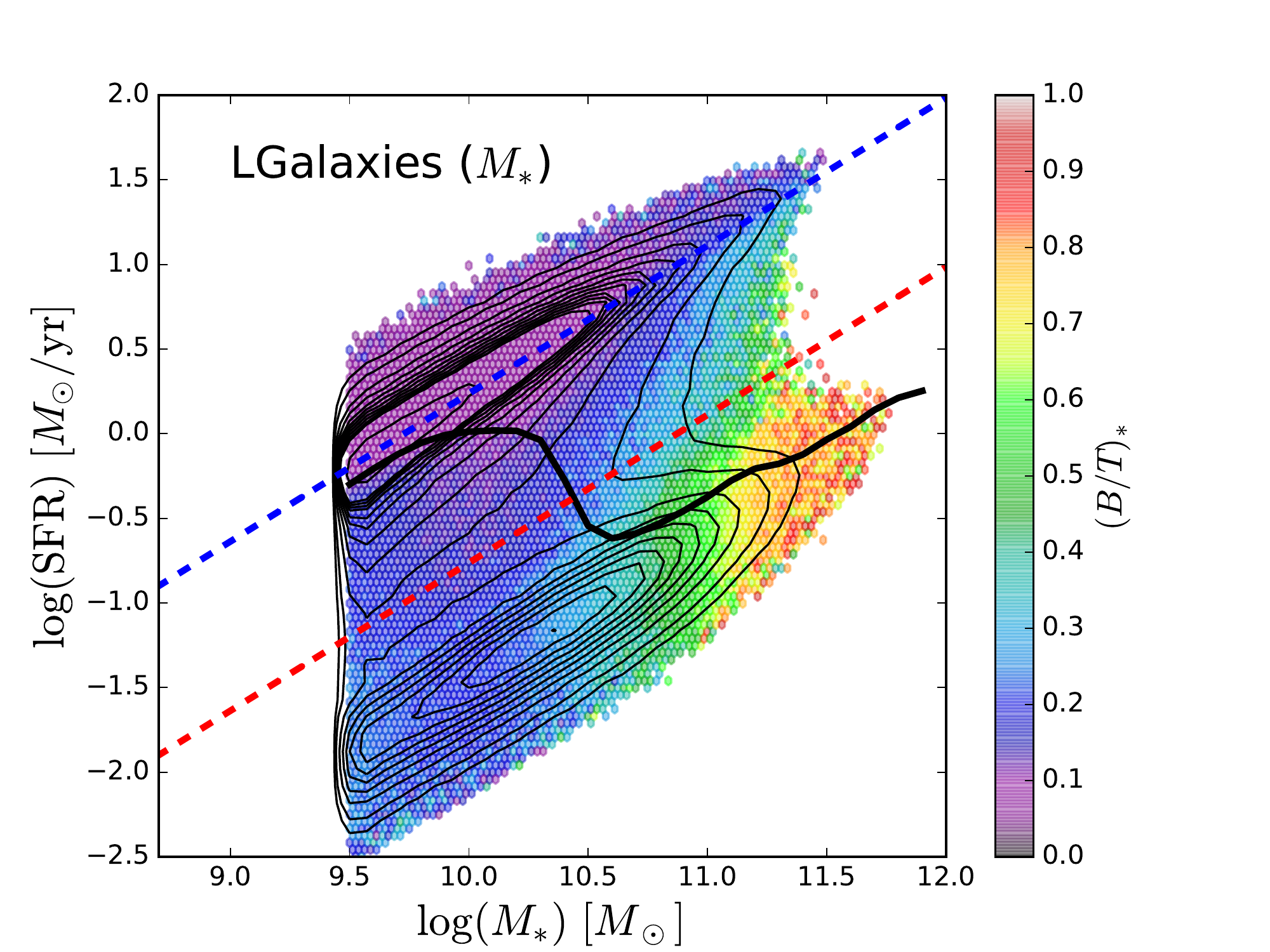}
\includegraphics[width=0.49\textwidth]{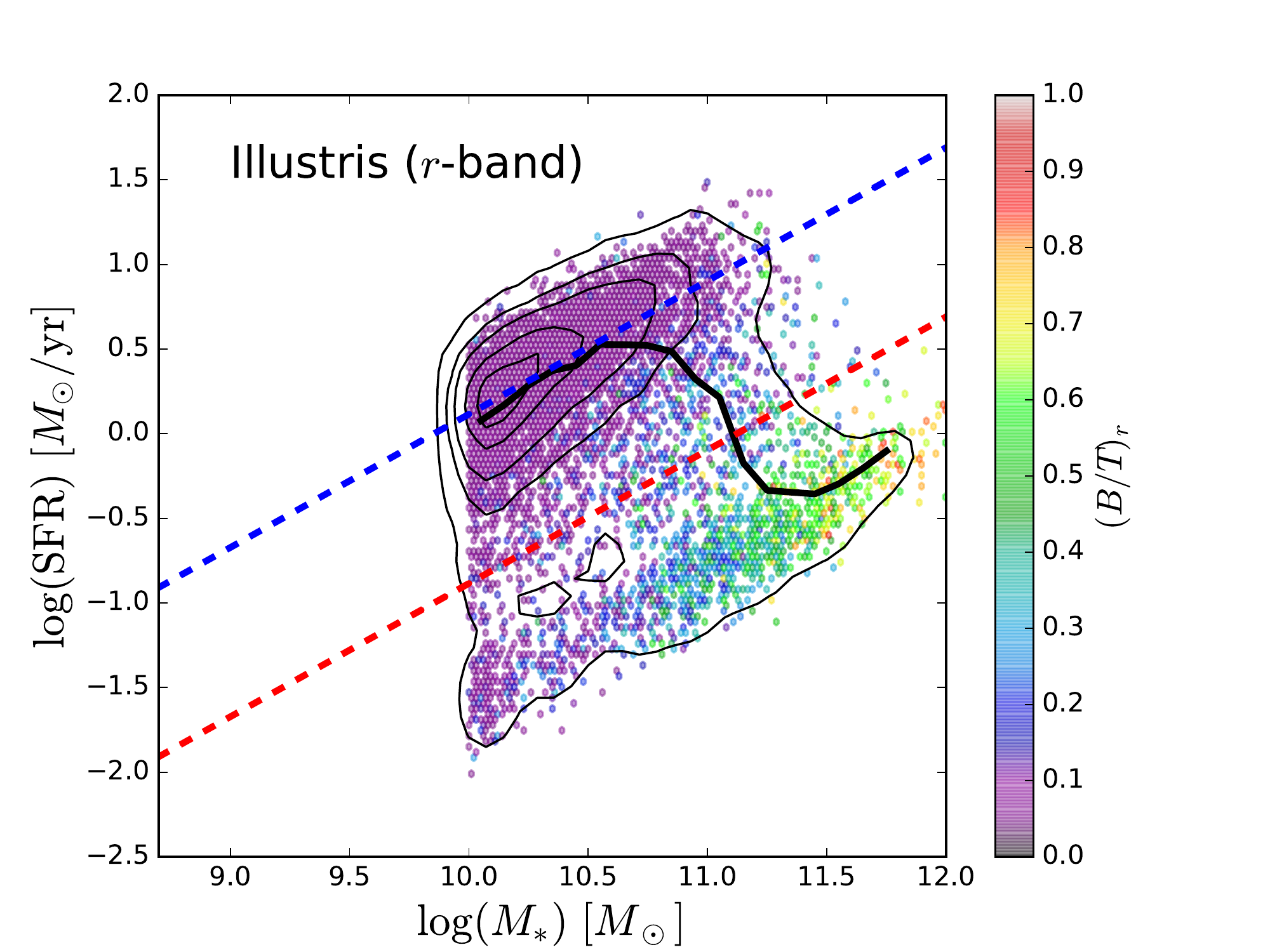}
\includegraphics[width=0.49\textwidth]{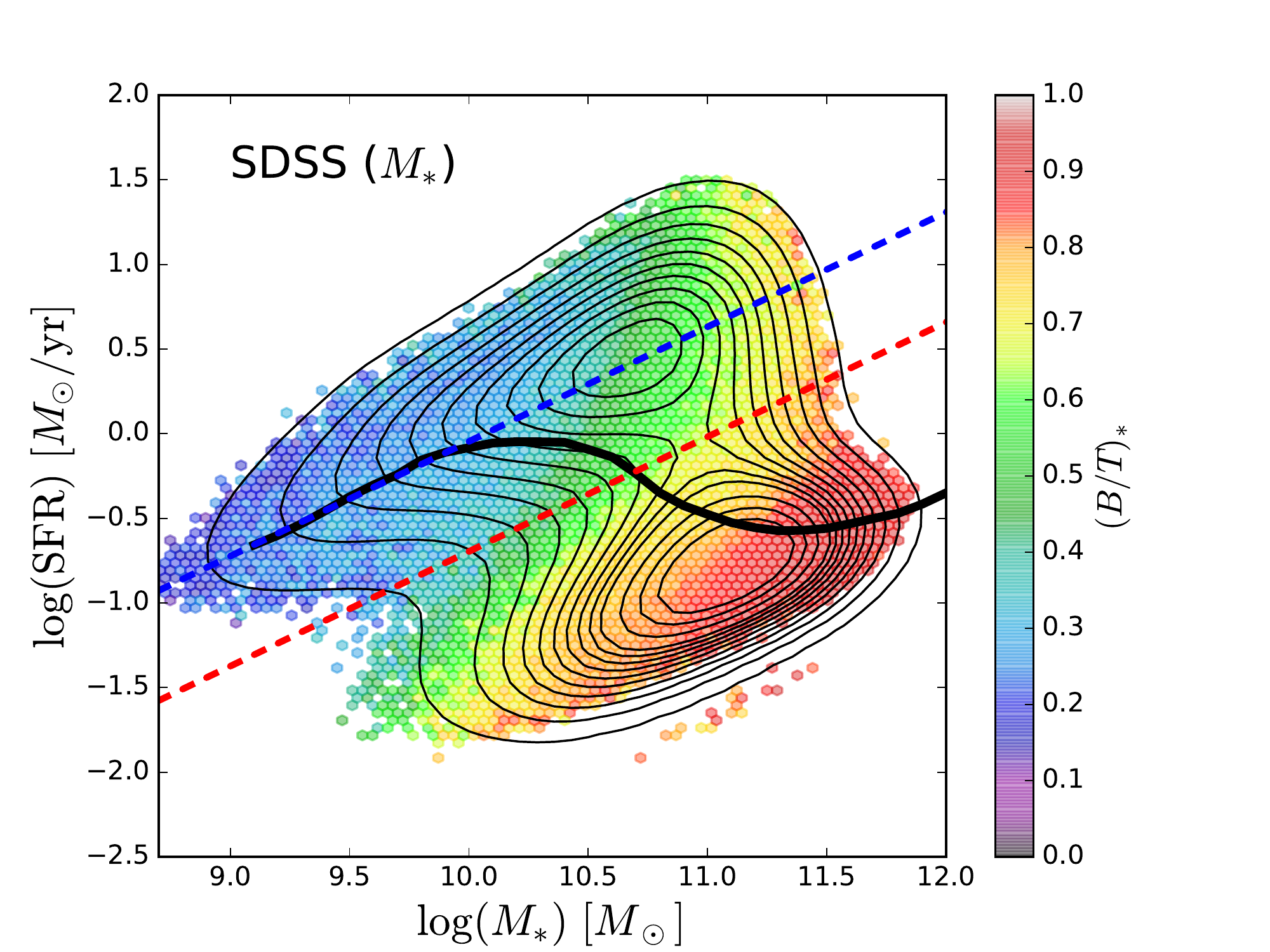}
\includegraphics[width=0.49\textwidth]{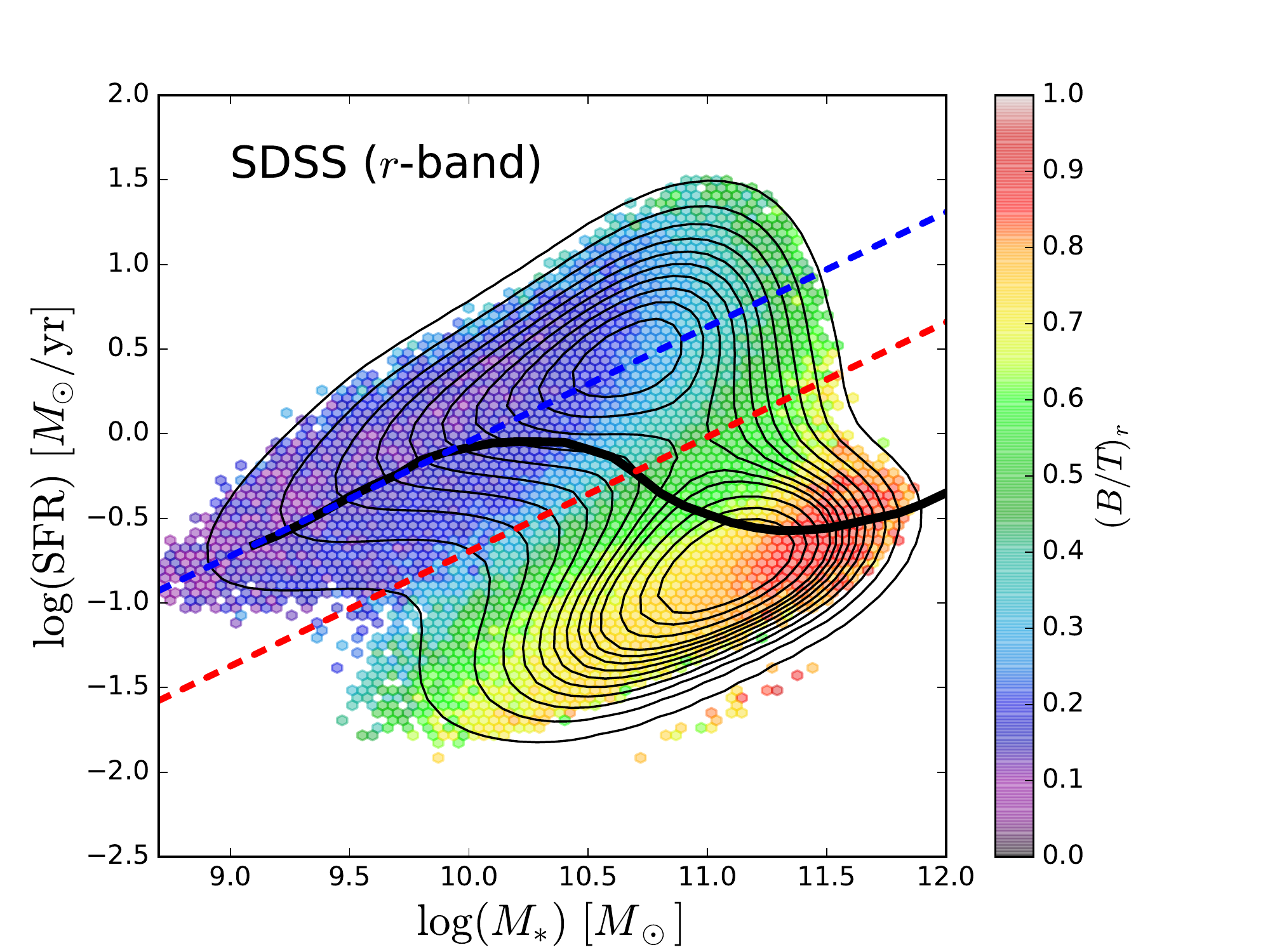}
\caption{The star forming main sequence (SFR - $M_{*}$ relation) shown colour coded by B/T ratio. From top left to bottom right: LGalaxies (B/T in stellar mass), Ilustris (B/T in $r$-band), SDSS (B/T in stellar mass), and SDSS (B/T in $r$-band). The median relation is shown as a thick black line on each panel, with light black lines indicating density contours. Blue dashed lines indicate fits to the star forming population; and red dashed lines indicate the minimum of the density contours (and hence the threshold of quenched galaxies). Each hexagonal bin is colour coded by the mean B/T of galaxies contained within a narrow range of $M_{*}$ and SFR. The panels in this figure should be compared column-wise, to ensure a fair comparison. }
\end{figure*}

In this subsection we consider two additional comparisons between the SDSS, Illustris and LGalaxies. Specifically, we investigate how well the simulations reproduce galaxy structure across the star forming main sequence (in Section 5.3.1), and consider the role of halo mass in determining galaxy structure in the models (in Section 5.3.2).

\subsubsection{Star Forming Main Sequence}

We show the star forming main sequence for LGalaxies and Illustris in Fig. 10. For comparison, the star forming main sequence for SDSS galaxies is reproduced below each simulation. In each panel the main sequence relationship is colour coded by mean B/T ratio, derived as a mass ratio in LGalaxies, and in $r$-band light for Illustris. The stellar mass weighted LGalaxies plot may be compared to the equivalent stellar mass weighted plot for the SDSS, and the $r$-band weighted Illustris plot may be compared to the equivalent $r$-band weighted plot for the SDSS. 

Quenched systems (with sSFR $< 10^{-2} {\rm Gyr}^{-1}$) extend down to an SFR = 0 in the simulations. Hence, a straightforward comparison is not immediately possible with the SDSS, which utilises indirect means to estimate low sSFRs that generally lead to a minimum SFR $\sim 10^{-2} M_{\odot}/{\rm yr}$ (infinitely far away in log-space). To combat this, we relocate the SFRs of quenched objects in the simulations to a random draw from a Gaussian fit to the quenched population's distance from the main sequence in the SDSS ($\Delta$SFR). For each simulation, quenched objects are inserted at a distance $\Delta$SFR from their own measured main sequence\footnote{This methodology accounts for differences in the normalisation and slope of the main sequence between the simulations and the observations, which are not the focus of this work. See Bluck et al. (2016) for a discussion on this issue.}. This process approximates the SFR values of quenched simulated galaxies as they would be measured in the SDSS. However, of course, the exact values of quenched objects (those lying below the dashed red lines) are not reliable in either the simulations or observations. Thus, what is important is the fact that they have {\it low} SFRs, not their specific values.

In general, galaxies in Illustris and LGalaxies are significantly more disk-dominated (and less bulge-dominated) than that observed. This is evident in stellar mass structural ratio (for LGalaxies) and in $r$-band photometry (for Illustris). Moreover, the transition from star forming to quenched galaxies engenders far less structural change in the simulated data than in the observed data. One surprising manifestation of this result is that the majority of quenched systems are disk-dominated in Illustris and LGalaxies (see also Table 1), whereas in the observed Universe the vast majority of quenched systems at all masses are spheroidal in structure. This fact suggests that the processes responsible for quenching galaxies in the simulations are not yet correctly modelling the associated structural changes in galaxies. We defer to further work the process of attempting to improve the SAM with updated physical prescriptions for quenching and structure formation (Bluck et al. in prep.). For Illustris, it will be highly interesting to see whether the new Illustris-TNG project improves on these problems or not (see Pillepich et al. 2018).

\begin{figure*}
\includegraphics[width=0.49\textwidth]{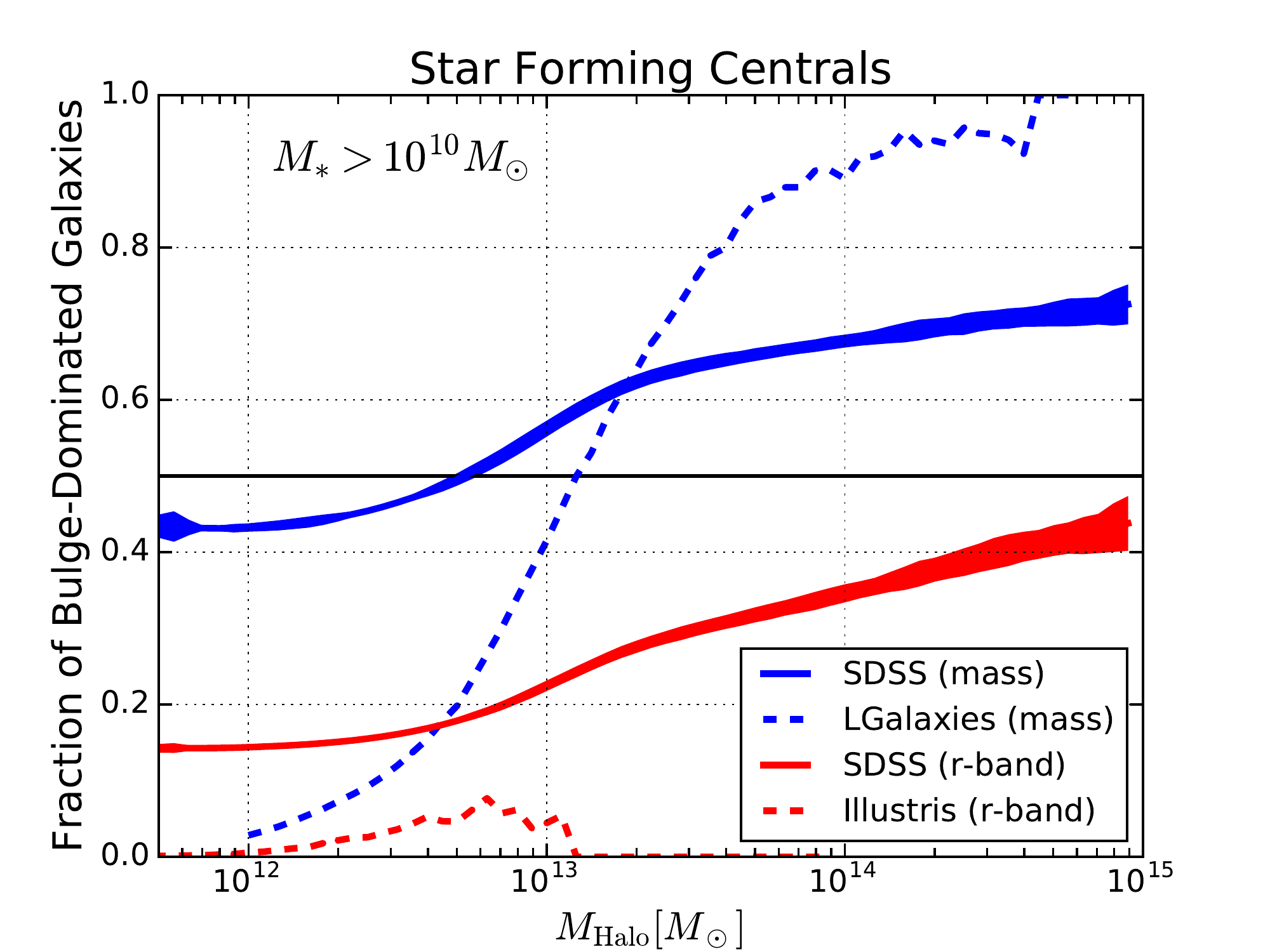}
\includegraphics[width=0.49\textwidth]{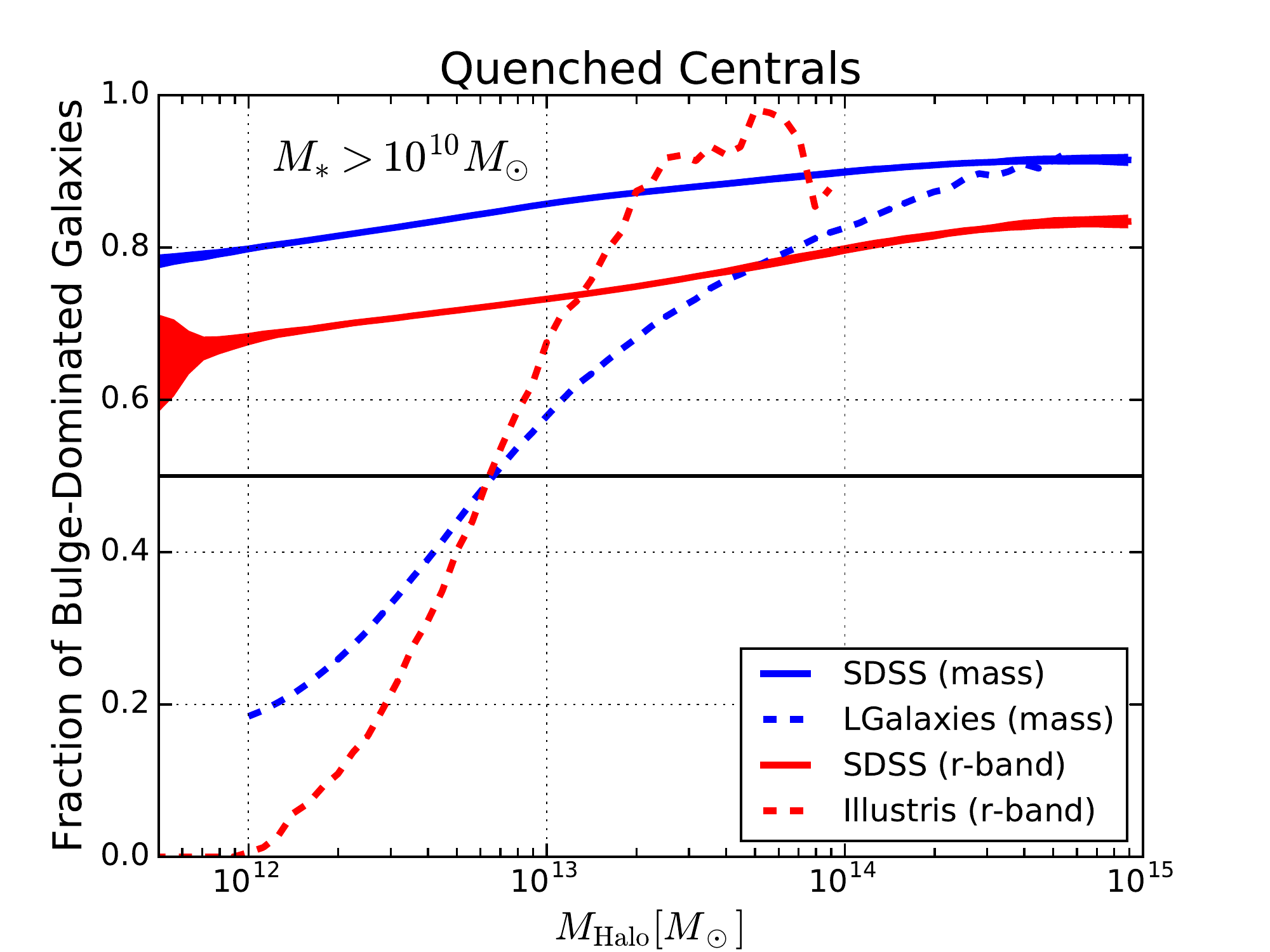}
\includegraphics[width=0.49\textwidth]{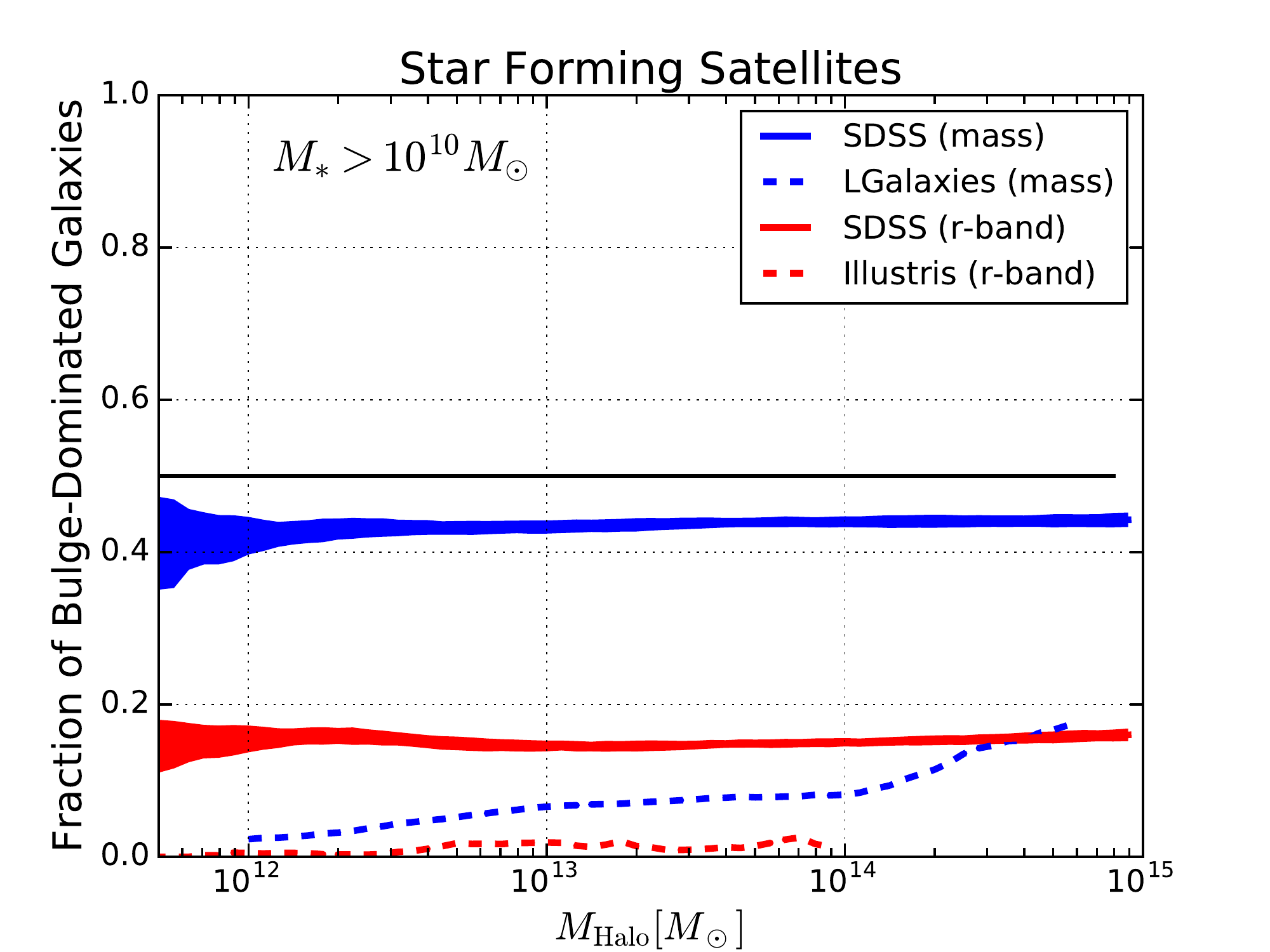}
\includegraphics[width=0.49\textwidth]{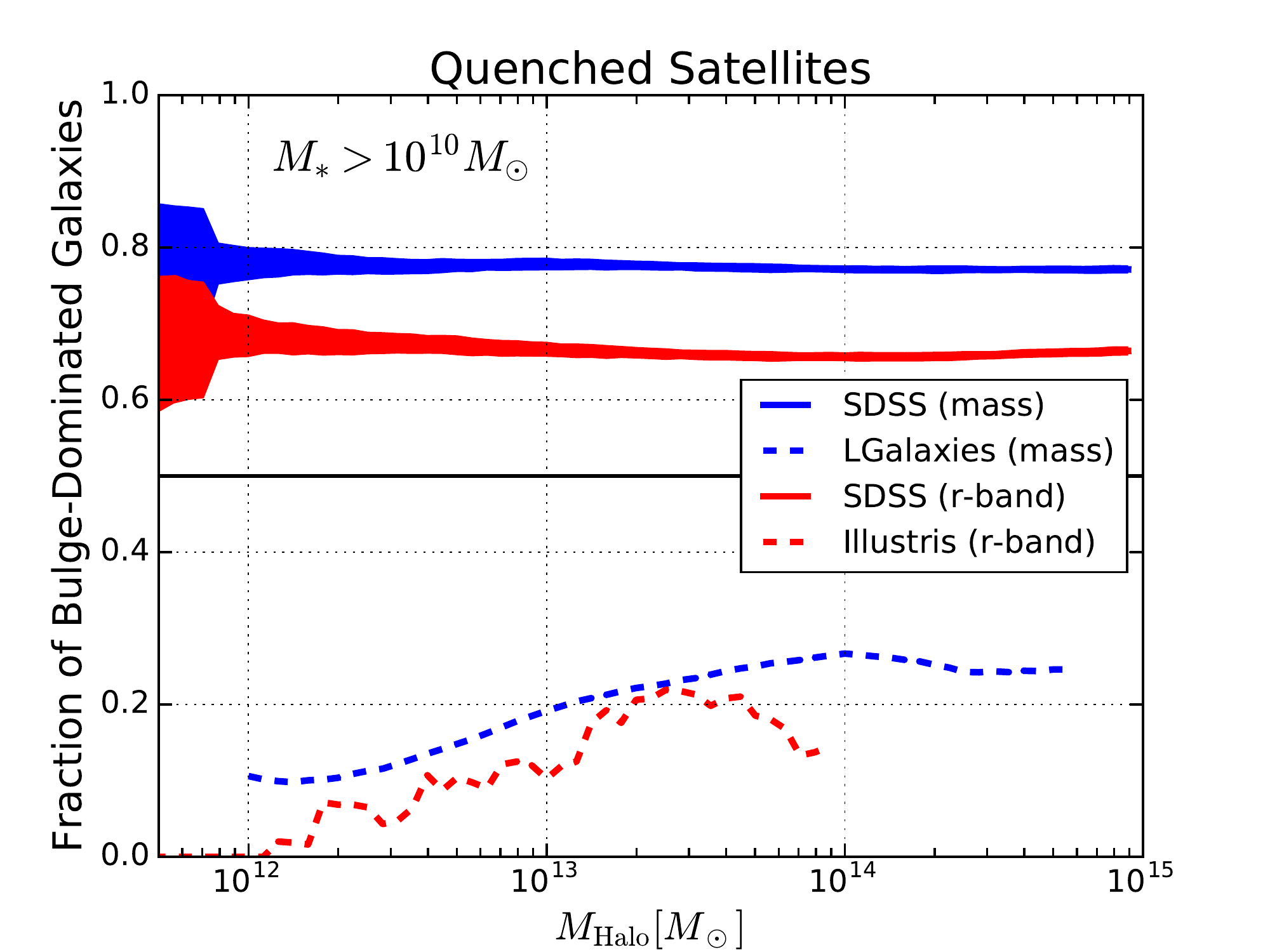}
\caption{The fraction of bulge-dominated galaxies as a function of group halo mass, separated into four galaxy populations (from top left to bottom right): star forming centrals, quenched centrals, star forming satellites, and quenched satellites. Results are shown separately for the SDSS (in mass and $r$-band light), LGalaxies (in mass), and Illustris (in $r$-band light). In each panel like colours should be compared with like colours. Between panels the same line type should be compared with the same line type. For the observational data, the width of each line corresponds to the 1$\sigma$ error on the bulge-dominated fraction, evaluated from an MC re-sampling of the $X$- and $Y$-axis variables.}
\end{figure*}

\subsubsection{Halo Mass}

We postpone a full analysis of the reproduction of structural trends with environment in the simulations (LGalaxies \& Illustris) to further work. However, in this appendix, we briefly explore how well the simulations reproduce the bulge-dominated fraction dependence on halo mass, for the four populations of galaxies discussed in Section 5.2.4.

In Fig. 11 we reproduce Fig. 9, replacing stellar mass with group halo mass. In the case of centrals, halo mass and stellar mass are highly correlated, whereas in the case of satellites they are only very weakly correlated (see Fig. 1, and associated text). For satellites in the SDSS, there is essentially no dependence of the fraction of bulge-dominated galaxies on halo mass, once the sample is separated out into star forming and quenched populations. This suggests that the relatively weak correlation between structural fraction and halo mass for satellites seen in Fig. 3 is primarily due to satellites in higher density environments being preferentially passive, and passivity correlating with structure. For centrals in the SDSS, there is a relatively strong correlation between halo mass and bulge-dominated fraction for star forming systems, which is noticeably weaker for passive galaxies. In general, quenched galaxies (both centrals and satellites) have a significantly higher fraction of bulge-dominated systems for their halo masses than star forming systems. This is seen in both stellar mass and $r$-band photometry defined B/T ratios.

As with stellar mass, for most ranges in halo mass both Illustris and LGalaxies seriously under-predict the fraction of bulge-dominated galaxies, compared to the SDSS. In general, the agreement between observations and simulations is even worse for quenched systems than star forming systems. One notable exception to this is for LGalaxies star forming centrals at high halo masses. These rare objects are predicted to be {\it more} bulge-dominated than seen in observations. In addition to the clear quantitative differences between the models and observations, the shapes of the relationships with halo mass are also very different to that observed. In light of these serious discrepancies (and those seen throughout this section), it is clear that the models must be significantly improved to better account for observed galaxy structure. In Section 6.2 we give some preliminary suggestions for how this may be done.

%
%

\section{Discussion}

\subsection{Interpretation of the ANN Rankings:  Quenching, Mergers, and the Role of Environment}

In this sub-section we discuss possible causal origins to the observed connections between galaxy structure and star formation, stellar mass, and environment. The ANN regression analysis of Section 4 reveals that $\Delta$SFR is the most predictive single variable for ascertaining B/T structure considered in this work (see Fig. 5). Furthermore, at a fixed stellar mass, adding SFR to the network leads to a higher improvement in predictive power than any other galactic or environmental parameter (see Fig. 6). As such, we find that there is a very close connection between star formation and structure in galaxies, for both centrals and satellites. This conclusion is qualitatively similar to much prior work (including, Cameron et al. 2009, Wuyts et al. 2011, Bell et al. 2012, Bluck et al. 2014, 2016, TBE, Brennan et al. 2017). However, we have expanded on these findings by fully quantifying this effect and, additionally, have shown that the dependence of B/T on star formation and stellar mass is separable (as hinted at in Bluck et al. 2014), with both parameters leading to substantial improvement over the other alone in the ANN analysis (see Section 4.3). 

The structural difference between star forming and quenched galaxies, at a fixed stellar mass (shown in Figs. 4, 6 \& 9) strongly suggests different evolutionary paths for star forming and quenched objects. In the literature, a number of mechanisms to link star formation and galaxy structure have been suggested. For instance, in Bluck et al. (2014) we argue that the strong empirical correlation between bulge mass and supermassive black hole mass (e.g., Ferrarese \& Merritt 2000, Haring \& Rix 2004, Saglia et al. 2016) implies that bulge dominated galaxies host more massive central black holes, at a fixed stellar mass (although see Simmons, Smethurst \& Lintott 2017 for evidence that stellar mass correlates with black hole mass more tightly than bulge mass, in disk-dominated systems). Consequently, there may be a greater potential for radio-mode AGN feedback (e.g., Croton et al. 2006, Bower et al. 2006) to affect the gas surrounding more bulge-dominated galaxies (e.g., Bluck et al. 2016). This mechanism shuts off star formation in galaxies by preventing cooling of gas in high mass haloes due to heating from jet triggered radio bubbles (e.g., Fabian et al. 2012). 

Models which incorporate radio mode AGN feedback as the dominant mechanism to quench massive galaxies lead to predictions which are generally in reasonable agreement with modern observations, whereas alternative models tend to be less successful (e.g., Vogelsberger et al. 2014a,b, Schaye et al. 2015, Henriques et al. 2015, Bluck et al. 2016, Terrazas et al. 2016, 2017, Brennan et al. 2017). Consequently, the clear connection between structure and star forming state seen in this work may have a casual origin in the sense that bulge dominated structures are more likely to host high mass central black holes, which quench star formation through ongoing maintenance-mode feedback episodes. However, testing this hypothesis in a more direct manner has proven highly challenging (see Bluck et al. 2016), and hence alternative explanations ought to be seriously considered as well. Examples of alternative mechanisms to AGN feedback for quenching galaxies discussed in the literature include: morphological quenching (Martig et al. 2009), halo mass quenching (Dekel \& Birnboim 2006), and progenitor bias (Lilly \& Carollo 2016). Ultimately, the extremely tight connection between star forming state and galaxy structure (as evidenced by the general trends in Section 3, offsets at a fixed stellar mass in Section 4, and through a sophisticated machine learning approach in Section 5) requires a compelling explanation. However, there is clearly still much work to be done in order to unambiguously resolve this issue. 

In Section 4, for all galaxy types, we find that stellar mass is the second most predictive variable for ascertaining galaxy structure (B/T), after $\Delta$SFR. Since more massive galaxies have a higher fraction of their mass accreted via mergers (e.g., Bluck et al. 2012, Ownsworth et al. 2014), the more massive the galaxy the higher the impact of major and minor merging on the system. This is as expected in the hierarchical model of structure formation (e.g., Coles et al. 2000, Springel et al. 2005). Furthermore, the merger of disk galaxies generally leads to the growth of a bulge component (e.g. Barnes \& Hernquist 1992, Stewart et al. 2009, Hopkins et al. 2010). Hence, the higher prevalence of mergers in high mass galaxies, coupled with the capacity for merging to grow a bulge component, offers a natural explanation for the close connection between stellar mass and B/T structure, quantified in this work. Perhaps more interesting than the presence of a strong correlation between structure and mass, is that this relationship is less strong than between structure and star forming state (as quantified by $\Delta$SFR). Observationally, quenched galaxies are spheroids and star forming galaxies are disks, almost invariably. This simple fact, however, is {\it not} reproduced by contemporary models (see Table 1 and Fig. 10). Once again, there is clearly much theoretical work needed in order to resolve these issues satisfactorily.

Although mergers are well established both observationally and theoretically to lead to bulge growth, in gas rich systems disks may re-grow in the postmerger phase, provided that the galaxy is not yet fully quenched (e.g., Stewart et al. 2009, Mitchell et al. 2014, Ellison, Catinella \& Cortese 2018). Thus, some combination of merger induced bulge growth plus disk quenching will be required to fully account for the connection between mass and structure as elucidated in this work. In addition to merging, violent disk instabilities are also more probable in high mass disks (e.g., Bouche et al. 2007, Zolotov et al. 2015, Somerville et al. 2015), which may redistribute a significant mass of stars and gas from a rotationally supported disk to a pressure supported bulge. Ultimately, it is likely that a combination of merging (major and minor), violent disk instabilities, and quenching of star formation is responsible for the high correlation between mass and structure in local galaxies. Disentangling the relative importance of each mechanism is best considered in terms of specific models and their relative success at reproducing the observed structural relationships (see Sections 5 above \& 6.2 below). Nevertheless, the greater dependence of structure on $\Delta$SFR than stellar mass suggests that quenching has a critical role to play in shaping galaxies (and / or vice versa).

In our machine learning analysis (presented in Section 4), we find that all quantitative environmental metrics considered (local density, halo mass, distance from central galaxy) are significantly less predictive of galaxy structure than internal parameters (e.g., $\Delta$SFR \& $M_*$, discussed above). Perhaps surprisingly, this is not only true for central galaxies but also for satellites and even inner satellites as well (see Fig. 5). At a fixed stellar mass, the addition of environmental parameters has no impact on the predictive power of the network for centrals (except for halo mass) and has only a small impact for satellites and inner satellites (see Fig. 6). Taken at face value, these results imply that environment is not strongly connected with the growth of galaxy structure. However, in Section 3.3 (particularly Fig. 4), we do find that there is a significant, yet subtle, impact of environment on structure, especially for satellites and inner satellites. The resolution of this apparent discrepancy is simply that environment only affects structure at high densities (or, equivalently, for satellites near the centre of high mass haloes). Thus, the majority of galaxies have structures which are completely unaffected by environment as clearly demonstrated in our ANN analysis.

Finally, the fact that in Section 4 we determine B/T values almost as accurately as the errors allow for the full set of input parameters suggests that we are not missing any fundamental parameters relevant to structure formation. To leading order, one would need only to know the stellar mass and SFR of a galaxy to predict it's B/T almost as reliably as actually performing a multi-wavelength decomposition. Environmental parameters, and additional metrics not considered in this work, would yield only marginal improvements over stellar mass and SFR. Of course, pathological cases do exist and simple measurements from spectra or SEDs will not fully replace the need for detailed photometric / stellar mass structural measurements. Furthermore, our study is restricted to z $\sim$ 0.1 and there may well be redshift evolution in these trends. Nonetheless, it is interesting to speculate about the possibility of using ANN to estimate structural parameters (with quantifiable errors and biases) as an alternative to image fitting for very large datasets.

\subsection{Why do the Simulations Fail to Accurately Reproduce Galaxy Structure?}

In Section 5 we find that the fraction of bulge-dominated galaxies is seriously underestimated in both the LGalaxies SAM and the Illustris hydrodynamical simulation, compared with the SDSS (see Figs. 7 - 10, and Table 1). Ultimately, these results highlights a profound failure in both Illustris and LGalaxies to approximate the structures of local galaxies in a reasonable manner. Thus, whilst the stellar mass functions, cosmic star formation history, and red - blue fractions are produced reasonably well in LGalaxies and Illustris (e.g., Vogelsberger et al. 2014a,b, Henriques et al. 2015), the structures of galaxies are not. Exposing and quantifying this fact is the most important contribution of this paper. 

Bottrell et al. (2017b) find a similar result for Illustris, whereby the distributions of B/T are significantly different to the SDSS, consistent with our findings. We expand on that work by investigating the stellar mass dependence of this trend and by investigating galaxy sub-populations (see Section 5.2). Additionally, Dickinson et al. (2018) find that low mass ($M_* < 10^{11} M_{\odot}$) Illustris galaxies have significantly different morphologies to that observed, via a very different methodology to the present work, utilising human classifications with the Galaxy Zoo approach (e.g., Lintott et al. 2008). Hence, there is now compelling evidence that Illustris fails to reproduce galaxy structures for the vast majority of the galaxy population. It is interesting that this is also the case in a leading SAM as well (see Section 5), and thus may point to a wider issue in contemporary galaxy evolution models.

The purpose of this sub-section is to explore the question of {\it why} leading contemporary simulations under-produce bulge dominated galaxies in comparison to appropriately analysed observations, and, furthermore, to offer some suggestions for improvements to the models moving forwards. As such, this discussion may be used as a starting point for further investigation into this important problem. Although the discrepancy with observations is similar in size and direction for Illustris and LGalaxies (see Figs. 7 - 9), the very different methodology between the two simulations will most probably require a different solution in each case. This is because, for LGalaxies, large scale physical processes are modelled indirectly, via tuneable free parameters; whereas, in Illustris, only the sub-grid physics (primarily star formation and feedback) is alterable a priori, by design. Hence, we will consider each simulation separately here. 

The Illustris simulation significantly over-predicts the abundance of very massive galaxies (as seen in Fig. 7). As such, the AGN radio-mode feedback prescription, utilised to quench massive galaxies, is not sufficiently efficient in the sense of quenching star formation early enough in the history of a given galaxy's evolution. However, in terms of impact on the gas in haloes, the Illustris feedback prescription is actually too efficient, yielding essentially gas-less cluster environments in stark contradiction to observations (e.g., Vogelsberger et al. 2014b, Genel et al. 2014). These well known problems inherent with the Illustris feedback model may be at the heart of the failure of Illustris to form bulge-dominated structures in the right abundance, as explained below.

For the evolution of galaxy structure, inefficient quenching will lead to excessive re-growth of disks after galaxy mergers, preventing the formation of pure spheroidal systems. The dearth of ellipticals in Illustris, as well as the under-production of high B/T composite systems, may thus be explained primarily due to the lack of successful quenching of high mass galaxies. In the new Illustris-TNG project (e.g., Springel et al. 2018, Nelson et al. 2018) considerable effort has been made to improve the AGN feedback model. A particular focus has been placed on getting the multi-epoch stellar mass function and hot gas fraction in haloes simultaneously correct. However, a key additional test of the success of the new feedback model will be to ascertain whether the structures of galaxies are better reproduced as a function of mass than in the original Illustris simulation. This structural test will also have the distinct advantage of not being correct by design.

In LGalaxies, the stellar mass function slightly {\it under}-predicts the abundance of massive galaxies (see Fig. 7); hence, increasing the strength of the AGN feedback model would not improve agreement with observations.  Thus, we must search for another avenue to improve the agreement with structures in the SAM. Bulge growth is driven primarily by mergers, where the merger rate of dark matter haloes is set externally by the Millennium Simulation and cannot be altered in LGalaxies. Galaxies merge in line with their dark matter haloes, although the timescales are different due to baryonic effects. More specifically, the merging of galaxies is modelled via the Chandrasekhar formula, with a tuneable free parameter fit via galaxy pair studies (e.g. Mundy et al. 2017). Consequently, the underlying galaxy and halo merger rate is highly constrained in the SAM, and not possible to vary freely. As such, the impact of merging on bulge formation is governed to a large extent by the threshold for major mergers ($\mu > 1/10$), which is already very low in value. In major mergers, {\it all} extant stars and gas (of both interacting galaxies) end up in a bulge/ spheroid. Hence, the impact of merging on bulge formation is already most likely at a maximum in LGalaxies. Therefore, additional mechanisms to mergers must be considered in order to increase the efficiency of bulge growth, to better agree with observations (e.g., Figs. 8 \& 9).

In the current implementation of LGalaxies, violent disk instabilities occur but only at a minimal level, i.e. when a disk becomes kinematically unstable, just enough matter is transferred from the disk to the bulge to achieve stability. Feasibly, up to the entire mass of the disk can be transferred to the bulge in a disk instability (e.g., Zolotov et al. 2015, Somerville et al. 2015). Therefore, an obvious place to start in improving the structural evolution in LGalaxies is to make the transfer mass in a violent disk instability a tuneable parameter, and to fit this to structural/ morphological data (Bluck et al. in prep.). Additionally, increasing the efficiency of star bursts can also give rise to slightly higher bulge masses and hence B/T values. Interestingly, the effect of increasing the potency of disk instabilities and star bursts will both impact intermediate mass galaxies the most (which have the right combination of high mass disks and high gas fractions). This is precisely the regime where the largest discrepancies with observations are found in Section 5, and hence is highly encouraging for this approach.

A broader issue, which affects both hydrodynamical simulations and SAMs, is that morphologies and structures are not frequently used to tune feedback models. Part of the reason for this omission is that, historically, measurements of galaxy morphology and structure have been quite subjective (e.g. dependent upon observed waveband or astronomers' visual judgment). Throughout this paper, the use of computationally determined multi-wavelength and stellar mass bulge-to-total ratios (B/T) allow us to significantly improve on the objectivity of measured galaxy structure, enabling a robust comparison to simulations. This process has revealed a significant tension between observations and simulations at present, which, in the final analysis, is perhaps unsurprising given that structure has not been used to inform the sub-grid physics of the models. Therefore, an obvious solution for the future is to carefully test the impact of sub-grid physics on the production of galactic structures, in addition to the more commonly used diagnostics (i.e. stellar mass, star formation and colour).

%
%

\section{Summary}

\noindent In this paper we perform a detailed analysis of the structures of $\sim$ 0.5 million galaxies at z $<$ 0.2 taken from the SDSS DR7. Throughout we use a definition of galactic structure as bulge-to-total stellar mass ratio, (B/T)$_*$, which indicates the ratio between bulge mass and total stellar mass. This parameter is superior to structural measurements by light (e.g., optical B/T ratios, S\'{e}rsic index and by-eye classification) since it traces the underlying mass distribution in galaxies and, more specifically, the segregation of stars into the dominant bulge and disk components. One particular advantage of using a structural parameter sensitive to mass rather than light, is that it mitigates concerns that correlations between star formation and structure are a trivial result of optically measured galaxy structures tracing star forming regions directly. \\

Our primary results are as follows: \\

\noindent $\bullet$ We confirm prior results by finding a close connection between galactic star formation (measured by sSFR \& $\Delta$SFR) and galaxy structure (measured by (B/T)$_*$). Our use of a mass weighted definition of structure further indicates that the tight connection between star formation and structure is not due to bright star forming regions dominating optical structural measurements (see Figs. 1 \& 2).\\

\noindent $\bullet$ For central galaxies, we find strong correlations between (B/T)$_*$ and stellar mass, halo mass, and sSFR. Similarly strong correlations are found for satellites with stellar mass and sSFR, but not for halo mass. For all galaxy types, there is a much weaker correlation between galactic structure and local density than with intrinsic galaxy properties (see Fig. 3).\\

\noindent $\bullet$  At a fixed stellar mass, we find subtle yet significant additional correlations with environment (local density, halo mass, location within the halo) for satellites. For centrals, significant deviation from the dominant stellar mass relationship is only seen with local density, in the most over-dense regions. For all galaxy types, there is a strong dependence of galaxy structure on sSFR at a fixed stellar mass (see Fig. 4).\\

\noindent $\bullet$ We rank the strength of connection between galaxy structure, (B/T)$_*$, and various other galaxy and environmental parameters via an ANN regression analysis in Section 4. We conclude that $\Delta$SFR followed by stellar mass are the most important variables for determining (B/T)$_*$ for all galaxy types. Environmental parameters are significantly less predictive of galaxy structure than intrinsic parameters for all types of galaxies. However, environment is slightly more predictive of (B/T)$_*$ for satellites than centrals, whilst intrinsic parameters are significantly less predictive of galaxy structure for satellites than centrals. In Appendix C we consider two alternative methods for ranking the parameters, which both lead to very similar conclusions. \\

\noindent $\bullet$  We compare the observed (B/T)$_*$ distributions and structural fraction relationships with stellar mass to predictions from the LGalaxies model and the Illustris hydrodynamical simulation in Section 5.  The main trend of increasing bulge-dominance with increasing stellar mass is reproduced in both simulations. However, both LGalaxies and Illustris seriously under-predict the fraction of bulge-dominated galaxies for the whole population, and as a function of stellar mass. Furthermore, the majority of quenched / passive galaxies in both simulations are disk-dominated, unlike in the SDSS (see Table 1). Thus, even though Illustris and LGalaxies successfully match many statistical properties of galaxies (e.g. Henriques et al. 2015; Vogelsberger et al. 2014a,b), there is still much work to be done with respect to structure and morphology. \\

In the appendices we present a number of additional checks, tests and resources. Specifically, in Appendix A we discuss the reliability of our bulge-disk decompositions; and in Appendix B we compare structures by mass with structures measured in optical bands. In Appendix C we explore the consistency of alternative ranking schemes for the observational parameters.

\section*{Acknowledgments}

We thank the anonymous referee for many insightful and constructive comments on our manuscript, which have helped to significantly improve the presentation of our results. We also thank Paul Torrey and Steven Bamford for fruitful conversations about this work. 

Support for JM is provided by the NSF (AST Award Number 1516374), and by the Harvard Institute for Theory and Computation, through their Visiting Scholars Program.

Funding for the SDSS and SDSS-II has been provided by
the Alfred P. Sloan Foundation, the Participating Institutions, the National Science Foundation, the U.S. Department of Energy, the National Aeronautics and Space Administration, the
Japanese Monbukagakusho, the Max Planck Society, and the
Higher Education Funding Council for England. The SDSS
Web Site is http://www.sdss.org/

The SDSS is managed by the Astrophysical Research Consortium for the Participating Institutions. The Participating
Institutions are the American Museum of Natural History, Astrophysical Institute Potsdam, University of Basel, University of Cambridge, Case Western Reserve University, University of Chicago, Drexel University, Fermilab, the Institute for Advanced Study, the Japan Participation Group, Johns
Hopkins University, the Joint Institute for Nuclear Astrophysics, the Kavli Institute for Particle Astrophysics and
Cosmology, the Korean Scientist Group, the Chinese Academy
of Sciences (LAMOST), Los Alamos National Laboratory,
the Max-Planck-Institute for Astronomy (MPIA), the Max-Planck-Institute for Astrophysics (MPA), New Mexico State
University, Ohio State University, University of Pittsburgh, University of Portsmouth, Princeton University, the United States Naval Observatory, and the University of Washington.


%
%

\appendix

\section{Reliability of the B/T Measurements}

\subsection{Summary of Prior Work -- Comparison to Simulated Galaxy Images \& The Impact of Pseudo-Bulges}

The original SDSS photometric bulge-disk decompositions for $g$- and $r$- band were performed in Simard et al. (2011) for $\sim$ one million objects. This work was expanded upon to compute bulge-disk photometric decompositions for the other SDSS bands in Mendel et al. (2014). Additionally, Mendel et al. (2014) compute the masses of galaxies and their component bulges and disks via SED fitting to the multi-wavelength bulge-disk decompositions. We refer the reader to these publications for a full explanation of the methods used, and for many detailed checks on the reliability of these structural measurements. Briefly, the accuracy of photometric B/T ratios is estimated to be $\sim$0.15-0.2 per galaxy (depending on wavelength), and the accuracy of bulge and disk masses is estimated to be on average $\sim$0.1 dex, not including issues of choice of IMF or stellar population synthesis code. Including these issues would increase the uncertainty of masses to approximately 0.2 dex.

In Mendel et al. (2014) $\sim$25,000 simulated bulge-disk galaxy images were constructed. The simulated galaxies were re-analysed through the same {\small GIM2D} package as the observational data. Simulated galaxies were placed randomly into realistic SDSS backgrounds, dimmed to their appropriate brightness for randomly selected redshifts drawn from the SDSS distribution, and the images were degraded to the resolution of the SDSS, with background noise introduced. In the appendix of Bluck et al. (2014, hereafter B14) a number of tests were performed using these simulations, which are particularly pertinent for testing the reliability of the bulge-disk decompositions, and their application to the science goals of this paper. We briefly summarise them in the remainder of this sub-section.

In B14, we find that the measured B/T values (in $r$-band) agrees with the input B/T of the simulated galaxy images with an average standard deviation of $\sigma$ = 0.17. Furthermore, the recovery of B/T values is not significantly biased in any direction, with a mean difference between measured and input values of $<$ 0.05. Given the lack of bias, and the very high number of galaxies used in this sample ($\sim$ 0.5 million), the accuracy of mean B/T for a given bin or sample is extremely high. Furthermore, throughout the analyses of this paper we often consider the fraction of galaxies above or below the B/T = 0.5 line (i.e. those which are bulge- or disk-dominated). The accuracy achieved by the bulge-disk decompositions is more than sufficient for this purpose.

In B14, we also test whether the accuracy of measuring model B/T values is significantly affected by galaxy size on-sky (in pixels) or by galaxy apparent magnitude. We find only a very weak dependence on these two observational features, with B/T recovered within an accuracy of $<$0.08 for the faintest and smallest imaged simulated galaxies. We also consider the possibility of measuring erroneously sub-components in single component systems, by analysing bulge-disk decompositions of single S\'{e}rsic images (performed in Mendel et al. 2014). We find that pure S\'{e}rsic model galaxy images with $n > 4$ often have a disk component added in erroneously by the code. These systems (which are pure ellipticals by design) are invariably still classed as bulge-dominated, although they have a mean B/T $\sim$ 0.7 instead of the desired B/T = 1. To combat this we designed a `false disk' correction routine (explained in Section 2, and in more detail in B14). Application of this technique corrects the vast majority of erroneous disks in the sample, and additional checks introduced in Thanjavur et al. (2016) ensure that virtually all real disks remain unchanged. It is important to emphasise here that the false disk corrections have no effect on the fraction of bulge- or disk-dominated systems, since only bulge-dominated systems can have their disks removed in the correction routine.

\begin{figure*}
\includegraphics[width=0.49\textwidth]{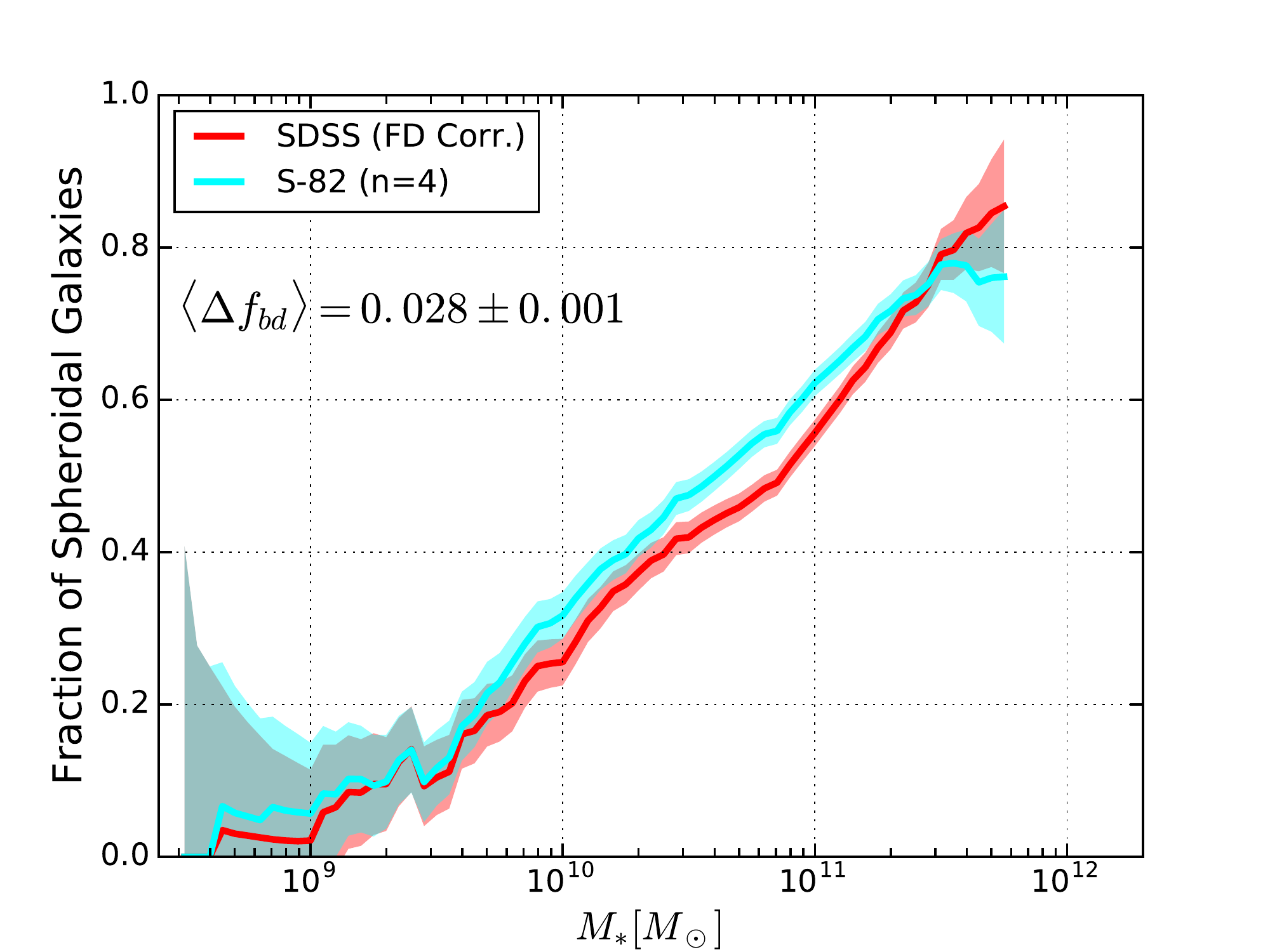}
\includegraphics[width=0.49\textwidth]{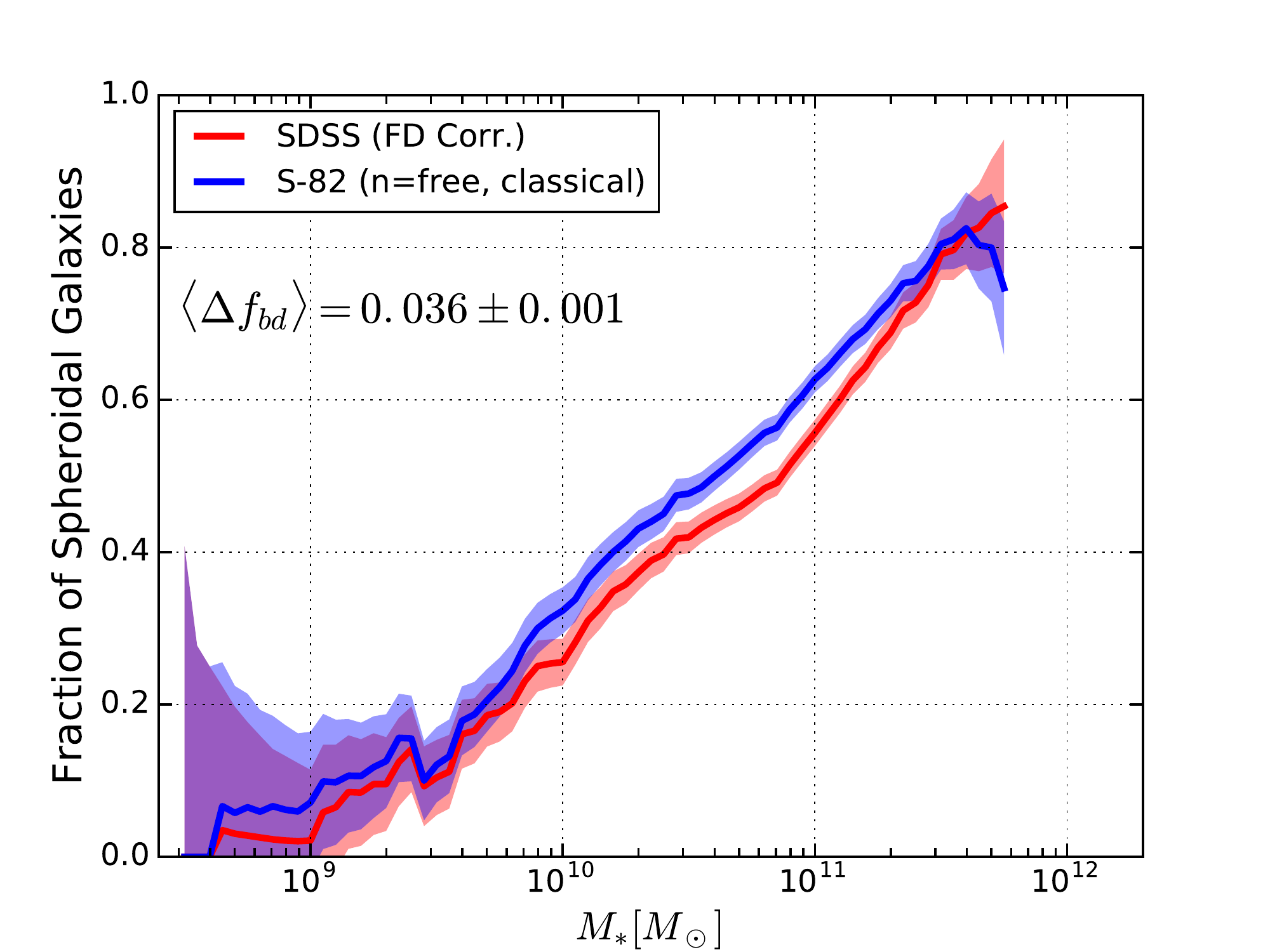}
\caption{The fraction of spheroidal (bulge-dominated) galaxies as a function of stellar mass in the Stripe-82 field. Bulge dominance is defined throughout this figure in $r$-band as: (B/T)$_{r} \geq 0.5$. On both panels, the red lines shows the results for the standard SDSS depth. The cyan and blue lines both show this relationship evaluated with the $\sim$ 2 magnitude deeper Stripe-82 survey. On the left panel the cyan line shows the results from an $n=4$ bulge model decomposition (as in our SDSS sample). On the right panel, the blue line shows the results from a free-$n$ bulge model, restricted to classical bulges (i.e. galaxies with $n_{\rm bulge} < 2$ are set to B/T = 0). The agreement between these relationships are generally very good, with an average difference in the fraction of spheroidal galaxies of $<$ 5\%.}
\end{figure*}

Finally in B14, we also test what impact a free-$n$ bulge (as opposed to $n=4$) model would have on the bulge-disk decompositions. In B14 Fig. D1 we find that at (B/T)$_{n=4} > 0.3$, the free-$n$ and $n=4$ models agree very closely in their measurements. Below this threshold the free-$n$ model yields in general higher B/T ratios than the $n=4$ model. However, crucially, the S\'{e}rsic indices of these galaxies are typically very low, indicating that they are pseudo-bulges not classical bulges (e.g., Kormendy et al. 2010). Ultimately, it is the latter we are concerned with in this paper. Relocating galaxies with a bulge component S\'{e}rsic Index $n<2$ to a B/T = 0 (i.e. restricting to classical bulges only) yields an excellent agreement between free-$n$ and $n=4$ measurements throughout the full range in parameter space (i.e. at all values of B/T, $M_{*}$ and $z_{\rm spec}$). Furthermore, introducing the false disk correction (mentioned above) improves the agreement between the models even more. Thus, for classical bulges in our false disk corrected sample (the sample used in this paper), the $n=4$ and free-$n$ models agree within a mean difference of 0.02 and standard deviation of $\sim$ 0.1.

In summary of this section, many tests of the bulge-disk decompositions used in this paper are already presented in the literature. Taken together, these studies show that our decompositions reliably recover the B/T values of simulated galaxies, imaged as if part of the SDSS. Furthermore, we find the use of a fixed $n=4$ bulge model accurately reproduces the B/T ratios of a free-$n$ bulge model throughout the full range of parameter space, provided we restrict our definition to classical ($n_{\rm bulge} > 2$) systems. As such, in the analyses of this paper we concentrate on classical bulges and ignore pseudo-bulges (treating these as effectively part of the disk). Moreover, given that we concentrate mainly on bulge {\it mass} ratio in this paper, the fact that pseudo-bulges are typically very blue and hence have low mass-to-light ratios further justifies our lack of concern with these objects in the present analysis. In any case, the effect of pseudo-bulges would have very little impact on our results since the B/T ratios of galaxies with $n_{\rm bulge} < 2$ are typically low, i.e less than B/T = 0.5, in which case they {\it are} dominated by a disk, exactly as we treat them to be.

\subsection{Measuring B/T with Deeper Imaging: Stripe-82}

Although we have established the reliability of the bulge-disk decompositions in prior work (discussed above), a persistent concern is the question of whether or not utilising different data would affect the measurement of structure. The most important variable is that of depth of the imaging. Fortunately, the SDSS has a region with substantially deeper imaging than the rest of the survey: Stripe-82. In this region, a follow-up survey to the legacy SDSS achieved an increase in depth of between 1.6-1.8 magnitudes. In Bottrell et al. (2018), a {\small GIM2D} bulge-disk decomposition of 16,700 galaxies is performed in $r$- and $g$-band. Bottrell et al. find measured B/T values from the SDSS-depth data and the deeper Stripe-82 data are similar, typically in agreement within $\sim$ 0.2 (consistent with the errors). Values of B/T near the extremes of the distribution are largely unaffected; however, galaxies with intermediate B/T values in the SDSS are generally scattered to higher values in the deeper data by on average $\sim$ 0.05 - 0.1. Full details on this analysis is provided in the upcoming Bottrell et al. (2018). Overall, most galaxies have similar B/T values in both surveys, and the distribution of B/T is also very similar (see Fig. 7). This is encouraging and suggests that our results should be largely unaffected by depth of imaging.

In this section we briefly consider whether the results of this paper would be significantly affected by increasing the depth of the SDSS. Specifically, we show in Fig. A1 the fraction of spheroidal (bulge dominated galaxies, with B/T $\geq$ 0.5) as a function of stellar mass (evaluated in $r$-band). On both the left and the right panel, the red lines show our fiducial SDSS result, restricted to the Stripe-82 region of the sky. On the left panel of Fig. A1 we additionally show the result from the Bottrell et al. analysis of the deeper Stripe-82 survey for an $n=4$ bulge model. The relationship between the fraction of spheroidal galaxies and stellar mass is very similar between these two data sets. In fact the mean difference is only $\sim$ 0.03. In the right panel of Fig. A1 we show the same comparison between the SDSS and Stripe-82, but this time comparing our $n=4$ bulge model to a free-$n$ model in Bottrell et al.. As with the $n=4$ model, the two relationships agree very well with a mean difference of only $\sim$ 0.035\footnote{Note that this is a positive shift in bulge dominance, which (although small) would slightly worsen the agreement with simulations and models. Given that our primary result is the poor reproduction of galaxy structure in modern simulations, it is important to stress that issues regarding depth do {\it not} help to alleviate this disagreement}.

Fig. A1, taken as a whole, indicates that: 1) deeper imaging would not significantly impact the measurement of structure in this work; and 2) the use of a free S\'{e}rsic bulge model would yield essentially identical results to our $n=4$ model, with the sole restriction that pseudo-bulges (with $n_{\rm bulge} < 2$) are treated exclusively as disk-dominated systems. Additionally, given that the recovery of B/T ratios in simulated galaxy images is performed equally well throughout the range of apparent (on-sky) galaxy sizes, it is unlikely that resolution effects would impact our results either (see Bluck et al. 2014). Thus, in summary, we find our bulge-disk decompositions to be robust to depth, resolution and model methodology. Furthermore, our analysis also benefits from the Simard et al. (2011) / Mendel et al. (2014) structural catalogues being the largest currently available in extragalactic astrophysics. This affords us a significant statistical advantage in analysing the structural dependence on other galactic and environmental properties in this work.

\section{Optical vs. Mass Structures}

\begin{figure*}
\includegraphics[width=0.49\textwidth]{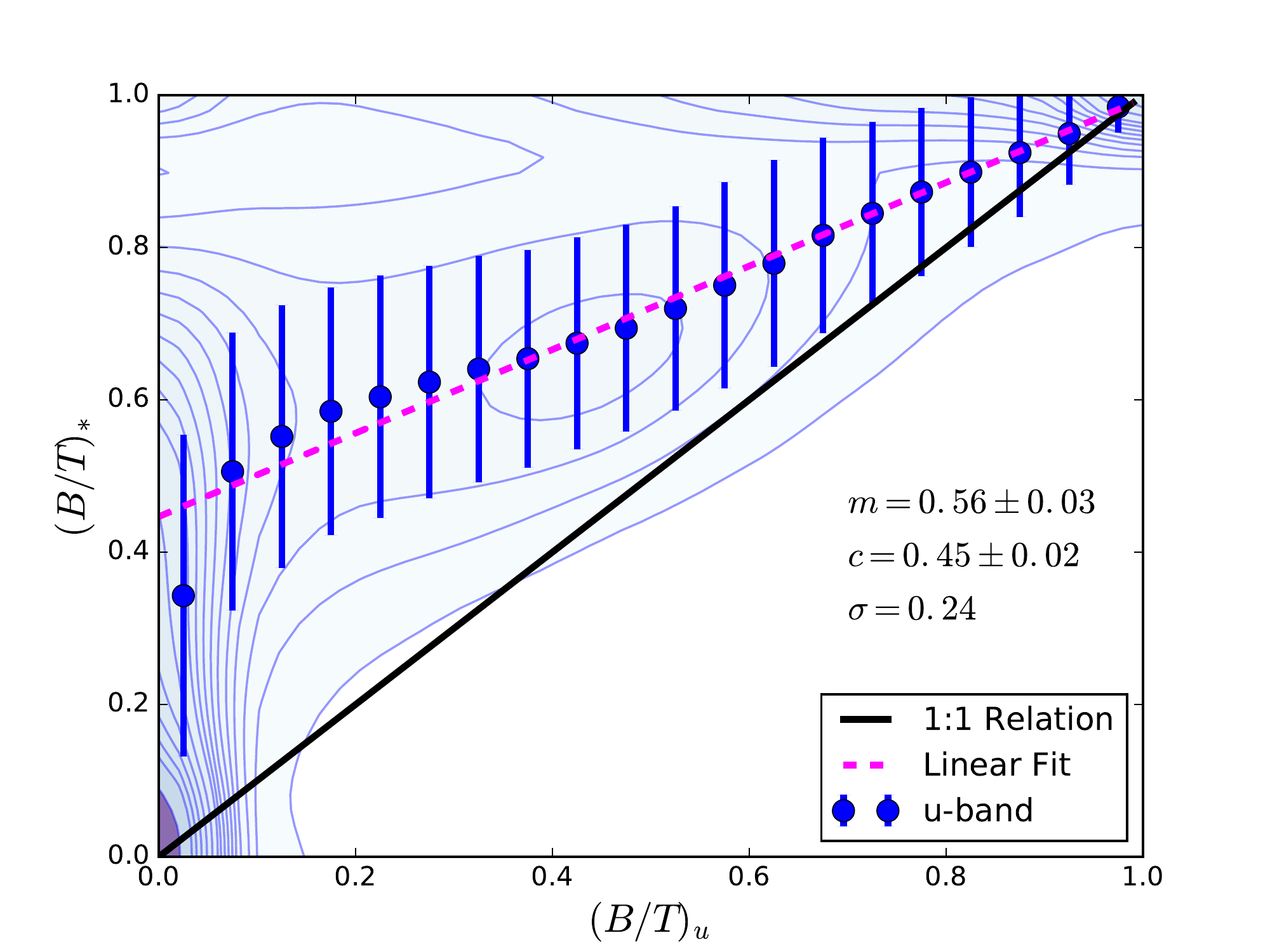}
\includegraphics[width=0.49\textwidth]{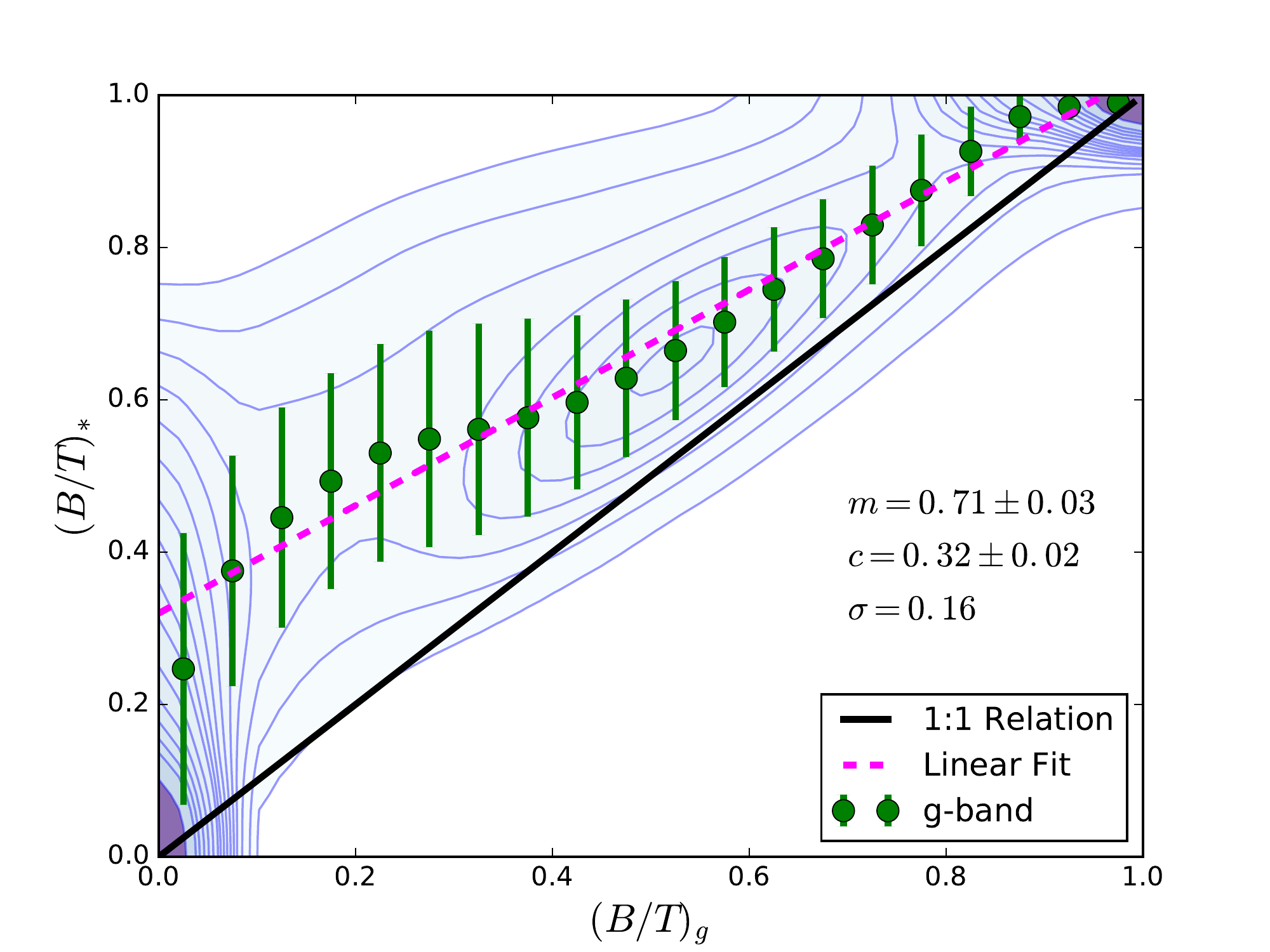}
\includegraphics[width=0.49\textwidth]{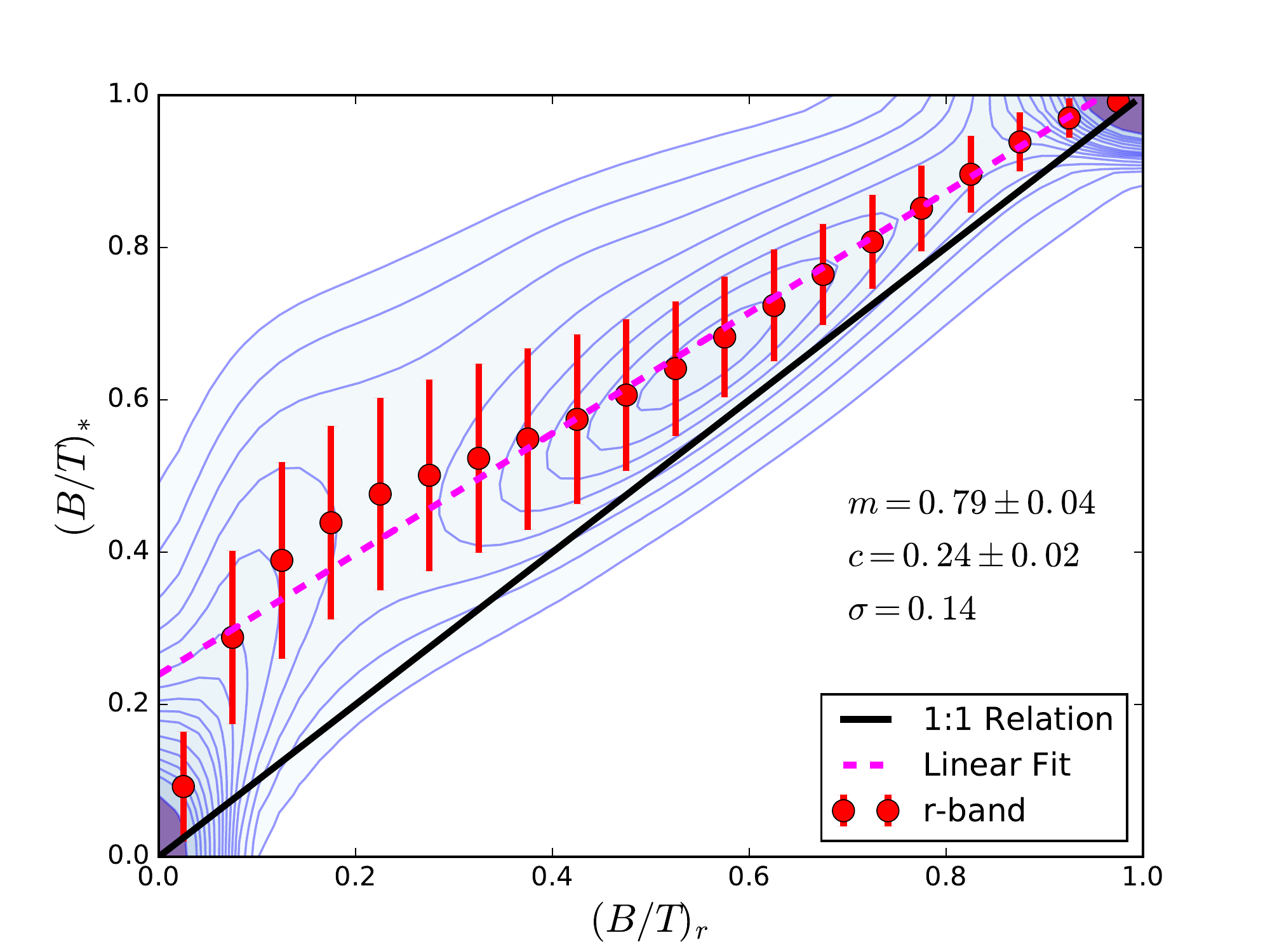}
\includegraphics[width=0.49\textwidth]{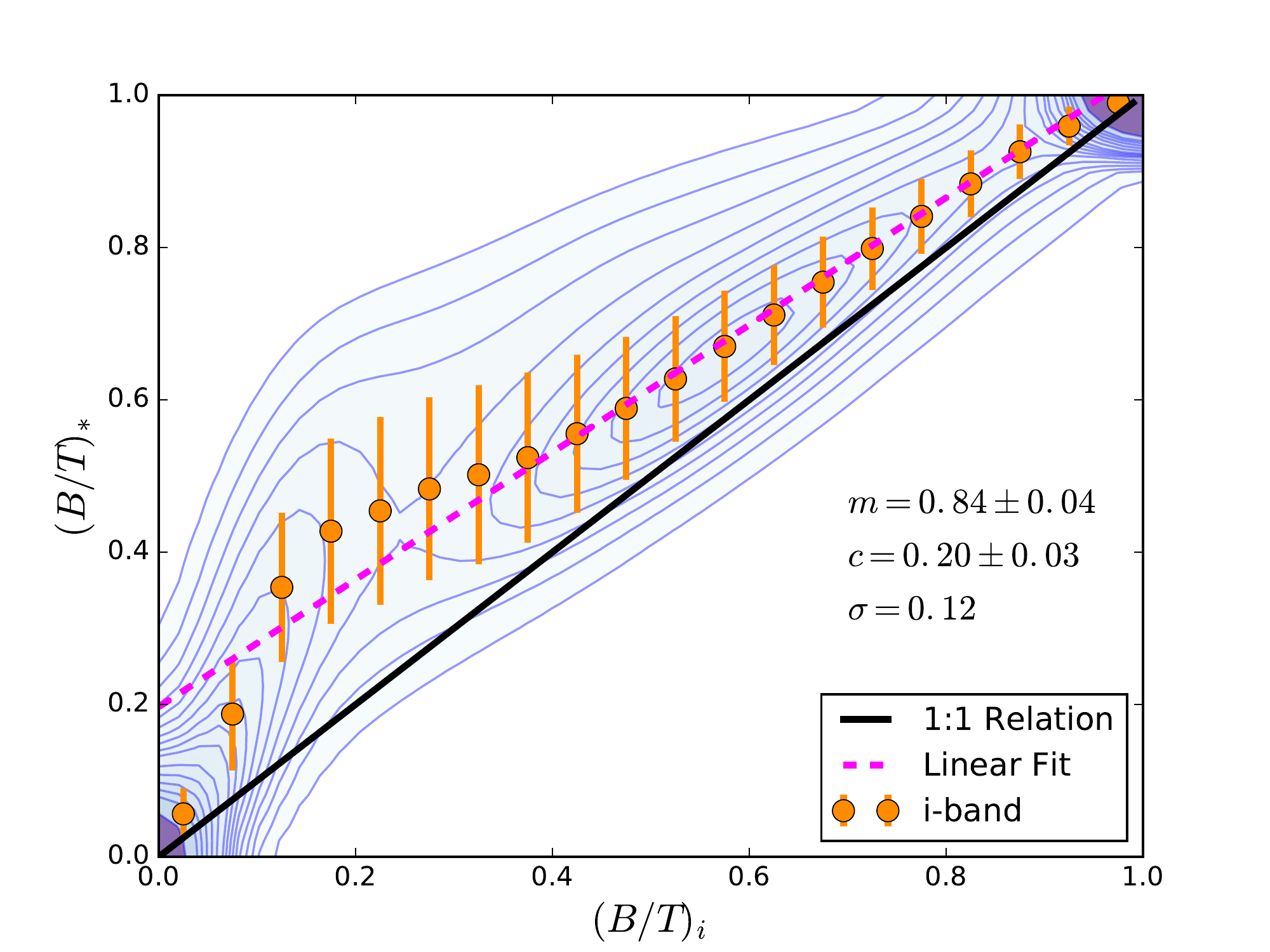}
\caption{The relationship between bulge-to-total stellar mass ratio and, from top left to bottom right: bulge-to-total light ratio measured in $u$, $g$, $r$ and $i$-band. In each panel density contours are shown as faint blue lines, the running median relation is shown as coloured circles, and the inter-quartile range is indicated by the error bars. A least squares linear fit, of the form $y=mx+c$, is produced for each relation (shown as a dashed magenta line, with the parameters shown on each panel). The dispersion ($\sigma$) is indicated on each panel for the linear fit. Additionally, the 1:1 relation is marked as a solid black line on each panel for comparison. In general, stellar mass B/T ratios are higher than photometric B/T ratios, particularly at low to intermediate values. Furthermore, the shift away from the 1:1 line increases with decreasing wavelength of optical waveband: the tightest agreement with stellar mass is found for $i$-band and the poorest agreement is found for $u$-band.}
\end{figure*}

\begin{figure}
\includegraphics[width=0.49\textwidth]{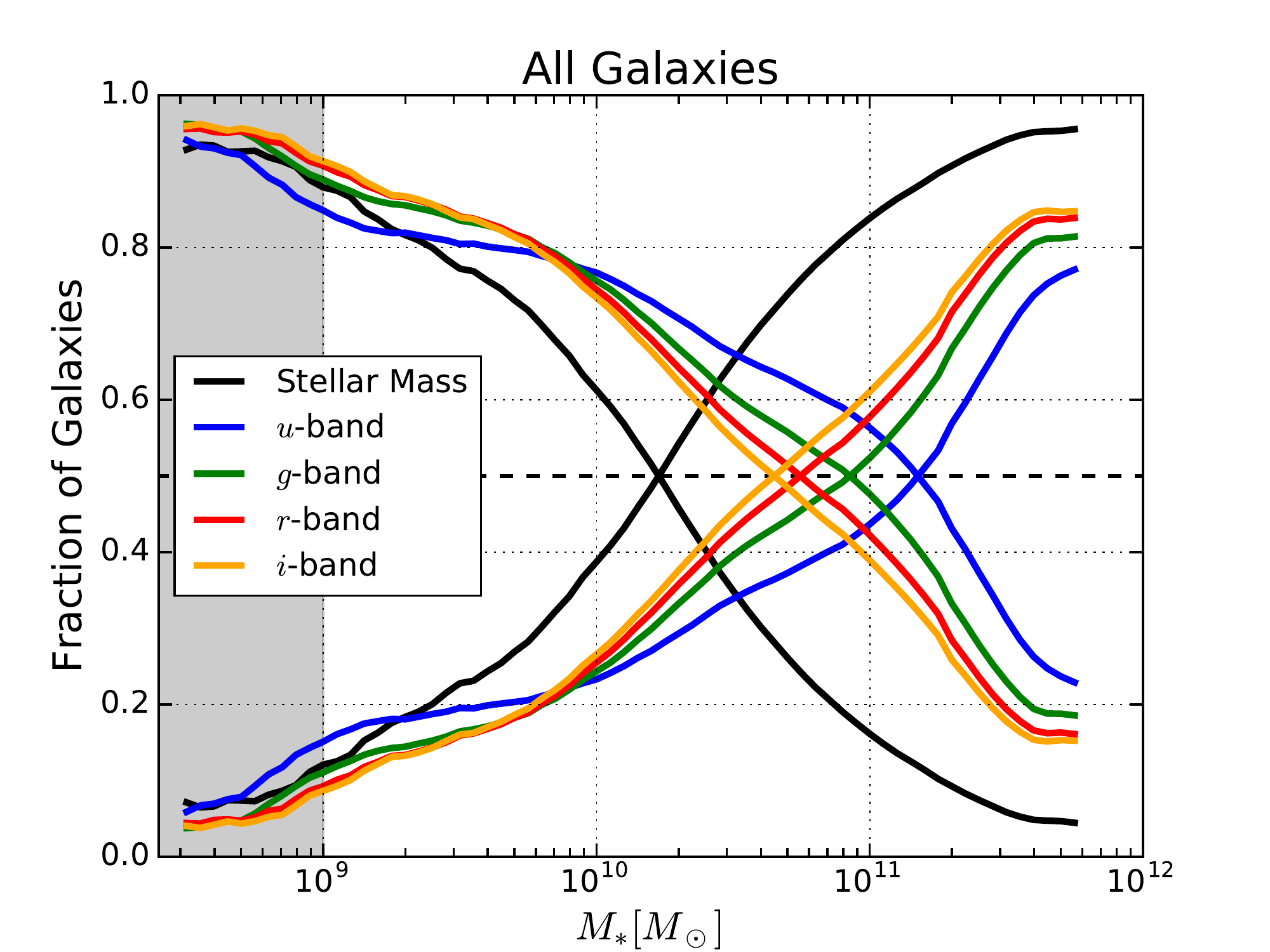}
\caption{The fraction of galaxies which are bulge- and disk-dominated plot as a function of stellar mass, for the full SDSS sample. Bulge dominance is defined as: (B/T)$_{X} \geq 0.5$, where $X$ = stellar mass, and each of the $u$, $g$, $r$, $i$ SDSS wavebands, as indicated on the legend. Typical errors on the structural fractions are $\sim$ 2-5\%, but are omitted on this plot for clarity.}
\end{figure}

In this appendix we investigate how our primary observational results (from Section 3) would change for a photometric definition of B/T instead of the stellar mass ratio we most frequently consider in the main body of the paper.

In Fig. B1, we plot stellar mass bulge-to-total ratio against the bulge-to-total light ratio in each of the $u,g,r,i-$bands. This figure highlights the differences between measures of structure by mass and light. On each panel, the 1:1 line is shown for comparison. In general, B/T by light is lower than B/T by mass throughout the optical spectrum. The magnitude of the discrepancy increases with decreasing wavelength of photometric band. In $i$-band there is a relatively tight relationship ($\sigma$ = 0.12) offset to higher B/T values in mass by $\sim$ 0.2. In contrast, in $u$-band the relationship is both significantly less tight ($\sigma$ = 0.24) and significantly more offset to higher B/T values in mass (by $\sim$ 0.45). At optical-to-near-infrared wavelengths, the redder the band the tighter the correlation with B/T by stellar mass. Although it would be possible to use these relationships to estimate structural parameters by mass from optical wavebands, the results of such an approach would be heavily SED dependent (e.g., different for star forming and quenched systems, as seen in Fig. 9), and hence is {\it not advised}.

Next, we investigate the relationship between the fraction of bulge- and disk-dominated galaxies as a function of stellar mass, for optical wavebands in comparison to mass defined structure. Specifically, in Fig. B2 we show this relationship, with bulge- and disk-dominance defined variously by stellar mass (as in the main body of the paper) and by each of the SDSS wavebands. That is, bulge-dominated galaxies are defined as having (B/T)$_{X}$ $\geq$ 0.5 and disk-dominated galaxies are defined as having (B/T)$_{X}$ $<$ 0.5, where $X$ represents each of the four photometric bands ($u$, $g$, $r$, $i$) and stellar mass as indicated on the legend of Fig. B2.

The crossover from disk to bulge dominance in light (photometric B/T bands) occurs at higher stellar masses than for B/T defined by stellar mass. Furthermore, there is a systematic shift to higher stellar mass at structural crossover with decreasing wavelength of photometric band, whereby the lowest stellar mass at structural crossover is found in $i$-band, and the highest stellar mass at structural crossover is found in $u$-band. This indicates that red light in a galaxy is preferentially originated in bulge structures, and blue light is preferentially emitted from disk structures. Given that the mass-to-light ratio is higher at redder wavelengths in a galaxy, the systematic increase in measured B/T with wavelength naturally explains the higher B/T values in stellar mass than in any optical waveband. The mass distribution of a galaxy is best approximated photometrically by $K$-band (e.g., Cowie et al. 1996, Bundy et al. 2006), and thus one might predict that at this wavelength the optical and mass determined structural fractions will closely coincide.
 
We have performed the optical structural analysis of Fig. B2 on the other parameters shown in Fig. 3 as well (halo mass, local density and sSFR), but we do not show these results here for the sake of brevity. The results for the other parameters are qualitatively similar to that of stellar mass. Specifically, the crossover halo mass of centrals occurs at progressively higher masses with decreasing wavelength, with stellar mass defined structure having the lowest halo mass crossover. For sSFR, the crossover from bulge to disk dominance occurs at progressively lower sSFR values for decreasing wavelength of optical band, and the stellar mass structure definition has a crossover at the highest values of sSFR. All of these results are qualitatively similar to what we have already seen with stellar mass in Fig. B2.
 
Taken together, our results on the structural fraction dependence on waveband highlight the importance of working with stellar mass derived structural quantities, since the wavelength of light analysed significantly affects the measurement of B/T ratio. When comparing to simulations and models, a mass definition of structure should be used in most cases, unless substantial effort is expended to mimic the photometric properties of realistic galaxies (as in Bottrell et al. 2017a,b). When comparing between optical surveys, care must be taken to make comparisons in the same rest-frame wavelength, or, where possible, to convert to a stellar mass defined structure rather than an optically dependent structural parameter. Failure to take the structural dependence on wavelength into account will lead to potentially significant systematic differences between observations at different rest-frame wavelengths.

\section{Alternative Rankings of Parameters}

\begin{figure*}
\includegraphics[width=0.8\textwidth]{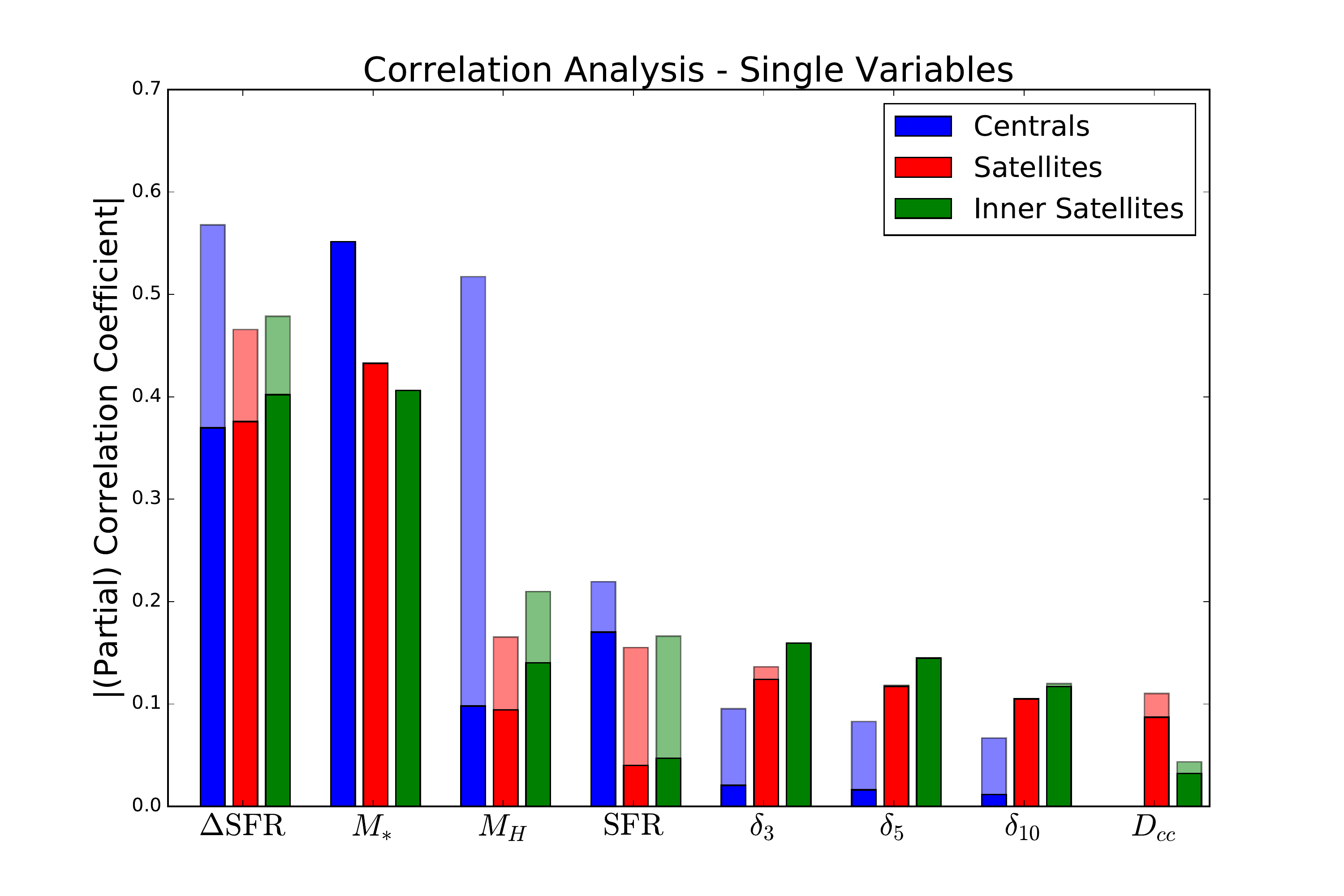}
\includegraphics[width=0.8\textwidth]{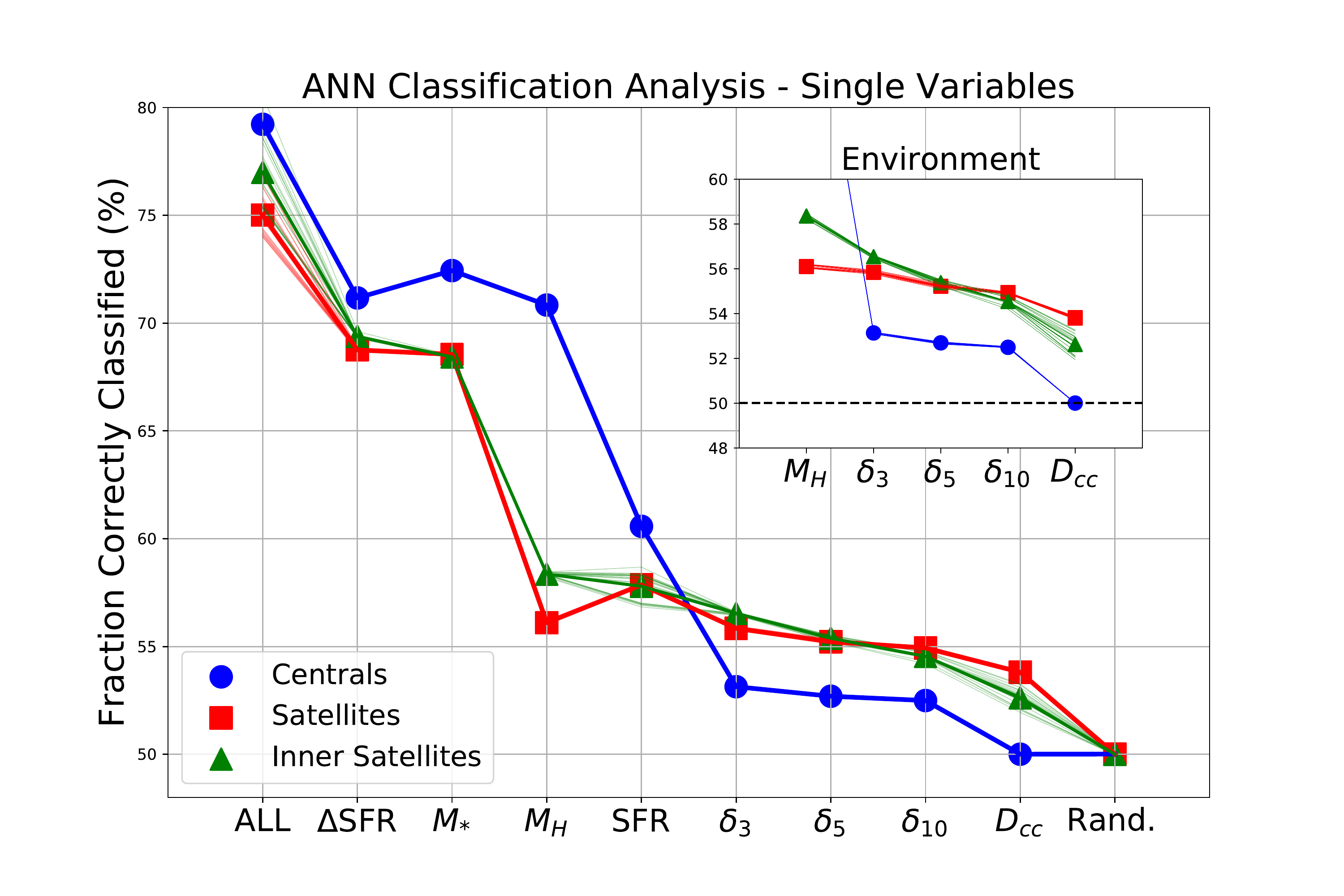}
\caption{{\it Top panel:} Strength of correlation between B/T structure and various galaxy and environmental parameters (as indicated on the $X$-axis). Translucent bars indicate the Spearman rank correlation coefficient (CC) and opaque bars indicate the partial correlation coefficient (PCC), evaluated at a fixed stellar mass. The ordering from most correlated to least correlated closely mimics that found for the regression analysis in Fig. 5. {\it Bottom panel:} Results from an ANN classification of galaxies into bulge- and disk-dominated systems. The $X$- axis shows the variables under consideration, and the $Y$-axis indicates the percentage of correct classifications achieved by the network using each parameter, or set of parameters, in turn. The ranking via classification is very similar to that by correlation (above) and to the ANN regression run, described in Section 4. }
\end{figure*}

\begin{table*}
\begin{center}
\caption{Comparison of Rankings of Variables for Predicting Galaxy Structure}
\begin{tabular}{c c c c | c c c | c c c}
\hline\hline
& & {\bf Centrals} & & & {\bf Satellites} & & & {\bf Inner Satellites} & \\
 & Regression  &  Classification  &  Correlation & Regression  &  Classification  &  Correlation  & Regression  &  Classification  &  Correlation \\
\hline \\
$\Delta$SFR                 & 1 & 2 & 1 & 1 & 1 & 1 & 1 & 1 & 1  \\\\

$M_{*}$             & 2 & 1 & 2 & 2 & 2 & 2 & 2 & 2 & 2  \\\\

$M_{\rm Halo}$ & 3 & 3  & 3 & 5 & 4 & =3 & 3 & 3 & 3  \\\\

SFR                   & 4 & 4  & 4 & 3 & 3 & =3 & 4 & 4 & =4  \\\\

$\delta_5$         & 5 & 5  & 5 & 4 & 5 & =6 & 5 & 5 & =4  \\\\

$D_{cc}$           & -  & -   & - & 6 & 6 & =6 & 6 & 6 & 6  \\\\

\hline
\end{tabular}
\label{tab-data}
\end{center}
\end{table*}

\noindent In this appendix we consider two alternative statistical methods for ranking the observed galaxy and environmental parameters as predictors of galaxy structure. Specifically, we show results from 1) a correlation strength analysis, and 2) an artificial neural network classification analysis. Both of these methods are not identical in methodology or interpretation to the main ANN regression analysis presented in Section 4, hence we ought not to expect identical results. Nonetheless, these approaches are similar in nature (all three aim to assess how closely connected galactic structure is to each parameter in turn) and thus a reasonable level of consistency in their rankings would be advantageous in establishing the underlying physical connection between the parameters.

\subsection{Correlation Analysis}

We perform a correlation analysis of galactic structure (B/T) with the same parameters as used in the regression analysis (Section 4). The input dataset is taken as the full volume weighted sample of galaxies at $9 < \log(M_{*}/M_{\odot}) < 12$. Two complementary approaches are used - 1) the Spearman Rank Correlation Coefficient (CC), which gives the strength of correlation between any two variables; and 2) the Partial Correlation Coefficient (PCC), which gives the strength of correlation between any two variables at a fixed third variable. The advantage of the second approach is that it allows us to hold fixed certain parameters while ascertaining how strongly other parameters affect galactic structure. In this sense the PCC analysis is similar to the $\Delta$(B/T)$_{*}$ analysis of Sections 3, and the CC analysis is closely related to the global trends of Section 3 and the ANN regression rankings of Section 4. 

The PCC is defined (e.g., Bait et al. 2017) as:

\begin{equation}
\rho_{XY.Z} = \frac{\rho_{XY} - \rho_{XZ}\rho_{YZ}}{\sqrt{1-\rho^{2}_{XZ}}\sqrt{1-\rho^{2}_{YZ}}}
\end{equation}

\noindent where, for example, the CC for $X,Y$ is defined as:

\begin{equation}
\rho_{XY} = \frac{{\rm Cov}(X,Y)}{\sigma_{X} \sigma_{Y}}
\end{equation}

\noindent and the weighted covariance is given most generally by:

\begin{equation}
{\rm Cov}(X,Y) = \frac{1}{\sum \limits_{i=1}^{N} w_{i}} \sum \limits_{i=1}^{N} w_{i}  (X_{i} - E(X)) (Y_{i} - E(Y)).
\end{equation}

\noindent $E(X)$ is the weighted expectation value of variable $X$, $\sigma_{X}$ is the weighted standard deviation of variable $X$, and the weight, $w_{i}$, is taken to be the inverse volume over which any given galaxy would be visible in the survey ($1/V_{\rm max}$) which is a function of stellar mass and colour (see Mendel et al. 2014).

In Fig. C1 (top panel) we present the results for our correlation analysis. The $X$-axis shows the same variables as used in the regression analysis (Fig. 5), ordered in the same way. Central galaxies are shown in blue, satellite galaxies are shown in red, and inner satellite galaxies are shown in green. Translucent bars (which reach in general higher than the opaque bars) represent the correlation coefficient (CC), with opaque bars indicating the PCC, evaluated at a fixed stellar mass. Given the definitions above, PCC $\leq$ CC in all cases. For all galaxy types, $\Delta$SFR and stellar mass are the most highly correlated with galactic structure (B/T).  Environmental parameters exhibit in general weaker correlations than intrinsic parameters, for all galaxy types. However, satellites and inner satellites have higher correlations between B/T and environmental metrics than seen in centrals. 

The rankings for the correlation analysis are compared to the rankings from the fiducial regression analysis in Table C1. Within the respective errors of the measurements, all parameters for all types of galaxies are identically ranked for these two statistical approaches. However, in the correlation analysis there is a lack of separation in performance witnessed for the middle ranked variables (i.e., halo mass, SFR and local density), particularly for satellites and inner satellites.  A possible explanation for this small difference, is that the correlations are implicitly assumed to be linear in the (P)CC analysis, whereas the ANN can fit any mathematical function to the data. As such, it is natural to expect that the trained network might be more discerning than a linear fit.

An additional insight we gain from the top panel of Fig. C1 is the strength of correlation at a fixed stellar mass. For all types of galaxies, $\Delta$SFR is highly correlated with B/T, even at a fixed stellar mass. Although much weaker in strength, the correlations between B/T and environment also remain when evaluated at a fixed stellar mass. These findings are qualitatively consistent with the results shown in Section 3.3. Taken together, these results imply that the star forming state and environment of galaxies have a significant impact on the structures of galaxies at a fixed stellar mass.

\subsection{ANN Classification}

In Fig. C1 (bottom panel) we show the results of an ANN classification analysis to predict galaxy structural types, from a set of other galaxy parameters, taken individually (as indicated in the legend). Specifically, we classify galaxies into early-types (with B/T $\geq$  0.5) and late-types (with B/T $<$ 0.5). Although this analysis is similar to the regression analysis, note that it may contain, in principal, different information. In the regression analysis we focus on predicting individual B/T values, whereas in the classification analysis our goal is to determine which category each galaxy ought to fall in, based on a set of parameters other than B/T itself. In the regression analysis, accuracy in B/T determination is equally important at all values of B/T, whereas, in the classification analysis, in general only whether B/T is high or low is important. As such, it is informative to see how the two analyses compare - one might expect them to be similar, but there is certainly no requirement for the rankings to be identical.

The $X$-axis of Fig. C1 (lower panel) shows the same set of variables as utilised in the ANN regression analysis of Section 4, ordered in the same way. The $Y$-axis here shows the percentage of correctly classified galaxies, into early- and late-types. The maximum performance of the network, run with all eight parameters, yields a an accuracy of 75 - 80\%, dependent on galaxy type. Given the intrinsic error in B/T in the training sample ($\sim$ $\pm$ 0.2), this is a reasonably good performance, and would formally be classed as `acceptable/ good' in the machine learning literature (e.g., TBE).  

The raking of parameters in terms of their effectiveness at predicting the structural class of galaxies is almost identical to the regression and correlation cases. These rankings are all shown together for comparison in Table C1. The most significant difference is that, for centrals, $\Delta$SFR and $M_{*}$ switch their ordering in classification compared to regression and correlation. However, both parameters remain superior to any other variable in all approaches. The explanation for this small difference is likely that, in both the regression and correlation analysis, all values of B/T are equally important, whereas in classification it is most important to get the range around B/T = 0.5 correct. Thus, the extremes can be inaccurately measured, so long as the network identifies that they are either high or low. If this is the cause of the discrepancy, most probably this indicates that $\Delta$SFR is more discriminating of B/T in the wings of the distribution, but around the centre of the distribution $M_*$ is more tightly coupled to B/T in centrals than $\Delta$SFR.

The high degree of consistency between all three statistical methods for ranking galaxy and environmental parameters in terms of how closely connected they are to galactic structure is encouraging, and suggests that this ordering may be physical in origin. That is, the most connected properties to B/T structure are potentially causally related in some manner (see Section 6 for further speculation on this point).

\end{document}